\documentclass{emulateapj}

\newcommand{\Lx}{\ensuremath{L_{\mathrm{X}}}}
\newcommand{\Yx}{\ensuremath{Y_{\mathrm{X}}}}

\newcommand{\Msol}{\ensuremath{\mathrm{M_{\odot}}}}

\newcommand{\rf}{\ensuremath{R_{\mathrm{500}}}}

\newcommand{\rc}{\ensuremath{r_{\mathrm{c}}}}
\newcommand{\rd}{\ensuremath{r_{\mathrm{d}}}}

\newcommand{\Mf}{\ensuremath{M_{\mathrm{500}}}}

\newcommand{\Zsol}{\ensuremath{\mathrm{Z_{\odot}}}}

\newcommand{\Mgas}{\ensuremath{M_{\mathrm{gas}}}}

\newcommand{\OM}{\ensuremath{\Omega_{\mathrm{M}}}}

\newcommand{\eg}{{\it e.g.\ }}
\newcommand{\egc}{{\it e.g.}}  
\newcommand{\etal}{et al.\ }

\newcommand{\Chandra}{\emph{Chandra}}

\newcommand{\ROSAT}{\emph{ROSAT}}
\newcommand{\XMM}{\emph{XMM-Newton}}

\newcommand{\MEKAL}{\textsc{MeKaL}}

\newcommand{\chisq}{\ensuremath{\chi^2}}

\newcommand{\gta}{\,\rlap{\raise 0.4ex\hbox{$>$}}{\lower 0.6ex\hbox{$\sim$}}\,}  
\newcommand{\lta}{\,\rlap{\raise 0.4ex\hbox{$<$}}{\lower 0.6ex\hbox{$\sim$}}\,}  

\newcommand{\nm}{\mbox{\ensuremath{\mathrm{~\nm}}}}

\newcommand{\cm}{\mbox{\ensuremath{\mathrm{~cm}}}}

\newcommand{\km}{\mbox{\ensuremath{\mathrm{~km}}}}

\newcommand{\kpc}{\mbox{\ensuremath{\mathrm{~kpc}}}}
\newcommand{\Mpc}{\mbox{\ensuremath{\mathrm{~Mpc}}}}
\newcommand{\s}{\mbox{\ensuremath{\mathrm{~s}}}}
\newcommand{\ks}{\mbox{\ensuremath{\mathrm{~ks}}}}

\newcommand{\keV}{\mbox{\ensuremath{\mathrm{~keV}}}}
\newcommand{\erg}{\mbox{\ensuremath{\mathrm{~erg}}}}

\newcommand{\degree}{\ensuremath{\mathrm{^\circ}}}
\newcommand{\arcm}{\ensuremath{\mathrm{^\prime}}}
\newcommand{\arcs}{\arcm\hskip -0.1em\arcm}

\newcommand{\C}{\mbox{\ensuremath{~\degree\mathrm{C}}}}

\newcommand{\pMpc}{\ensuremath{\mathrm{\Mpc^{-1}}}}
\newcommand{\ps}{\ensuremath{\mathrm{\s^{-1}}}}

\newcommand{\ergps}{\ensuremath{\mathrm{\erg \ps}}}

\newcommand{\kmpspMpc}{\ensuremath{\mathrm{\km \ps \pMpc\,}}}

\newcommand{\YM}{\mbox{\ensuremath{\mathrm{Y_{X}-M_{500}}}}}

\newcommand{\LCDM}{$\Lambda$CDM~}

\begin{document}


\title{Images, structural properties and metal abundances of galaxy
clusters observed with \Chandra\ ACIS-I at $0.1<z<1.3$.}

\author{B. J. Maughan\altaffilmark{1}$^,$\altaffilmark{$\dagger$}}
\author{C. Jones\altaffilmark{1}}
\author{W. Forman\altaffilmark{1}}
\author{L. Van Speybroeck\altaffilmark{1}}
\affil{Harvard-Smithsonian Center for Astrophysics, 60 Garden St, Cambridge, MA 02140, USA.}
\altaffiltext{$\dagger$}{\Chandra\ fellow}
\email{bmaughan@cfa.harvard.edu}

\shorttitle{Structural properties and metal abundances of clusters}
\shortauthors{B. J. Maughan \etal}


\begin{abstract}
We have assembled a sample of 115 galaxy clusters at $0.1<z<1.3$ with
archived \Chandra\ ACIS-I observations. We present X-ray images of the
clusters and make available region files containing contours of the
smoothed X-ray emission. The structural properties of the clusters were
investigated and we found a significant absence of relaxed clusters (as
determined by centroid shift measurements) at $z>0.5$. The slope of the
surface brightness profiles at large radii were steeper on average by
$15\%$ than the slope obtained by fitting a simple $\beta$-model to the
emission. This slope was also found to be correlated with cluster
temperature, with some indication that the correlation is weaker for the
clusters at $z>0.5$. We measured the mean metal abundance of the cluster
gas as a function of redshift and found significant evolution, with the
abundances dropping by $50\%$ between $z=0.1$ and $z\approx1$. This
evolution was still present (although less significant) when the cluster
cores were excluded from the abundance measurements, indicating that the
evolution is not solely due to the disappearance of relaxed, cool core
clusters (which are known to have enhanced core metal abundances) from the
population at $z\ga0.5$.
\end{abstract}


\section{Introduction}
The study of galaxy clusters provides important insight into the formation
of structure in the universe and allows tight constraints to be placed on
cosmological parameters. Observations at different wavelengths probe
complementary aspects of clusters' properties. In optical bands, the
properties of individual cluster galaxies are studied and the
measurement of their velocities enable the estimation of the total binding
gravitational mass, via the application of the virial theorem
\citep[e.g.][]{zwi37}. Observations at optical wavelengths also allow
cluster mass estimates via strong and weak lensing analyses
\citep[e.g.][]{lyn89,tys90}.

At longer wavelengths, radio observations of clusters have important
applications. The inverse Compton scattering of cosmic microwave background
photons by the energetic electrons in the intra-cluster medium (ICM) gives rise
to the Sunyaev Zel'dovich effect \citep[SZE; e.g.][]{bir99}. Observations
of this effect can be used to detect clusters and probe the properties of
the ICM. Radio-emitting jets from active galactic nuclei (AGN) interact
with the ICM, inflating cavities and driving shocks in the X-ray gas, and  
are likely to have an important impact on ICM properties \citep[e.g.][]{mcn00}.

The most sensitive measurements of the ICM properties are currently made at
X-ray wavelengths. The ionised gas of the ICM is the dominant baryonic
component of galaxy clusters, and X-ray observations yield fundamental
properties of the ICM such as X-ray luminosity and temperature, and permit
mass estimates (assuming the gas is in hydrostatic equilibrium) even for
clusters at high redshifts \citep[see][for a review]{sar86}. In the local
universe it has been found that most ($\sim2/3$) clusters host cool cores
\citep{per98,vik06c}. The high gas densities in these core regions give
rise to efficient radiative cooling with corresponding bright
peaks in the X-ray emission and suppressed temperatures. Early models
predicted that large amounts of gas should cool out of the ICM in this
process, although the fate of the condensing material was unknown
\citep[e.g.][]{fab94b}. More recent \XMM\ observations have found that the
majority of the core gas does not cool out of the X-ray
emitting regime, requiring some form of heating to regulate the process
\citep{pet01}. AGN feedback is a strong candidate for providing this energy
input \citep[e.g.][]{nul05}. The gas cooling in cool cores is still
significant however; core temperatures of $\sim30\%$ of the global value
are typical, with detectable cooling extending to $\sim0.15\rf$
\citep[e.g.][]{vik05a,san06a}\footnote{\rf\ is the radius enclosing a mean
overdensity of 500 with respect to the critical density at the cluster's
redshift.}.

While cool cores are common features in relaxed systems, clusters are
observed in a variety of dynamical states, and X-ray observations allow
cluster morphologies to be studied. Early work categorised clusters into
different classes based on their appearance \citep{jon84,jon99}, while more
quantitative methods have been developed enabling cluster morpholgies to be
correlated with other properties \citep[e.g.][]{buo95,moh93}. Using such
methods, evidence has been found that clusters are generally less relaxed
at $z>0.5$ \citep{jel05}. X-ray imaging analysis has also been used
recently to demonstrate that the fraction of cool core clusters in the
population drops from $\approx65\%$ locally to $\sim15\%$ at $z>0.5$
\citep{vik06c}. These two observational findings are consistent with the
higher merger rates expected at high-redshifts.

Further insight into the ICM properties can be gained from the emission
lines in the X-ray spectra of clusters. These are due to significant
amounts of heavy elements in the gas. The observed metal abundances are
consistent with models of the enrichment of the ICM by supernovae based on
observed star formation rates \citep{ett05}. X-ray observations have
enabled spatially resolved abundance measurements in clusters, showing
sharply central peaks in cool-core clusters, and only mild abundance
gradients in non cool-core clusters \citep[e.g.][]{deg01}. Recent work has
also found lower average ICM metal abundances at high redshifts than
locally
\citep{bal06a}.

The \Chandra\ X-ray observatory, with its high spatial and spectral
resolution, is well suited to the study of galaxy clusters. Imaging analyses
and spatially resolved spectroscopy of the ICM can be performed without the
complications of point source contamination and deconvolution of the
telescope point spread function (PSF). An important contribution of
\Chandra\ to the study of galaxy clusters is its large public data
archive. This enables the construction of large samples of clusters whose
properties are determined in a consistent manner, allowing powerful
statistical studies of cluster properties, their correlations, and their
evolution.  In this {\it paper} we present a catalogue of 115 galaxy
clusters observed by \Chandra. The construction of the sample and the
data reduction methods are detailed in  \textsection \ref{s.cat} and
\textsection \ref{s.reduction}. Our image  analysis and X-ray images of the
clusters are presented in \textsection \ref{s.img} and the spectral
analysis and spectral properties of the clusters are given in \textsection
\ref{s.spec}. Notes on individual clusters are given in \textsection
\ref{s.notes} and the results are presented and discussed in \textsection
\ref{s.results}. A \LCDM cosmology of $H_0=70\kmpspMpc$
($\equiv100h\kmpspMpc$), and $\OM=0.3$ ($\Omega_\Lambda=0.7$) is adopted
throughout and all errors are quoted at the $68\%$ level.

\section{The sample}\label{s.cat}
The sample was assembled from all publically available \Chandra\ data as of
November 2006. The positions of all \Chandra\ pointings were
correlated with the NASA/IPAC Extragalactic Database (NED), and observations
with a galaxy cluster or group listed in NED within $10\arcmin$ of the
\Chandra\ aimpoint were kept. This list of observations was then refined by
selecting only those for which the detector was ACIS-I, a galaxy group or
cluster was the target of the observation, and the object had a published
redshift greater than $0.1$. No lower mass limit was applied to the target
objects, but the lower redshift cutoff and the choice of the ACIS-I
detector limits the number of galaxy group size objects in the sample. This
combination of detector and lower redshift cutoff was chosen to ensure that
the cluster emission within the radius \rf\ (our measurement of
\rf\ is discussed in
\textsection \ref{s.spec}) was within the field of view, allowing
properties to be measured to that radius without
extrapolation. Quantitatively, we required that $>50\%$ of the area in the
annulus $(0.9-1.0)\rf$ intersected with the active ACIS-I CCDs. All
clusters at $z>0.1$ met this requirement, with the exception of
$PKS0745-191$ (ObsID 6103) at $z=0.103$, which was rejected from the sample
accordingly.  Observations were then examined individually and rejected if
insufficient useful data were left after lightcurve filtering (described in
\textsection \ref{s.reduction}), if they were affected by low level
background flares (identified by inspection of the spectrum of the local
background, as described in \textsection \ref{s.spec}), or if no
significant extended emission was detected at the cluster position. The
latter was the case for a few candidate high-redshift clusters for which
detections in earlier data (generally from \ROSAT) were resolved into
distinct point sources by \Chandra.

\begin{figure}
\begin{center}
\scalebox{0.33}{\includegraphics*[angle=270]{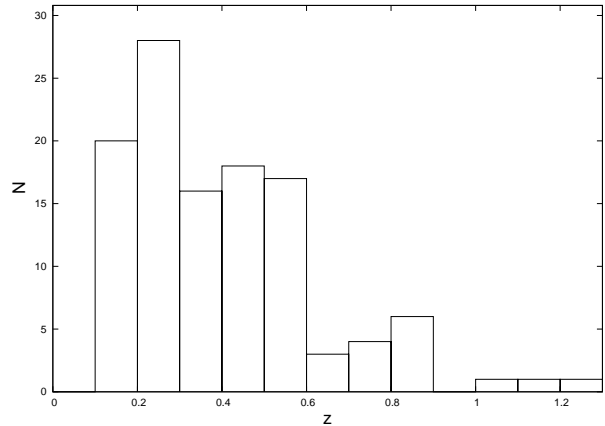}}\\
\caption{\label{f.zhisto}Redshift histogram of the final sample.}
\end{center}
\end{figure}

The final sample then consisted of 149 observations of 115 galaxy clusters.
The individual observations are summarised in Table \ref{t.cat}. The
clusters are listed in order of right ascension, and the positions given
are the ICRS equatorial coordinates of the X-ray centroids. Column 2
gives the \Chandra\ observation identifier for each observation and column
5 gives the redshift of each cluster (redshifts are taken from NED or BAX,
the X-Rays Clusters Database). In columns 6 and 7 the
date of each observation and corresponding ``blank-sky'' background
period\footnote{See \url{http://cxc.harvard.edu/contrib/maxim/acisbg/}} are
given. Finally, column 8 lists the cleaned exposure time of each
observation (the data cleaning process is described in \textsection
\ref{s.reduction}). The redshift histogram of the final sample is shown in
Fig. \ref{f.zhisto}.

The clusters form a heterogeneous collection, having been observed for
different purposes. Nearly half of the clusters (55) were observed in the
guaranteed time observations of Leon Van Speybroeck as part of a program to
combine X-ray data with SZE observations in order to place constraints on
$H_0$ and $\OM$ \citep{bon06}. One of the main goals in the current {\it
paper} is to present a catalogue 
of the properties of this rich sample of clusters. The standardised
methods that were used to analyse the data ensure that the results
presented are self-consistent and can be duplicated in, for instance,
comparative analyses of ``mock observations'' derived from numerical
simulations. In follow-up papers we will investigate the X-ray scaling
relations for this sample, including their evolution and scatter, and look
in detail at clusters that deviate strongly from the trends measured for
the sample as a whole.

\section{Data Reduction}\label{s.reduction}
Initially, the level 1 event files were reprocessed with the latest
calibration (as of September 2006).  If the focal plane temperature of the
observation was cooler than $-118.5\C$ then the charge-transfer
inefficiency and time-dependent gain corrections were applied. For early
observations with warmer focal plane temperatures, only the standard gain
correction was used.

Background flare filtering was then performed with {\it lc\_clean} in the $0.3-12\keV$
band for data from the ACIS-I CCDs, excluding the target CCD and all
sources detected in the standard \Chandra\ X-ray centre pipeline
processing. Lightcurves were cleaned by $3\sigma$ clipping and periods with
count rates $>20\%$ above the mean rate were rejected. This is consistent
with the cleaning applied to the blank-sky background datasets. The
background dataset appropriate for the date of each observation was chosen,
and if background period D was used, and the observation was taken in
VFAINT telemetry mode, then the additional VFAINT cleaning procedure was
applied to the source and background
datasets\footnote{\url{http://asc.harvard.edu/ciao/threads/aciscleanvf}}.

Finally, if the observation was not taken in the same detector
configuration as the background files (ACIS-01236), then CCDs other than
ACIS-0123 were removed from both source and background datasets. Otherwise
CCD 6 was left in both the source and background files.

The use of blank-sky background data means that background data is
available for the same detector regions as the source data. This eliminates
systematic differences between the source and background data in both
imaging and spectral analysis. However, the background consists of a
particle-induced component which dominates at $>2\keV$ and varies with
time, and a soft Galactic X-ray component which varies spatially over the
sky. The contributions of these components in the blank-sky background will
be different from those in the source dataset. These differences are
accounted for in the imaging and spectral analysis methods described below.

An additional background dataset was produced for each observation to
enable the subtraction of the ACIS readout artifact. While a CCD is being
read out, source photons are still detected, creating a streak of source
photons along the readout direction. The ratio of the readout time to the
exposure time of one frame is $1.3\%$, so $1.3\%$ of the source
photons will be affected. Following the method described by \citet{mar00c},
we account for this effect by generating an additional background dataset
from the source dataset, with the photon CHIP-Y positions randomised and
coordinates and energies recalculated. This normalisation of this readout
background is set to $1.3\%$ of that of the source, and the blank-sky
background normalisation is reduced accordingly. We note that the readout
artifact was only significant for the clusters with very bright central
peaks in their surface brightness distribution, but this correction
procedure was applied to all data for consistency.

\section{Image Analysis}\label{s.img}
\subsection{Image preparation}
Exposure maps were generated at an energy of $1.5\keV$, chosen to
correspond to the peak ACIS effective area. Point and extended
sources in the field were detected using the wavelet detection algorithm of
\citet{vik98b}. A broad band ($0.3-7\keV$) exposure-corrected image was
used for point source detection, while the $0.7-2\keV$ band was used for
the detection of extended emission. The latter energy band was chosen to
optimise signal to noise for typical cluster spectra and was used for all
of the following imaging analysis. The centroid of the cluster emission was
then located by refining its position with 5 iterations within $150\arcs$
and then 5 iterations within $50\arcs$ centered on the previous centroid
(with point sources masked out). The target coordinates for each
observation were used as the starting point for the iterations. This
centroid was then used to define the position of the cluster throughout the
analysis.

Extended sources that were detected as distinct separate sources compared
to the target cluster were excluded, but any substructure that was detected
in the cluster was left untouched. The division between separate extended
sources and cluster substructure is not always simple, and in many cases
redshifts were not available for different emission components. The thirteen
clusters with excluded extended sources within \rf\ (in projection) are
noted in \textsection \ref{s.notes}.

Exposure-corrected radial profiles of the source and background images were
then produced, with point sources excluded. An iterative process was
followed to define the outer detection radius of the cluster emission and
correctly normalise the background image in the imaging energy band. We
defined the detection radius ($\rd$) as the outer radius of the last bin
with a width between $5\arcs$ and $150\arcs$ in which emission was detected
at $\ge0.5\sigma$. While this is a weak criterion for detection, our
purpose here is to be conservative in defining a source-free region in each
observation, and the detection radii were confirmed by visual inspection of
the source and background radial profiles. In each step of the iteration,
the detection radius was determined, and a background scaling factor was
defined as the ratio of source and background count rates in the imaging
band, with the emission within \rd\ excluded. The background image was
renormalised by this factor, and a new detection radius was computed. The
process was repeated until \rd\ converged to within $1\%$. This method
provides a robust normalisation of the background in the imaging band, and
definition of the background emission in the source image.

The key assumption of this method is that some of the field of view
is free of cluster emission. Visual inspection of the source and background
radial profiles showed that this was the case even for the brightest clusters
in the sample.

\begin{figure}
\begin{center}
\scalebox{0.33}{\includegraphics*[angle=270]{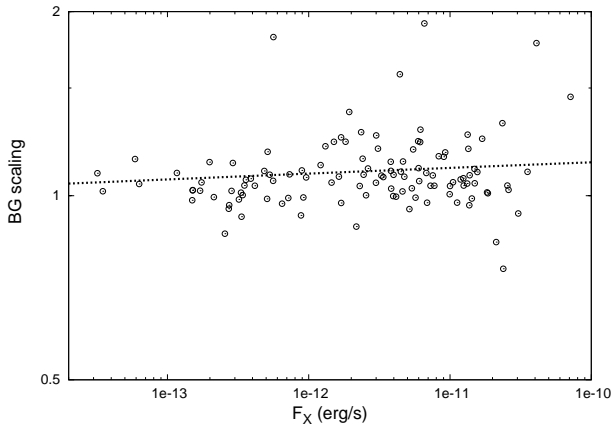}}\\
\caption{\label{f.bgfx}The $0.7-2\keV$ background scaling factor is plotted
against cluster flux. The dotted line is the relation expected due to a
selection effect of fainter clusters being prefferentially observed in low
background fields. The clusters with very high scale factors are those in
regions of high Galactic foreground emission.}
\end{center}
\end{figure}

If the cluster emission were not fully excluded in this process, then the
remaining cluster emission would cause the soft background level to be
overestimated. This would result in a correlation between the derived
background scaling and the cluster flux. This is plotted in
Fig. \ref{f.bgfx}, with the background scaling defined as the ratio of the
source to blank-sky count rates in the background region in the imaging
band ($0.7-2\keV$). The cluster fluxes were measured from the spectral
analysis, using a different background correction (a scale factor derived
at high energies; see \textsection \ref{s.spec}) and so should be
independent of the soft background scaling factor used here. A Spearman's
rank test gives a $\sim1\sigma$ correlation between background scaling and
flux. However, this can be caused by a selection effect whereby fainter
clusters are preferentially observed in fields with lower soft X-ray
background levels.

This was tested by correlating the cluster fluxes and background scaling
factors with the $0.75\keV$ X-ray background flux measured in the \ROSAT\
all-sky survey \citep{sno97} in an annulus of $0.5\deg-1\deg$ centered on
each cluster. We found significant positive correlations between the soft
X-ray background and both the cluster flux (at $93\%$ probability) and our
background scaling factor ($99.99\%$). These two relations were then
combined to predict the expected relation between background scaling and
cluster flux under the hypothesis that any relation between those two
variables is only due to the association of fainter clusters with lower
background regions. This predicted relation is plotted in Fig. \ref{f.bgfx}
and reproduces the weak trend found in the data. When this artifical
correlation is factored out, any remaining trend is extremely weak, placing
a limit of $<1\%$ on the level of contamination of the background region by
cluster emission. Note also that the clusters with
large scale factors ($\ga1.4$) in Fig. \ref{f.bgfx} are those with very
high levels of Galactic foreground emission, and are not affected residual
cluster emission. This is illustrated in \textsection \ref{s.spec} and is
corrected for in the imaging analysis by the simple scaling of the
background image.

\subsection{Surface brightness modelling}
The background-subtracted, exposure-corrected X-ray surface brightness
($S_X$) profiles were then binned so that all radial bins were significant
at the $3\sigma$ level and were fit with a simple one dimensional (1D)
$\beta$-model of core radius \rc\ and outer slope $\beta$ \citep{cav76},
\begin{eqnarray}
S_X(r) & = &
S_X(0)\left(1+\left(\frac{r}{r_c}\right)^2\right)^{-3\beta+1/2}.
\end{eqnarray}

Images of the clusters were then reduced in size to a square of side $4\rf$
($\rf$ is defined in \textsection \ref{s.rv}) centred on the cluster
centroid and fit with a two-dimensional (2D) elliptical $\beta$-model. The
ellipticity of the model was defined as $e=(1-b/a)$, where $a$ and $b$ are
the semi-major and -minor axes respectively. The background was modeled
with a constant level, which was fit to the normalised background
image. All models were convolved with an image of the telescope PSF
generated for the detector position of each cluster centroid at an energy
of $1.5\keV$, and multiplied by the exposure map.

It is well known that the $\beta$-model is not generally a good description
of cluster surface brightness profiles, with additional components required
to match the sharp surface brightness peaks found in the cores of many
clusters \citep[\egc][]{pra02,jon84} and evidence for steepening at large
radii \citep{vik99}. The model retains some appeal, however, as it is
simple to deproject analytically to estimate the gas density
distribution. As discussed in \textsection \ref{s.mgas}, we use a more
complex model to fit the cluster emission and derive gas density
profiles. The $\beta$-model is used solely to allow us to investigate the
steepening of the profiles at large radii, and provide a first measurement
of the cluster morphologies via their ellipticity.

In order to measure the local surface brightness slope at \rf\
($\beta_{500}$), the background-subtracted, exposure corrected profiles
were grouped into logarithmic bins. A straight line was fit in log space to
the data in the radial range $(0.7<r<1.3)\rf$ giving the best fit slope and
its uncertainty. This slope is independent of the core properties of the
clusters. For 8 of the
faintest clusters, the emission was not detected significantly at $1.3\rf$, so
$\beta_{500}$ was not be measured for these clusters. The logarithmic slope is related to the slope of the standard
$\beta$-model by $d\log(S_X)/d\log(r) = 1-6\beta$ (for $r>>r_c$). We thus
defined $\beta_{500}$ as
\begin{eqnarray}
\beta_{500} & = & \frac{1}{6}\left(1-\frac{d\log(S_X)}{d\log(r)}\bigg|_{r=\rf}\right).
\end{eqnarray}
The same procedure was used to measure $\beta_{mod}$, the slope at \rf\ of
the 1D $\beta$-model that was fit to the entire profile. The measured
ellipticities and surface-brightness profile slopes are are given in Table
\ref{t.morpho}.

\subsection{Gas density profile}\label{s.mgas}
The X-ray emissivity of the intra-cluster medium depends strongly on the
gas density and only weakly on its temperature. This means that the
observed projected emissivity profile can be used to accurately measure the
gas density profile. For each annular bin in the surface brightness
profile, the observed net count rate was corrected for area lost to chip
gaps, bad pixels and excluded regions, and converted to an integrated
emission measure. The conversion factor was calculated for each bin
assuming an absorbed \MEKAL\ \citep{kaa93} plasma model folded through an
ARF generated for that region, and an on-axis RMF. The absorption of the
spectral model was set at the galactic value and the metal abundance was
set at $0.3$. As the data are not sufficient to allow temperature profiles
to be measured for most of the clusters in the sample, the temperature of
the spectral model in each radial bin was obtained from the mean
temperature profile found by \citet{vik05a}, normalised to the global
temperature measured for each cluster (see \textsection \ref{s.rv}). As
noted above, the dependence of the conversion from count rate to emission
measure on the assumed temperature is weak. The conversion factor decreases
by $\sim15\%$ when the model temperature is increased from $3\keV$ to
$10\keV$, with most of the decrease occurring below $\sim4\keV$. Our
analysis was repeated assuming the cluster was isothermal for the purpose
of deriving the emission measures, and the results were not significantly
changed.

The resulting profiles were fit with a modified version of the standard 1D
$\beta$-model as proposed by \citet[][herafter V06]{vik06a} \citep[see][for other variations on the $\beta$-model]{pra02};
\begin{eqnarray}\label{e.emis}
n_pn_e & = & \frac{n_0^2 (r/r_c)^{-\alpha}}{(1+r^2/r_c^2)^{3\beta-\alpha/2}} \times (1+r^\gamma/r_s^\gamma)^{-\epsilon/\gamma}.
\end{eqnarray}
The first term modifies a $\beta$-model to include a power-law cusp with
slope $\alpha$ at $r<<r_c$. The second term allows a change of slope by
$\epsilon$ at radius $r_s$ ($r_s>r_c$) across a transition region whose
width is controlled by $\gamma$. Following V06, $\gamma$ was fixed at 3 for
all fits. In order to simplify the model slightly, so that it could be used to
fit the range of high and lower quality data in the sample we
ignore a second $\beta$-model component included in the V06 models for some
of their clusters. 

This model was then projected along the line of sight and fit to the
observed projected emission measure profile. The effect of the small
\Chandra\ PSF was neglected in this analysis. There are strong correlations
between the parameters in the emission measure model so the individual
parameters are not reported here. However, the gas masses derived from the
best fitting models are given in Table \ref{t.ap}, with uncertainties
determined from Monte-Carlo randomisations of the observed emission measure
profiles.

\subsection{Centroid shifts}
Various methods have been used to quantify the amount of substructure in
cluster images. Power ratios
derived from different moments of the surface brightness distribution have
been found to be a useful diagnostic of cluster morphology
\citep{buo95}. Centroid shifts measured from the variation in separation
between the peak and centroid of cluster emission with aperture size have
similarly been used to measure cluster morphologies
\citep[\egc][]{moh93,oha06}. \citet{poo06} used numerical simulation of cluster
mergers to test the effectiveness of these methods and found that the
centroid shift method was the most sensitive to the dynamical state of
clusters and the least sensitive to noise in the cluster images. We thus
adopt the centroid shift as our preferred statistic.

Centroid shifts were measured following the method of \citet{poo06}. The
centroid of the cluster emission was determined in a series of circular
apertures centred on the cluster X-ray peak. The radius of the apertures
were decreased in steps of $5\%$ from \rf\ to $0.05\rf$ and the centroid
shift, $\langle w \rangle$, was defined as the standard deviation of the
projected separations between the peak and centroid in units of \rf. To
increase the sensitivity of this statistic to faint structure, the central
$30\kpc$ was excluded for the centroid (but not the X-ray peak)
measurements. The measured centroid shifts are summarised in Table
\ref{t.morpho}. The errors quoted for the $\langle w \rangle$ are the
statistical uncertainties on a standard deviation calculated from $n$
measurements ($\langle w \rangle/(2n-2)$).

Centroid shifts were measured from exposure-corrected images in order to
eliminate the effects of chip gaps, and vignetting. The high spatial
resolution of \Chandra\ means that point sources could be excluded without
significantly biasing the measured centroid position. As the purpose of the
statistic is to quantify substructure, all extended sources were left
untouched. The redshifts of many of the extended sources detected around
the target clusters are unknown, so those sources known to be foreground or
background objects were not excluded in order to treat all clusters
consistently. In fact, as detailed in \textsection \ref{s.notes}, only $13$
of the 115 clusters had distinct extended sources that are projected within
\rf\ of the cluster core and just one of those (V1701+6414) is known to be
at a redshift different from the target cluster.

\subsection{Contour maps}
In order to examine the morphologies of the clusters, contour maps of the
X-ray emission were generated using images that were adaptively smoothed to
$3\sigma$ significance with the smoothing algorithm {\it asmooth}
\citep[][we modified the original algorithm slightly to include exposure
correction]{ebe06a}. The raw images were produced in the $0.7-2.0\keV$ band and
were binned by a factor of 2, giving approximately $1\arcsec$
pixels. The images were then cropped to a box $3\Mpc$ on a side
centred on the cluster's X-ray centroid. To correct for
vignetting, CCD gaps and exposure variations in combined observations,
exposure maps were provided to the smoothing algorithm.

To enable the statistical significance of morphological features to be
determined from the contour plots, we selected contour levels to bound
regions that were detected at a significant level above the local
emission. The methodology is explained in detail in
\citet{mau07a}, but briefly, the emission enclosed between a contour and
the one above it is detected at $>3\sigma$ significance (after exposure
correction) with respect to
the emission between the contour and the one below it. In the case of the
lowest contour, the emission is detected at  $>3\sigma$ above the
background emission in the same region (measured in the blank-sky images).
Note that this contouring scheme means that contour levels
cannot be interpreted in the same way as in standard contour plots; separate
contours that are the same number of levels above the background are
unlikely to correspond to the same flux level in the smoothed image.

\begin{figure*}
\includegraphics[angle=0,scale=0.60, bb=50 210 460 565]{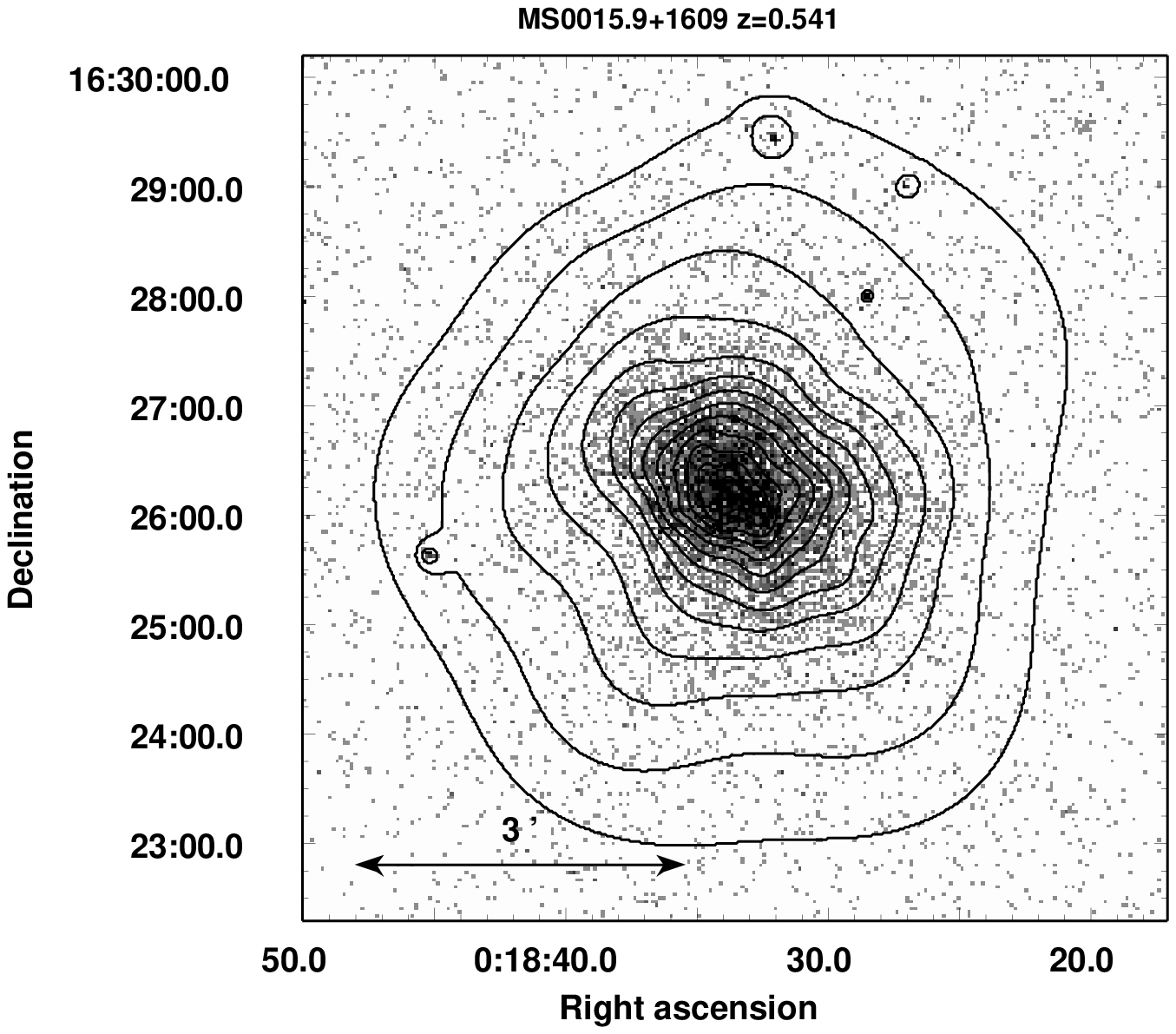}
\includegraphics[angle=0,scale=0.60, bb=50 210 460 565]{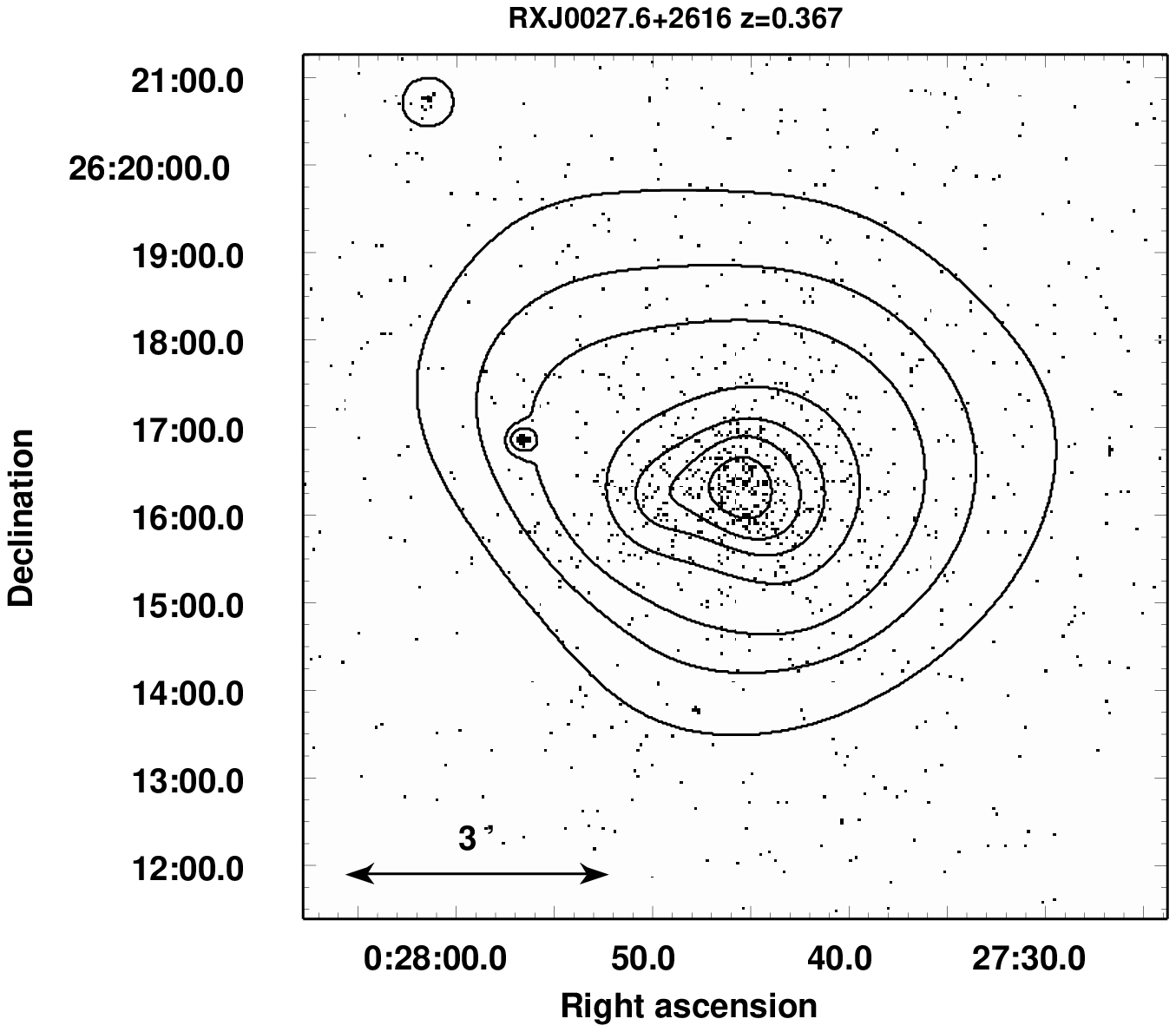}\\
\includegraphics[angle=0,scale=0.60, bb=50 210 460 565]{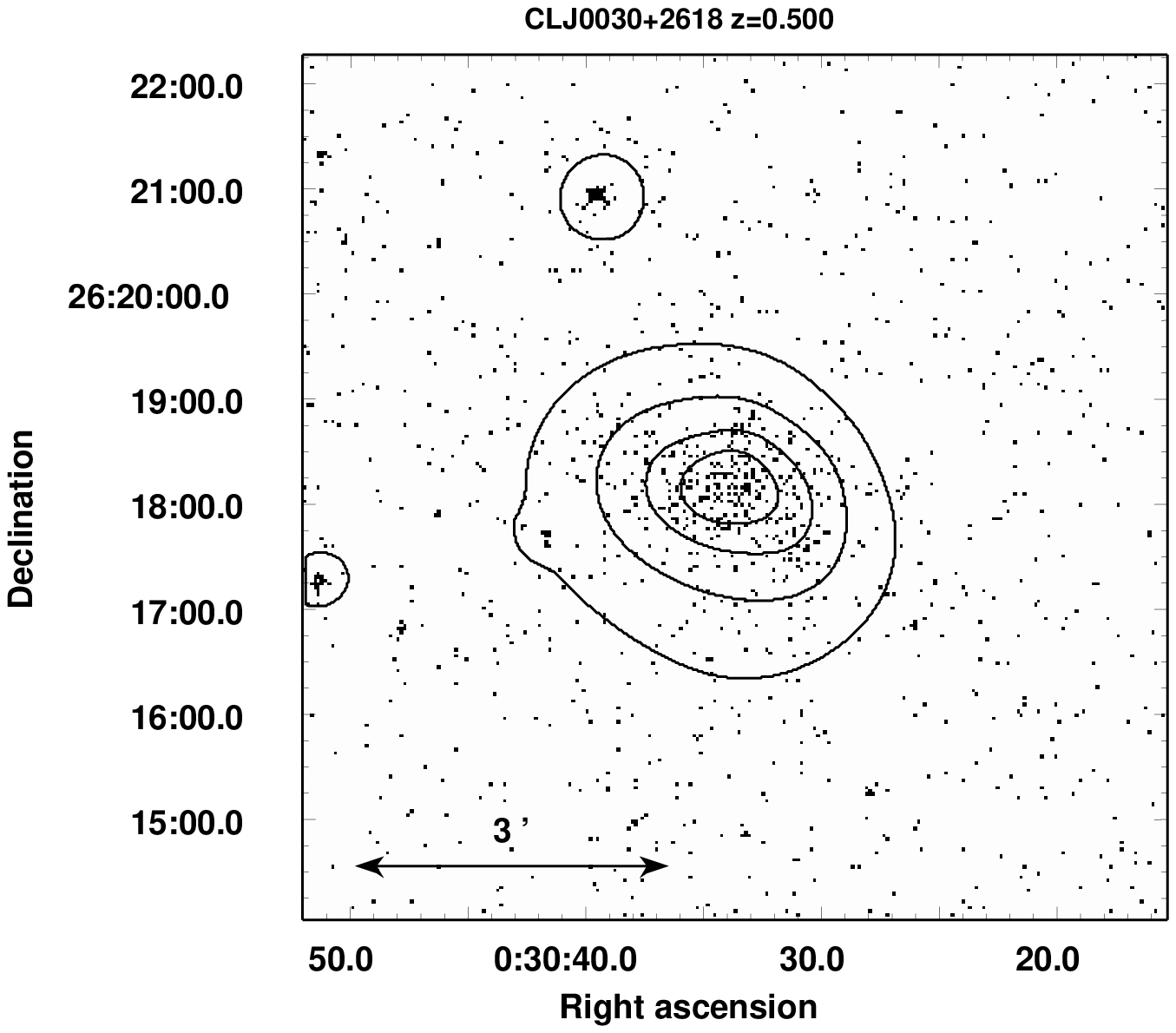}
\includegraphics[angle=0,scale=0.60, bb=50 210 460 565]{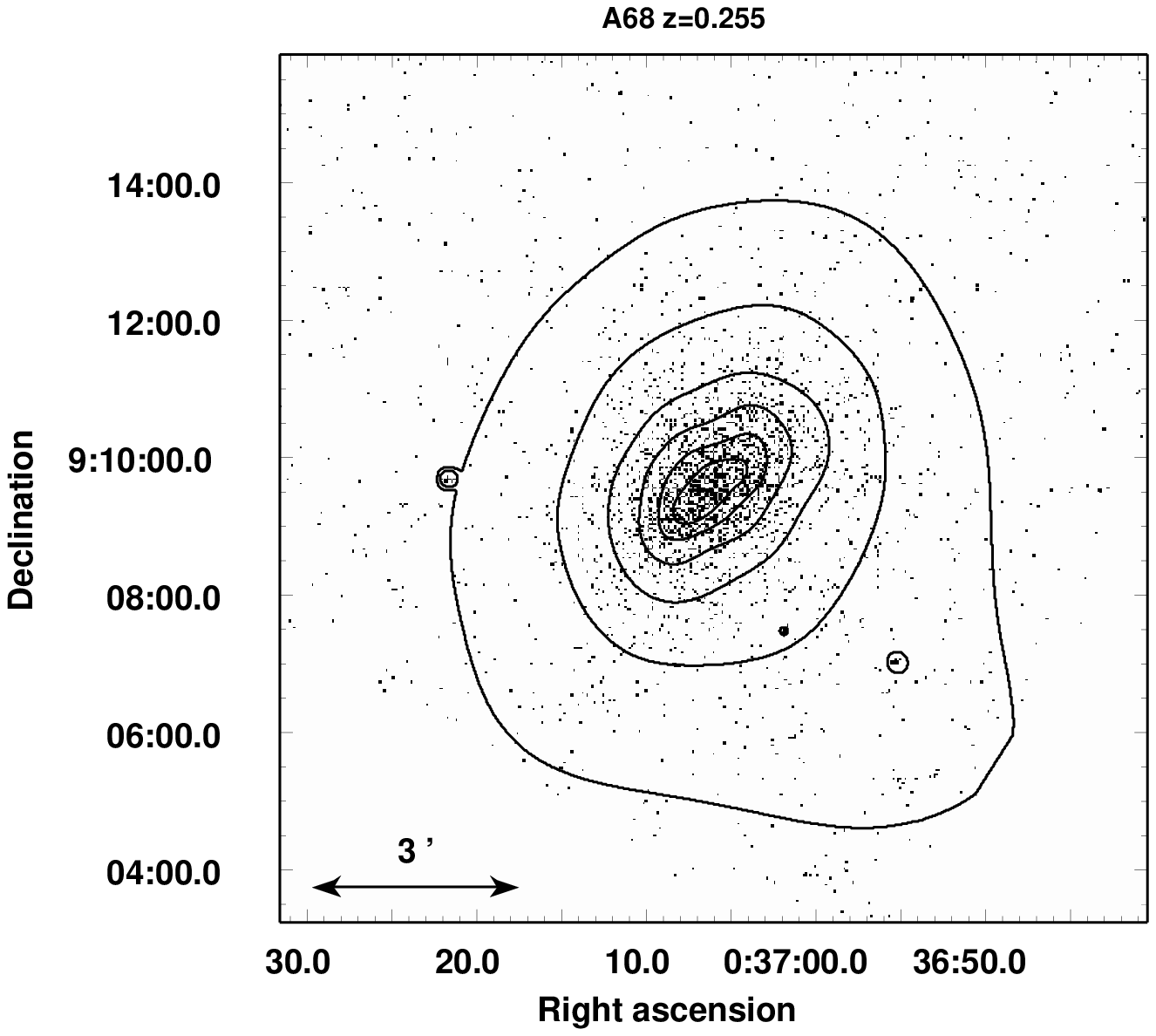}\\
\includegraphics[angle=0,scale=0.60, bb=50 210 460 565]{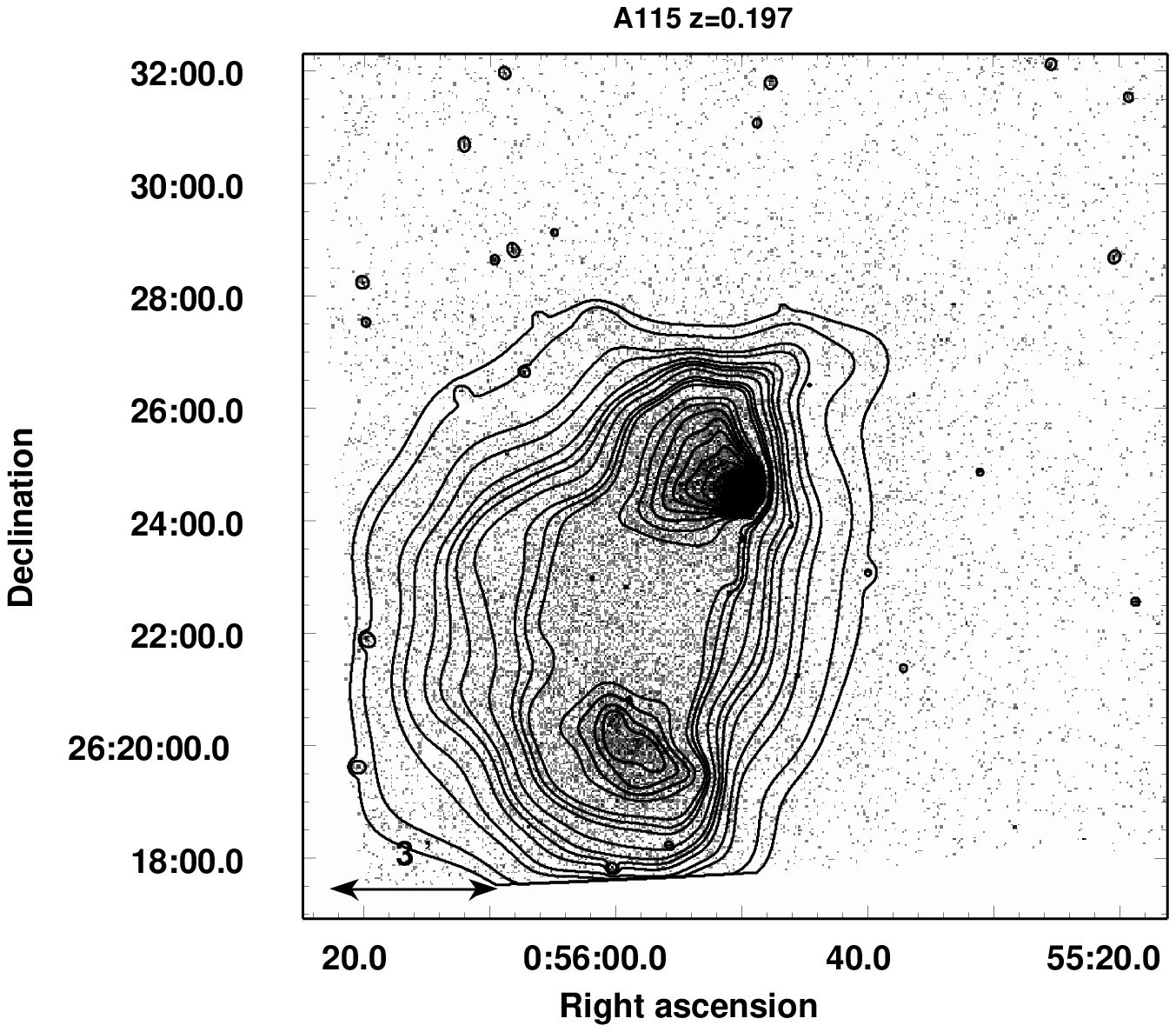}
\includegraphics[angle=0,scale=0.60, bb=50 210 460 565]{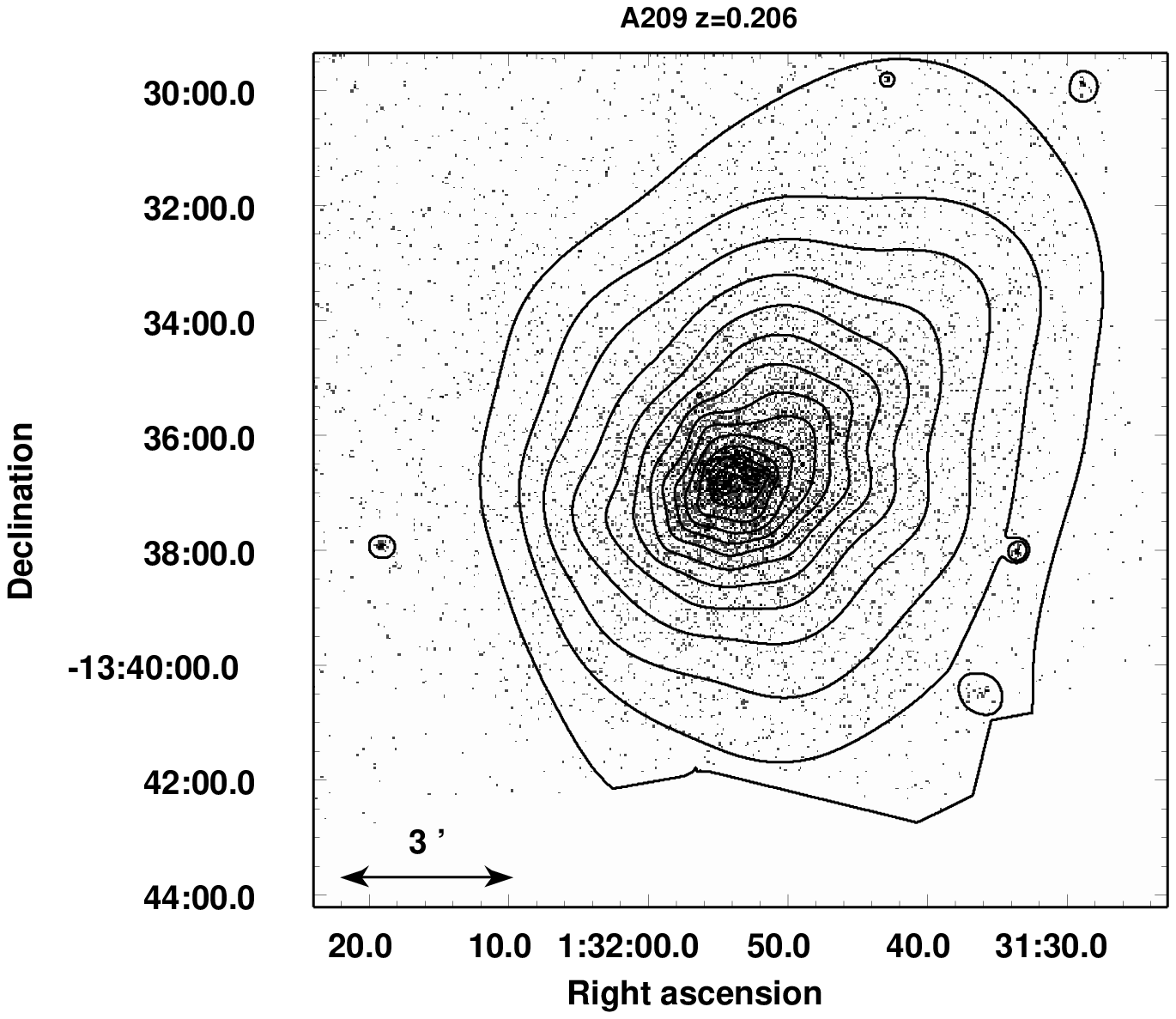}\\
\caption{\label{f.contour}X-ray images of the clusters in the $0.7-2\keV$ band. The images are $3\Mpc$ on a side and are overlaid with contours of the same emission adaptively smoothed to $3\sigma$. The contour levels $c_i$ and $c_{i+1}$ are set so that the emission bounded by those levels is detected at a significance of $>3\sigma$ above the emission between levels $c_{i-1}$ and $c_i$ (see text for details).}
\end{figure*}

Fig. \ref{f.contour} shows example contour plots for a sample of the
clusters. The plots for the full set of clusters are available in the
electronic version of this paper. In addition, for each cluster, a region file containing the
contours is available as online data\footnote{\url{http://cxc.harvard.edu/cda/Contrib/2007/MAUG1}}. The
intention is that these provide a useful resource enabling contours of the
X-ray emission to be easily overlaid on images obtained from other
detectors. The region files also contain the flux levels of each contour
to aid in their interpretation.

\section{Spectral Analysis}\label{s.spec}
\begin{figure}
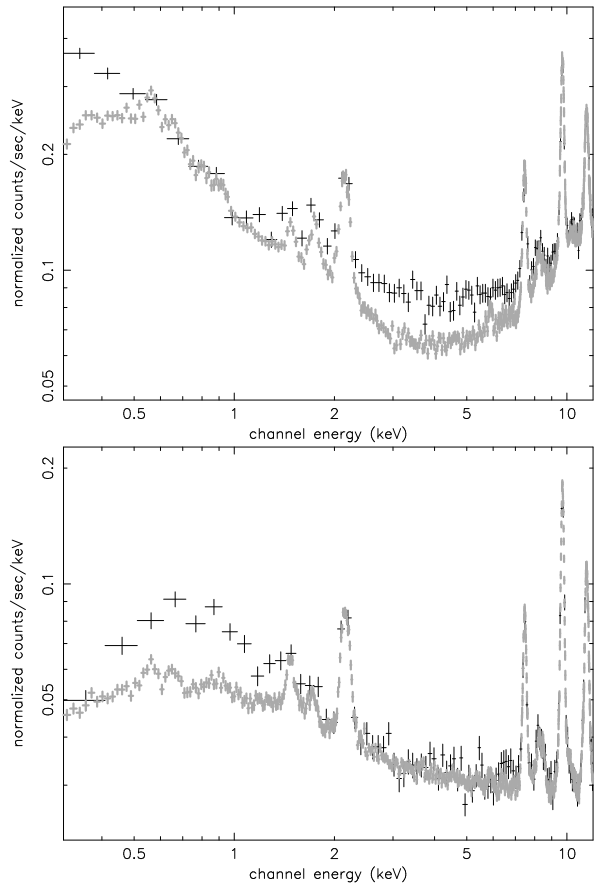

\rotatebox{-90}{\scalebox{0.33}{\includegraphics{f4a.eps}}}\\
\rotatebox{-90}{\scalebox{0.33}{\includegraphics{f4b.eps}}}\\
\caption{\label{f.bgspec}{Blank-sky (grey) and local (black) background
spectra are plotted for V1701+6414 (top) and for A1942 (bottom). The
observation of V1701+6414 was affected by low level background flares, the
effect of which can be seen in the elevated $2-7\keV$ continuum. The blank
sky backgrounds could not be used for spectral analysis here. The
observation of A1942 is a more typical example where the blank-sky and
local background spectra agree well except for the Galactic foreground emission at $<2\keV$.}}
\end{figure}

In order to monitor the spectrum of the background emission in each
observation, a spectrum was extracted from the source-free region and from
the same region in the blank-sky background. These spectra were compared to
test the validity of using the blank-sky background spectrum for the
analysis of that dataset. In order to account for differences in the
particle-induced hard background, the background spectra were normalised to
match the count rate in the $9.5-12\keV$ band in the source datasets. After
this normalisation, an excellent agreement was generally found, except
below $\sim2\keV$ due to differences in the soft Galactic foreground
emission between the target and blank-sky fields. In rare cases,
significant differences were found in the level of the continuum emission
at $2-7\keV$ due to low level background flares that were not detected by
the standard lightcurve filtering. If there were other, unaffected
observations of the same target, then the data with low flares were
rejected. For two clusters (A1682 and V1701+6414) there was no alternative
data, and so a local background spectrum was used for the spectral
analysis. Background spectra are shown in Fig. \ref{f.bgspec} for
V1701+6414 and for A1942, a more typical example where the blank-sky and
local background spectra agree well except for the Galactic foreground
emission at $<2\keV$.

In order to account for the differences in the soft background, the blank
sky and readout background spectra were then subtracted from the local
background spectrum to give a ``soft residual'' spectrum. These residuals
were generally positive, indicating a higher soft background flux in the
target pointings than the blank-sky fields, but in some cases were
negative. The soft X-ray background is thermal in origin, and while the
residual spectrum is the difference between thermal spectra, it is
dominated by the oxygen lines at $\sim0.6\keV$ and is adequately described
by a $0.18\keV$ APEC \citep{smi01} model with zero redshift and positive or
negative normalisation \citep{vik05a}. The best fitting model to the soft
residuals was included as an additional background component in all fits to
cluster spectra, with the normalisation scaled for the differences in
extraction area. 

For all spectral fits, the source emission was modelled with an absorbed
APEC model in the $0.6-9\keV$ band. The temperature, metal abundance and
normalisation of the hot component were free to vary, while the absorbing
column was fixed at the Galactic value inferred from $21\cm$ observations
\citep{dic90}. The model was refit with the soft Galactic component
normalisation set to $\pm1\sigma$ and again with the overall blank-sky
background renormalised by $\pm2\%$, and the resulting systematic
temperature uncertainties were added in quadrature to the statistical
uncertainties. For each source spectrum that was fit, the amount of flux
lost to any excluded point and extended sources was estimated using the
azimuthal average of the exposure-corrected, background subtracted flux at
that cluster-centric radius. While this assumes azimuthal symmetry for the
clusters, it is an improvement over a simple geometric flux correction that
would not take into account the radially varying surface brightness
distribution of the cluster.

\subsection{Determination of \rf}\label{s.rv}
Cluster properties were measured within the radius \rf. This radius
generally matches the detected extent of the clusters' emission, and
corresponds to approximately half of the virial radius. \rf\ was measured
following the approach recommended by \citet{kra06a} to estimate \rf\ from
\Yx, where \Yx\ is defined as the product of the cluster temperature and
gas mass. \Yx\ provides a reliable method for estimating cluster masses
(and hence \rf), as the \YM\ relation has low scatter, and the shape and
scatter in the relation is insensitive to merging clusters
\citep{kra06a}. This makes it ideal for our sample, which contains clusters
with a wide range of dynamical states. The low-scatter of the \YM\ relation
and its self-similar evolution have been verified for a subset of the
clusters in this sample \citep{mau07c}. \rf\ and \Yx\ were computed
iteratively, measuring the temperature in the aperture $(0.15<r<1)\rf$ and
the gas mass within \rf, computing a new \Yx, and hence estimating a new
value of \rf. The process was repeated until \rf\ converged to within
$1\%$. An initial temperature estimate of $5\keV$ was used to estimate the
first value of \rf\ to begin the iteration.

In order to estimate \rf\ from \Yx\ we used the \YM\ relation measured for
the \citet{vik06a} sample of clusters
\begin{eqnarray}\label{e.ym}
\Mf & = & \frac{h}{0.72}^{\frac{5B_{YM}}{2}-1}C_{YM}E(z)^{a_{YM}}\frac{Y_X}{6\times10^{14}\Msol\keV}^{B_{YM}},
\end{eqnarray}
with $B_{YM}=0.564$, $C_{YM}=7.047\times10^{14}\Msol$ and
$a_{YM}=-2/5$ (A. Vikhlinin, priv. comm.). $E(z)$ describes the
redshift evolution of the Hubble parameter, and is given by
\begin{eqnarray}\label{c4eqn_ez}
E^2(z) & = & \OM(1+z)^3 + (1-\OM-\Omega_\Lambda)(1+z)^2 + \Omega_\Lambda.
\end{eqnarray}
By definition, \rf\ is then given by $4/3\pi\rf^3=500\rho_c(z)/\Mf$. This
observed \YM\ relation has a normalisation $\sim15\%$ lower than that found
in the simulations of KVN06, which may be due to the effect of turbulent
pressure support which is neglected in the mass derivations for the
observed clusters (KVN06). Adopting the normalisation from the simulations
would have the effect of increasing \rf\ by $\sim2.5\%$ ($\sqrt[3]{15}\%$),
increasing \Mgas\ by $\sim15\%$ with a negligible effect on other
properties.

The aperture of $(0.15<r<1)\rf$ within which temperatures were measure in
defining \rf\ was chosen to exclude any effects of cool core emission in
the clusters, and is consistent with the \YM\ relation used. Once \rf\ was
determined, the cluster properties were also measured with the core
emission included, and the values obtained for both apertures are given in
Table \ref{t.cat}. We found that exclusion of the core ($r<0.15\rf$)
removed $\sim30\%$ of the flux in non cool-core clusters (and a larger
fraction from clusters with cool cores). This meant that for 4 of the
faintest clusters (RXJ0910+5422, CLJ1216+2633, CLJ1334+5031
and RXJ1350.0+6007), we were unable to measure a temperature in that
aperture. For those clusters, all of the cluster emission ($r<\rf$) was
used for the temperature measurements in estimating \rf. None of these
clusters showed evidence for the presence of cool cores, so the effect of
including the core emission on biasing the temperature should be
negligible.

\subsection{Combining observations}
Twenty nine of the clusters in this sample were observed more than once by
\Chandra. In these cases, the observations were first analysed separately
as described above, and then the data were combined for certain stages of the
analysis. Source, background and exposure map images were projected onto a
common coordinate system and summed. The source lists from the individual
observations were combined and those sources were excluded. The summed
images were then used for all imaging analysis.

Multiple observations were also combined for the spectral analysis. Source
and background spectra were extracted as above for each individual
observation. The spectra were then fit simultaneously as before with the
temperature and metal abundance of the hot APEC components tied
together. This method was followed when determining \rf\ and the global
temperature.

\section{Notes on individual clusters}\label{s.notes}
In this section we note any peculiarities or points of interest about
individual observations or clusters that in some cases required a departure
from the analysis procedure described above.

\noindent {\bf A115 (3C28.0):} This cluster is undergoing a major off-axis
merger, with the two subclusters separated by $300\arcs$ ($1\Mpc$) in
projection. The fainter southern subcluster was manually excluded from our
analysis. The \Chandra\ observation of this system is discussed in detail by
\citet{gut05}.

\noindent {\bf CLJ0152.7-1357:} This system is dominated by a north-south
merger between equal mass clusters whose cores are separated by $95\arcs$
($720\kpc$) in projection. The cluster is also associated with a network of
large scale structure \citep{kod05short,mau06b}. The northern and southern
subclusters (hereafter CLJ0152.7-1357N and CLJ0152.7-1357S) were analysed
separately with the emission from each excluded during the analysis of the
other.

\noindent {\bf MACSJ0329.7-0212:} ObsIDs 3257 and 6108 were rejected due to
background flaring, leaving ObsID 3582.

\noindent {\bf MACSJ0404.6+1109:}  An extended source at
$\alpha[2000.0]=4^{\rm h}4^{\rm m}42.85^{\rm s}$
$\delta[2000.0]=+11^{\circ}09\arcm 39.6\arcs$, $170\arcs$ ($0.9\Mpc$) from the
the cluster core, was manually excluded.

\noindent {\bf RXJ0439.0+0715:} ObsID 526 was rejected due to a very high
background level, leaving ObsID 3583.

\noindent {\bf 1E0657-56:} This cluster, a spectacular merger known as
the ``bullet cluster'' \citep{mar04}, has been observed 10 times with
\Chandra\ ACIS-I for a total of $>500\ks$. For simplicity we use only two
of the observations in our analysis which total $\sim100\ks$, which is more
than sufficient for our purposes.

\noindent {\bf MACSJ0717.5+3745:} ObsID 1655 was rejected due to a very high
background level, leaving ObsID 4200. 

\noindent {\bf MS0906.5+1110:} This cluster is part of a structure
that includes A750 ($\alpha[2000.0]=09^{\rm h}08^{\rm m}59.25^{\rm s}$
$\delta[2000.0]=+11^{\circ}02\arcm 39.9\arcs$) at the same redshift. The
centre of A750 is $300\arcs$ ($0.9\Mpc$) from the core of
MS0906.5+1110. The emission from A750 falls
partially within the \Chandra\ field of view in this observation and also
within the \rf\ of MS0906.5+1110 and was manually excluded.

\noindent {\bf RXJ1234.2+0947:} This system \citep[also referred to as Z5247
in][]{ebe98} apparently consists of a binary merger between two clusters of similar
mass. However, only one redshift is available in \citet{ebe98} for this
system, and the position given is that of the southern subcluster. The
northern subcluster was thus manually excluded from the current
analysis.

\noindent {\bf A1682:} This cluster is part of a large scale filament
comprising at least two other extended X-ray sources at the cluster
redshift. These sources are at $\alpha[2000.0]=13^{\rm h}06^{\rm
m}58.52^{\rm s}$ $\delta[2000.0]=+46^{\circ}31\arcm 37.5\arcs$, and
$\alpha[2000.0]=13^{\rm h}07^{\rm m}13.33^{\rm s}$
$\delta[2000.0]=+46^{\circ}29\arcm 08.4\arcs$, and are $145\arcs$
($0.5\Mpc$) and $227\arcs$ ($0.8\Mpc$) south east of the cluster core,
respectively. These sources were manually excluded for the current
analysis. Furthermore, the only ACIS-I observation of this cluster was
affected by long, low level flares causing the background spectrum to
differ significantly from the blank-sky background. For this reason, a
local background was used for the spectral analysis.

\noindent {\bf CLJ1334+5031:} The only ACIS-I observation of this cluster was
affected by long, low level flares causing the background spectrum to
differ significantly from the blank-sky background. For this reason, a
local background was used for the spectral analysis.

\noindent {\bf A1763:} An extended source at $\alpha[2000.0]=13^{\rm
h}34^{\rm m}53.21^{\rm s}$ $\delta[2000.0]=+40^{\circ}56\arcm 55.34\arcs$,
$326\arcs$ ($1.2\Mpc$; $\sim\rf$) from the the cluster core, was manually excluded.

\noindent {\bf A2069:} An extended source at $\alpha[2000.0]=15^{\rm
h}24^{\rm m}27.70^{\rm s}$ $\delta[2000.0]=+30^{\circ}01\arcm 44.29\arcs$,
$530\arcs$ ($1.1\Mpc$; $\sim\rf$) from the the cluster core, was manually excluded.

\noindent {\bf RXJ1524.6+0957:} This cluster is in a region of high
Galactic X-ray emission at the base of the north polar spur, and thus had
an extremely strong background residual spectrum. A cool APEC model at
$0.30\keV$ was required to fit this soft foreground component.

\noindent {\bf A2111:} An extended source at $\alpha[2000.0]=15^{\rm
h}39^{\rm m}32.70^{\rm s}$ $\delta[2000.0]=+34^{\circ}28\arcm 07.37\arcs$,
$210\arcs$ ($0.8\Mpc$) from the the cluster core, was manually excluded. 

\noindent {\bf A2163:} The lightcurve of ObsID 1653 was cleaned by
hand to remove a long, low-level background flare.

\noindent {\bf MACSJ1621.3+3810:} The lightcurve of ObsID 3594 was cleaned by
hand to remove several periods of very high background.

\noindent {\bf MS1621.5+2640:} An extended source at $\alpha[2000.0]=16^{\rm
h}23^{\rm m}48.25^{\rm s}$ $\delta[2000.0]=+26^{\circ}34\arcm 22.11\arcs$,
$189\arcs$ ($1.1\Mpc$; $\sim\rf$) from the the cluster core, was manually excluded. 

\noindent {\bf RXJ1701+6414:} This cluster (at $z=0.453$) is separated by
$285\arcs$ in projection from the foreground cluster A2246 at
$z=0.225$. A2246 was manually excluded for this analysis. The observation
was also affected by long, low level flares causing the background spectrum
to differ significantly from the blank-sky background. For this reason, a
local background was used for the spectral analysis.

\noindent {\bf A2261:} An extended source at $\alpha[2000.0]=17^{\rm
h}22^{\rm m}12.15^{\rm s}$ $\delta[2000.0]=+32^{\circ}06\arcm 54.0\arcs$,
$200\arcs$ ($0.7\Mpc$) from the the cluster core, was manually excluded.

\section{Results and discussion}\label{s.results}

\subsection{Cluster morphologies}
\begin{figure}
\hspace{-1.5cm}\rotatebox{0}{\scalebox{0.45}{\includegraphics{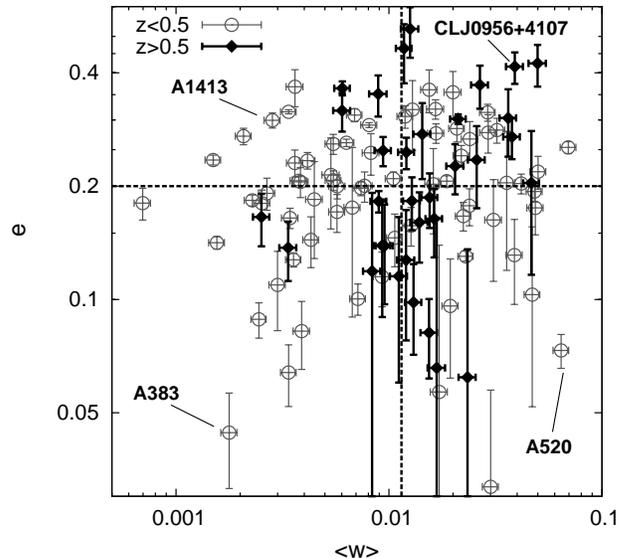}}}
\caption{\label{f.morpho}{The clusters are plotted on the plane of centroid
shift and ellipticity. The dashed lines show the median values of the full
sample. Hollow and solid points show low and high-redshift clusters respectively.}}
\end{figure}

\begin{figure}
\scalebox{0.42}{\includegraphics{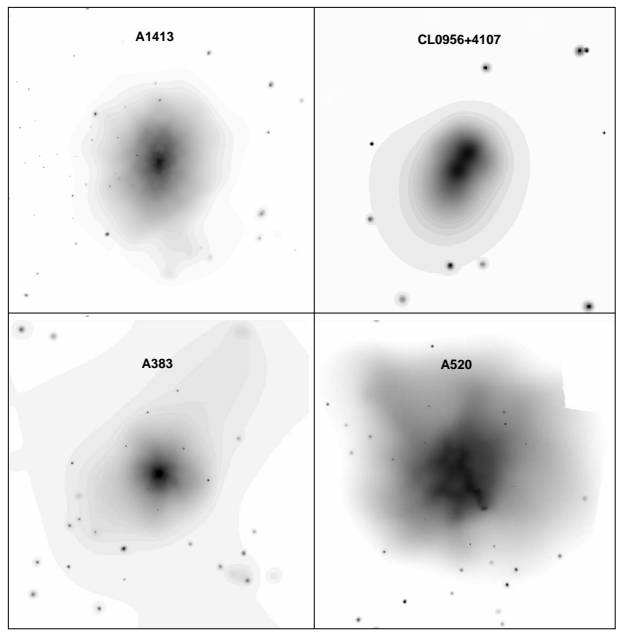}}
\caption{\label{f.morphoimg}{Adaptively smoothed images of clusters
exemplifying the morphological characteristics potted in Fig
\ref{f.morpho}. The clusters are plotted here in the same quadrants in
which they reside in Fig \ref{f.morpho}.}}
\end{figure}

We have used two simple quantities (ellipticity and centroid shift) to
measure the morphology of the clusters in the sample. In
Fig. \ref{f.morpho} we plot the clusters on the plane of these two
quantities. The dashed lines indicate the median values for the entire
sample. For some nearby clusters, \rf\ was not completely contained within
the field of view, so $R_w$ (the maximum radius within which $\langle w
\rangle$ could be measured) was less than \rf. The $10$ clusters for which
$R_w<0.9\rf$ were rejected from this morphological study to ensure
uniformity.  Broadly speaking, clusters to the bottom left of
Fig. \ref{f.morpho} are the most relaxed, with undisturbed, circular
morphologies (\eg A383) while those to the top left are elliptical clusters
without significant substructure (\eg A1413). The top right quadrant is
home to clusters that are both elliptical and disturbed (\eg CLJ0956+4107)
and finally the bottom left quadrant contains disturbed clusters with
broadly circular geometries (\eg A520). Adaptively smoothed images of these
four example clusters are plotted in Fig. \ref{f.morphoimg}

\begin{figure}
\rotatebox{-90}{\scalebox{0.33}{\includegraphics{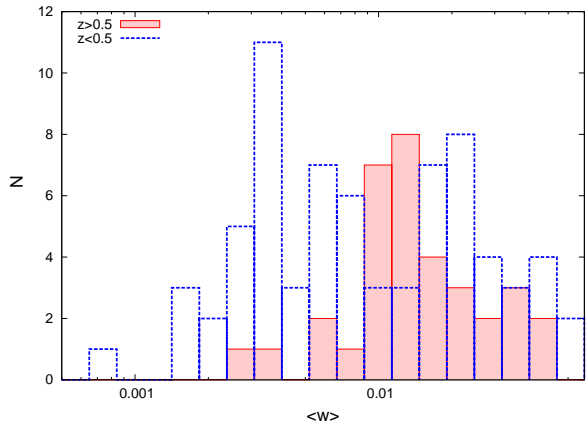}}}
\caption{\label{f.whisto}{Histograms of the centroid shift parameter $\langle w \rangle$
for the low and high-redshift subsets of our sample.}}
\end{figure}

The sample was split into low and high-redshift subsets using a redshift
cutoff of $z=0.5$ and the two subsets are marked separately on
Fig. \ref{f.morpho}. There is a clear absence of very relaxed clusters at
$z>0.5$, although the distribution of ellipticities does not vary significantly
with redshift. The distribution of $\langle w \rangle$ alone is plotted in
Fig. \ref{f.whisto} for the low and high-redshift subsets. Again the
absence of very relaxed clusters in the high-redshift subset is clear. A
Kolmogorov-Smirnov test gave a probability that the two $\langle w \rangle$
subsets came from the same parent distribution of $4.5\times10^{-4}$
($>3.5\sigma$). This reinforces, at greater significance, the results of
\citet{jel05} who found more substructure in distant clusters than local
clusters by using power ratios on \Chandra\ images \citep[see also][]{vik06c}.

\subsection{Evolution of metal abundance}
\begin{figure}
\rotatebox{-90}{\scalebox{0.33}{\includegraphics{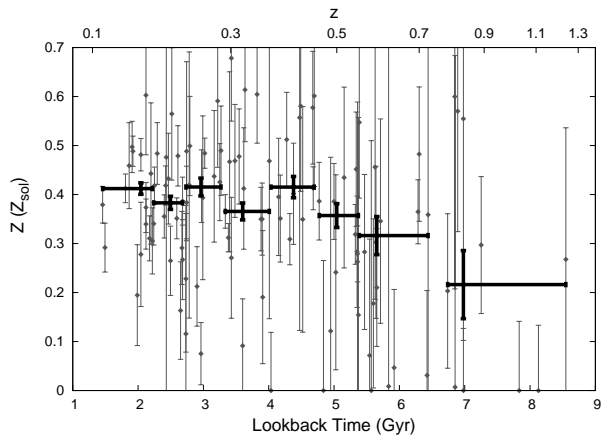}}}
\caption{\label{f.Z500}{ICM metal abundances measured within \rf\ are
plotted against lookback time and redshift for each cluster. All emission
within \rf\ was used when measuring these abundances. The thick lines show the
abundances measured from a joint fit to the clusters in different redshift
bins. The bins contain $\sim15$ clusters and the points are at the median
redshift of each bin.}}
\end{figure}

The abundance of metals in the ICM is the signature
of star formation activity (and the ensuing supernovae) in
the member galaxies, and the observed evolution of metal abundances can be
used to trace the history of these processes
\citep[\egc][]{ett05,bal06a}. Measuring metal abundances in distant clusters is
challenging, and requires a higher quality of data than measuring
temperature alone (see \textsection \ref{sec:depend-spectr-uncert}). In
Fig. \ref{f.Z500} we plot metal abundances measured within \rf\ as a
function of redshift and lookback time. There are large uncertainties on
the individual measurements, and for 9 clusters, the abundance measured was
an upper limit. The abundance values are iron abundances relative to the
solar values of \citet{and89}. While these have been superceded by more
recent measurements \citep[e.g][]{gre98}, they allow straightforward
comparison with other works. A simple scaling of $0.676$ converts from the
\citet{and89} iron abundance to values relative to the \citet{gre98}
abundances.

To improve the precision of the measurements, the clusters were grouped by
redshift, and the spectra were fit simultaneously with the Galactic
absorption, redshifts and soft foreground component fixed at their measured
values for each cluster. The model temperatures were free to fit
independently for each cluster (but tied together for multiple observations
of the same cluster) and the metal abundances were tied for all clusters in
the group. The metal abundance measured for each group of clusters is then
effectively a weighted mean, with the higher signal-to-noise (S/N) spectra
naturally having a stronger weighting due to the \chisq\ fitting
method. The clusters were binned into groups of 15 clusters (with the
highest redshift bin containing 11 clusters) and the abundances measured
for each group are plotted with the unbinned data in Fig. \ref{f.Z500}. The
locus of each bin is at the median redshift of the clusters in that bin.

The grouped abundances decrease with redshift from $\sim0.4\Zsol$ locally
to $\sim0.2\Zsol$ at $z\approx1$, with most of the decrease occurring at
$z>0.5$. The significance of the decrease was investigated by fitting a
constant level to the grouped abundances. This model was rejected at the
$99.9\%$ level. Note that fitting a model to the unbinned data is
problematic because of the upper limits at higher redshifts; a survival
analysis that can include the upper limits correctly is not appropriate
here, as it is not possible to include measurement errors in such analyses.

Clusters with cooling cores have been observed to have central peaks in
their abundance profiles within $r\approx150\kpc$, with abundances
approaching solar values \citep{deg01,vik05a}. Outside of the core the
abundance profiles flatten significantly to values around
$0.3\Zsol$. \citet{deg01} also found significant (although much weaker)
abundance gradients in non cool-core clusters.  As the global abundances
measured here are emission weighted, the sharp surface brightness peaks in
cool-core clusters cause their central abundance peaks to be
overrepresented in the global value, biasing it high. A similar, but much
milder bias will also be present in the non cool-core clusters. It is
therefore instructive to remove this effect by excising the core regions
from the metal abundance measurements. 

\begin{figure}
\rotatebox{-90}{\scalebox{0.33}{\includegraphics{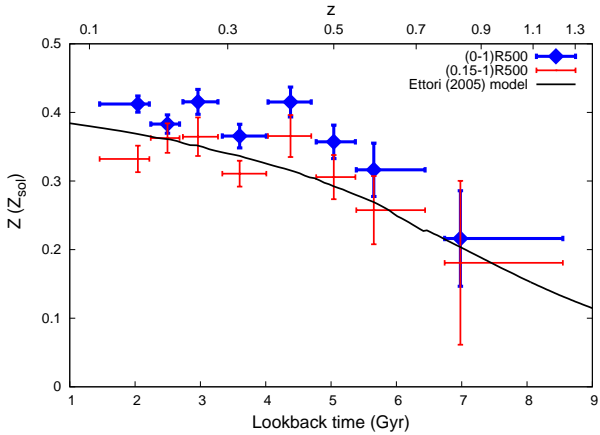}}}
\caption{\label{f.Zcf}{Mean ICM metal abundances measured from joint
spectral fits to clusters grouped into redshift bins. Abundances measured
in the $(0.15-1)\rf$ and $(0-1)\rf$ apertures are shown. The latter
points are the same as those plotted in Fig. \ref{f.Z500}. The solid line
shows the supernova enrichment model of \citet{ett05}.}}
\end{figure}

To this end, abundances were measured from spectra extracted in the annulus
$(0.15-1)\rf$, although this leads to significantly lower signal to noise
and hence larger uncertainties. The clusters were grouped as before and the
spectra were fit simultaneously. The abundance evolution measured in this
way is plotted in Fig. \ref{f.Zcf} along with the binned data from
Fig. \ref{f.Z500} (which show the abundances measured when the core region
was not excluded). When the core regions are excluded, the data show the
same trend of decreasing abundance with redshift, but at lower
significance; the constant abundance model could only be rejected at the
$87\%$ level. The metal abundances are lower on average at all redshifts
when the core was excluded, but the difference becomes less significant at
$z>0.5$. This can be partially explained by the decrease with redshift of
the fraction of cool core clusters in the population
\citep{vik06c}. However, the data suggest that weak abundance gradients
such as those found by \citet{deg01} in non cool-core clusters are still
present at $z\sim1$.

Our results are consistent with those of \citet{bal06a} based on a sample
of 56 clusters observed in with \Chandra\ (including many clusters in
common with our sample). Key differences between these studies are the
sizes of our samples, and the choice of spectral aperture. \citet{bal06a}
used circular apertures in the approximate range $(0.3-0.6)\rf$ to maximise
the S/N for each cluster, and included the core emission. We used the
apertures $(0-1)\rf$ and $(0.15-1)\rf$ for all clusters. The \citet{bal06a}
method has the advantage of yielding the most precise measurements for each
cluster, while our method is consistent for all clusters, and allows us to
address the effects of cool cores on the abundance measurements.

The overall decline in abundance with redshift that we observe is in
qualitative agreement with the theoretical predictions of
\citet{ett05}; the line in Fig. \ref{f.Zcf} is the solid line taken from
their Fig. 5. The abundances measured with the cluster cores excluded agree
well with the theoretical prediction, while those measured from the entire
cluster are systematically higher than the model. This is likely to be due
to the cool-core bias on our emission weighted global abundances. On the
other hand, the abundances measured with the cores excluded will somewhat
under-represent the true global value, because the higher metallicity gas
in the cores is then ignored completely. It would be interesting to compare
the evolution of mass-weighted abundances or total iron mass with
theoretical predictions, but such measurements are not possible for the
distant clusters with the current data.

\subsection{Surface brightness profile slopes at large radii}
The slope of the X-ray surface brightness profile is an important
quantity because hydrostatic cluster masses derived within some
radius depend on the slope of the gas density profile at that radius, which
is measured from the surface brightness profile. The slope is often
measured by fitting a simple $\beta$-model to the data, but this has been
found to underestimate the slope of the data at large radii as the fit is
driven by the higher S/N data in the inner parts of the profile
\citep{vik99}. More recently, V06 demonstrated that much of the
discrepancy between the normalisation of the mass-temperature relation from
observational studies and simulations can be explained by underestimates
of the gas density profile slope at large radii.

\begin{figure}
\rotatebox{-90}{\scalebox{0.33}{\includegraphics{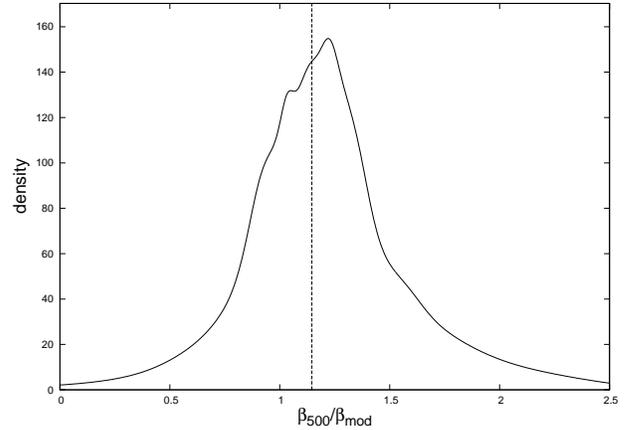}}}
\caption{\label{f.bratio}{Kernel density plot of the ratio of the
logarithmic slope of the observed surface brightness profiles at \rf\ to
that of the best fitting $\beta$-model at the same radius. The dashed line
shows the weighted mean.}}
\end{figure}

The ratio $\beta_{500}/\beta_{mod}$ was calculated for each cluster, and a
kernel density plot of the population is show in Fig. \ref{f.bratio}. To
form this plot a Gaussian kernel was computed for each measurement with the
central position and $\sigma$ given by the measured
$\beta_{500}/\beta_{mod}$ and uncertainty. The kernels were normalised to
have an area of unity and summed. The advantage of this plot over a
histogram is that the uncertainties in the variable are included; well
measured values contribute a high, narrow kernel while those with large
uncertainties contribute a broad, low kernel. The resulting curve peaks at
$\beta_{500}/\beta_{mod}=1.2$ and the weighted mean of the values is
$1.15$. Thus, on average, the slope of the cluster surface brightness
profiles at \rf\ is steeper by $\sim15\%$ than the best-fitting
$\beta$-model. This is consistent with the findings of \citet{vik99} and
V06 and further reinforces the need for careful modelling of surface
brightness profiles at large radii.

\subsection{Similarity of surface brightness profiles}
A simple and useful model of the structure of galaxy clusters is that they
are self-similar. That is to say that clusters are identical when scaled
appropriately by their mass. This would be the case if gravity were the only
important factor in the formation, growth, and evolution of clusters. There
are many examples of cluster properties deviating from self similarity
\citep[e.g.][]{mar98a,pon99,san03}, but  it remains a useful baseline
model, and departures from self-similarity provide useful clues to
non-gravitational processes that should be incorporated into cluster
models. In the self-similar model, the surface brightness profiles of
clusters should be identical, once scaled by mass. However, it has been
established that there is a trend for cooler systems to exhibit shallower
surface brightness profiles \citep[\egc][]{jon99,vik99,san03}. This
demonstrates the increasing relative contribution of non-gravitational
processes in clusters of lower masses.

\begin{figure}
\rotatebox{-90}{\scalebox{0.33}{\includegraphics{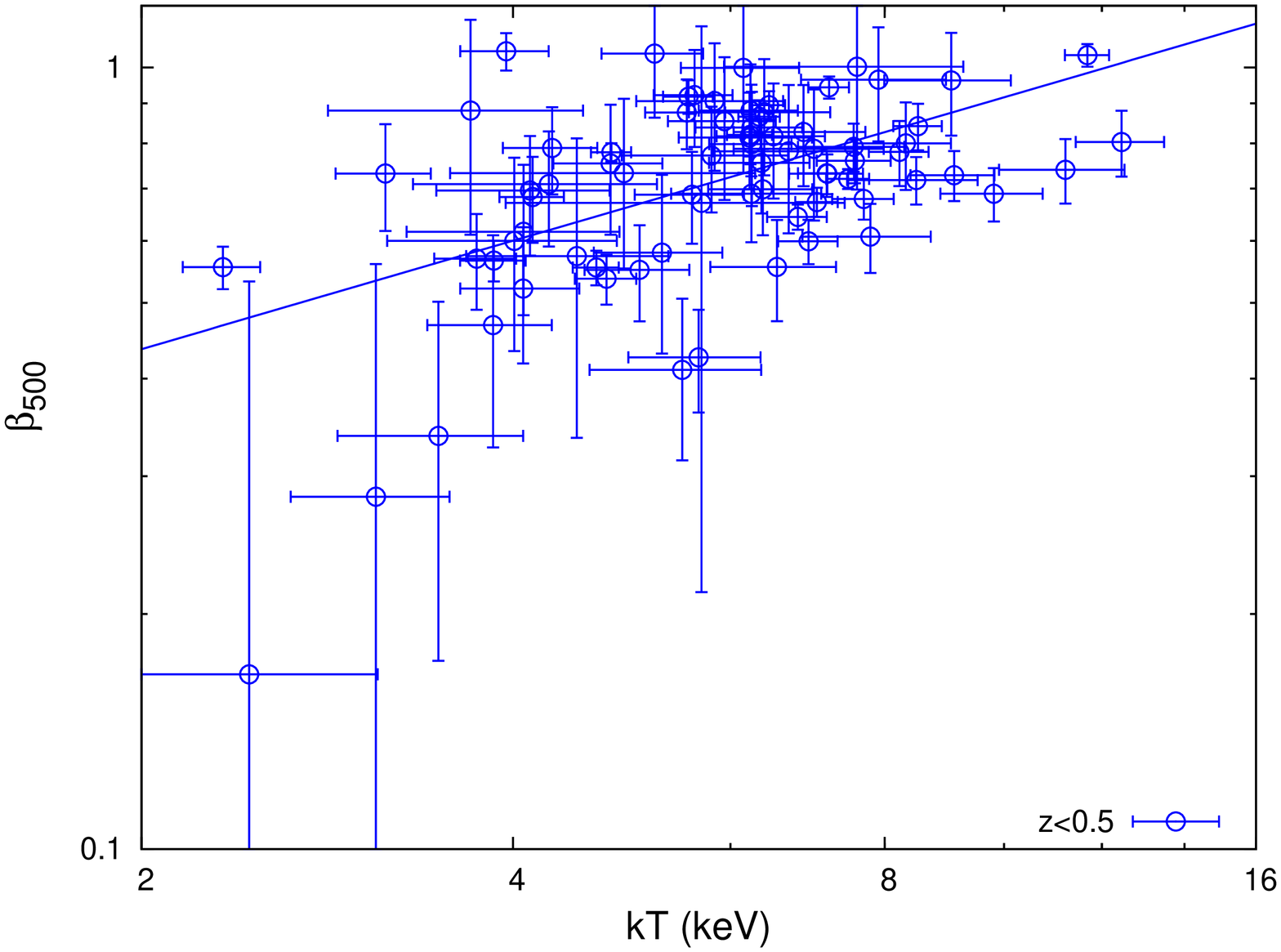}}}
\rotatebox{-90}{\scalebox{0.33}{\includegraphics{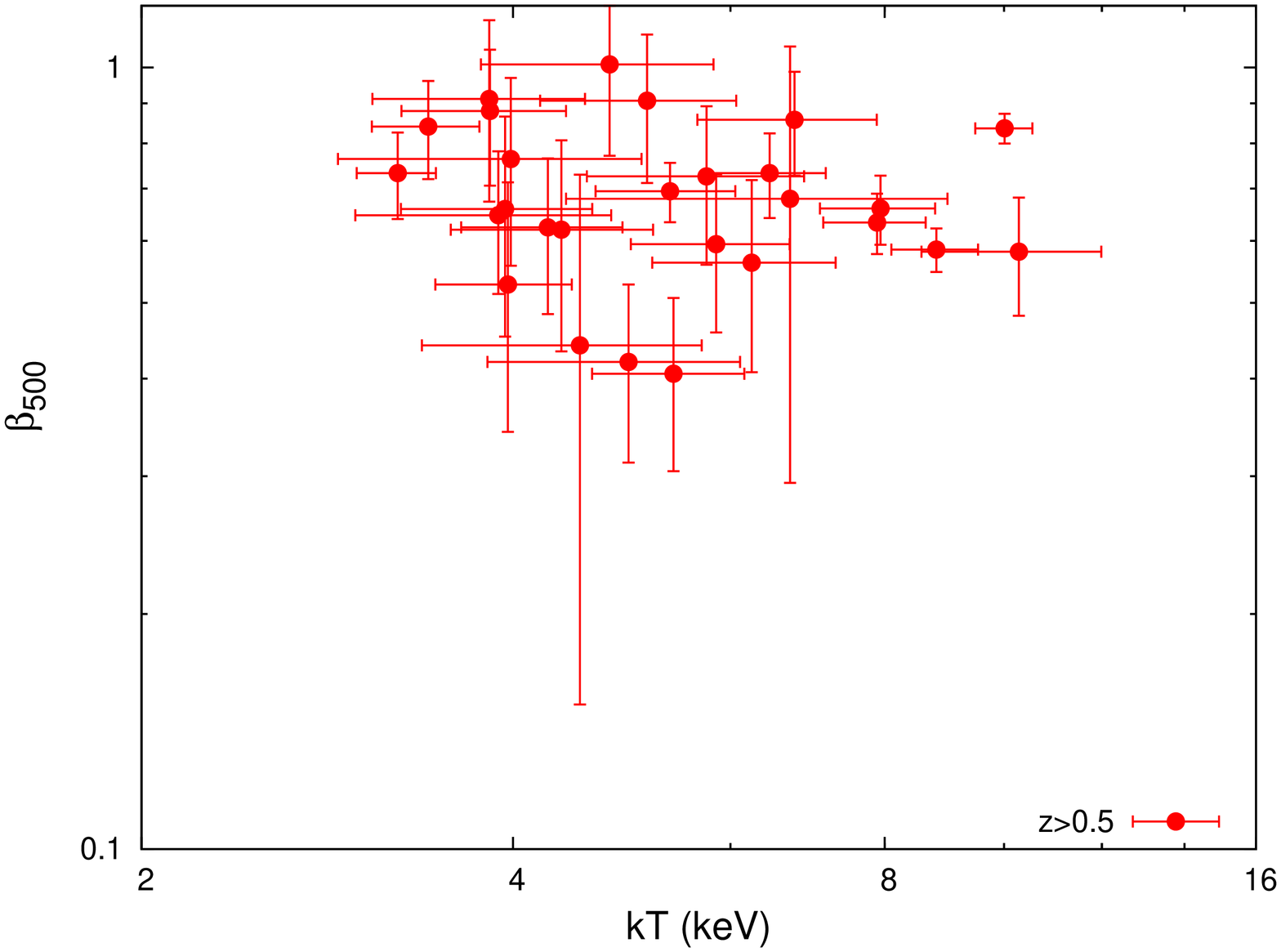}}}
\caption{\label{f.bt}The $\beta_{500}-kT$ relations for the low-redshift
(top) and high-redshift (bottom) subsets of the sample. The solid line in
the top panel shows the best-fitting power law to that data.}
\end{figure}

The $\beta_{500}-kT$ relation was investigated for our sample and is
plotted in Fig. \ref{f.bt}. The best-fitting relations were measured with
an orthogonal, weighted ``BCES'' regression \citep[as described
by][]{akr96}, on the data in log space. The sample was divided at
$z=0.5$ into low and high-redshift subsets, and these were fit separately
and are plotted in Fig. \ref{f.bt}. There was a significant trend for
cooler clusters to have lower values of $\beta_{500}$ in the low-z subset;
the best fitting relation had a slope of $0.46\pm0.09$. However, for the
high-resdhift subset, there was no significant correlation, the
best-fitting slope was $-0.1\pm0.2$. The low-redshift slope is thus steeper
than the high-redshift slope at the $\approx2.5\sigma$ level. This is not
simply due to the absence of cooler systems in the distant subset; the
significance was unchanged when the systems cooler than $4\keV$ were
excluded from the local subset. This result, while not strongly
significant, could indicate that the self-similarity breaking is due to
processes that become more important at lower redshifts. Alternatively, the
possible flattening of the $\beta_{500}-kT$ relation with redshift could be
related to the morphological evolution of the clusters. However, we find no
correlation between $\beta_{500}$ and $\langle w \rangle$ to support this
hypothesis.

\subsection{Dependence of spectral uncertainties on data
quality}\label{sec:depend-spectr-uncert} 

\begin{figure}
\rotatebox{-90}{\scalebox{0.33}{\includegraphics{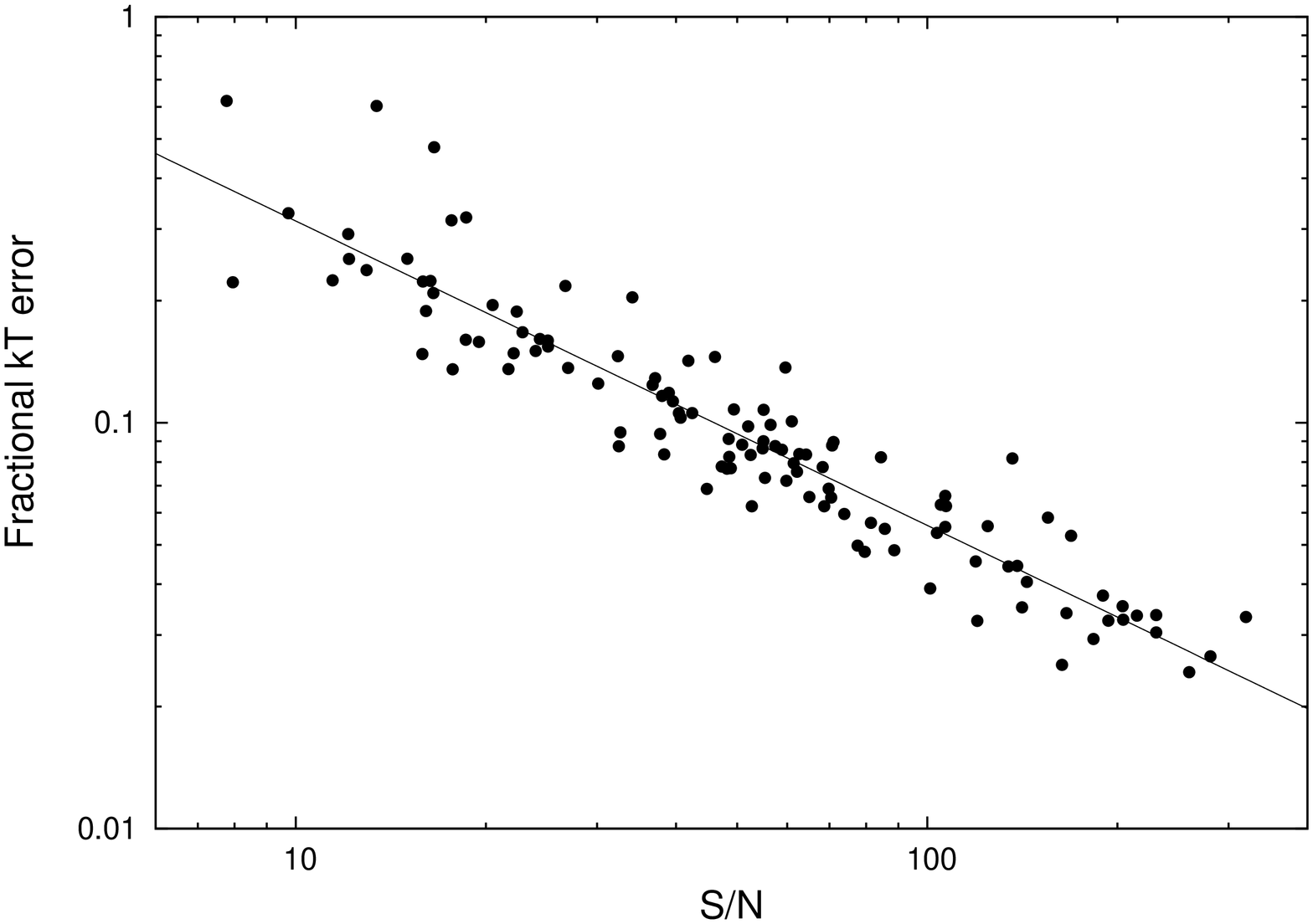}}}
\rotatebox{-90}{\scalebox{0.33}{\includegraphics{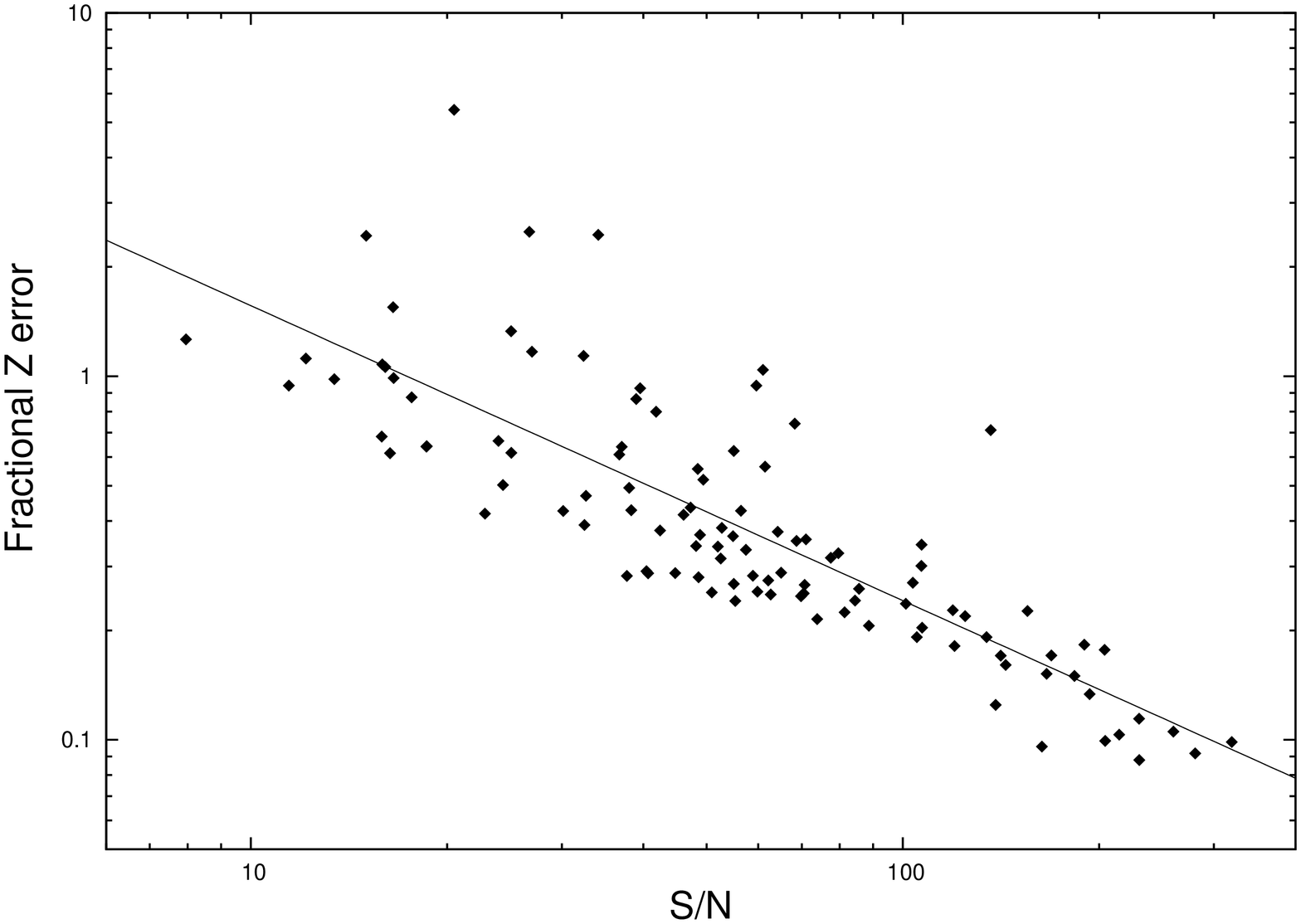}}}\\
\caption{\label{f.sn}{\it Top:} Fractional $kT$ errors versus spectrum signal
to noise measured in the spectral fitting band. {\it Bottom:} Fractional $Z$
errors versus spectrum signal to noise.}
\end{figure}

Finally, the large size of the sample enabled us to examine how the
uncertainties on measured spectral properties depend on the signal to noise
ratio of the spectra. Fig. \ref{f.sn} shows the fractional uncertainties
on the temperature and metal abundance measured for the clusters plotted
against the signal to noise (S/N) ratio in the spectral fitting band. There
are certainly factors other than data quality that influence the
measurement uncertainties, such as multiple temperature and abundance
components and different levels of absorption. However, fairly tight power
law correlations exist in both cases allowing us to define approximate
relations that can be used for a ``rule of thumb'' assessment of data
quality. We find $\sigma(kT)/kT=1.77(S/N)^{-0.75}$ and
$\sigma(Z)/Z=10.0(S/N)^{-0.81}$.

\section{Summary and Conclusions}\label{s.conc}
We have performed a systematic analysis of the largest available sample of
galaxy clusters at $z>0.1$ observed with \Chandra\ ACIS-I. Images and
contour files are made available for each cluster. The structural
properties and metal abundances of the clusters were investigated, and the
main results can be summarised as follows:
\begin{itemize}
\item There is a significant absence of relaxed clusters at high
redshift as measured by the centroid shifts of the cluster emission.
\item The metal abundance of the ICM decreases with redshift, in line with
theoretical predictions based on supernova rates.
\item The abundance evolution is still present (although less significant)
if the core regions of the clusters are excluded. This indicates that the
observed evolution is not a result of the disappearance of cool-core
clusters from the population at high redshifts.
\item Metal abundances are lower at all redshifts when the cluster cores
were excluded. This suggests that weak abundance gradients are present in
the ICM even at high redshifts where the influence of cool-cores is absent.
\item The surface brightness slope at \rf\ ($\beta_{500}$) is steeper on
average by $\sim15\%$ than a simple $\beta$-model fit to the entire profile.
\item There is a significant correlation between $\beta_{500}$ and cluster
temperature for local clusters, and an indication that the slope of this
relation is shallower at high redshifts.
\end{itemize}

An important factor in these results is likely to be the higher
cluster merger rate at high redshifts. This is predicted by cosmological
simulations \citep[\egc][]{coh05} and explains the absence of relaxed,
cool-core clusters at $z\ga0.5$ \citep{jel05,vik06c}. The
changing cool-core population in turn has an effect on the metal abundance
evolution, both because the metal abundances in the cool cores is higher
and because the high emissivity of those regions then biases the the global
emission-weighted abundances towards higher values. The fact that the
evolution is also present when the cores are excluded suggests that the
influence of cool cores is not the only effect. Indeed, the observed
evolution agrees with that predicted by \citet{ett05} based on models of
the rates star formation and subsequent supernova feedback with
redshift. Furthermore, enhanced star formation has been found in the
central galaxies of some cool core clusters
\citep[\egc][]{hic05}. This suggests that cool cores could be directly
influencing the rate of enrichment of the ICM, although the amount of
cool-core induced star formation is not well constrained. Finally, if the
slope of the $\beta_{500}-kT$ relation were confirmed to be shallower at
$z\ga0.5$, then this would suggest that the breaking of self-similarity in
the surface-brightness profiles was due to processes that become important
at lower redshifts. This could be related to the onset at lower redshifts
of cooling cores and their regulation by AGN outbursts \citep[\egc][]{nul05}.

As a final point, we remind the reader that the sample is not statistically
complete, and is subject to the different selection processes by which the
target clusters were originally chosen to be observed. As such,
unquantifiable biases could be present, and so some caution is required in
the interpretation of the results based on this sample. However, the
agreement between the results presented here and other studies suggest that
the impact of the heterogenous selection on our results is small.  In
subsequent papers we will use this large sample to study the cluster
scaling relations, their evolution, departures from self-similarity, and
the sources of scatter in the relations \citep[\egc][]{mau07c}.

\acknowledgments
This research has made use of the X-Rays Clusters Database (BAX) which is
operated by the Laboratoire d'Astrophysique de Tarbes-Toulouse (LATT),
under contract with the Centre National d'Etudes Spatiales (CNES). We also
used the NASA/IPAC Extragalactic Database (NED) which is operated by the
Jet Propulsion Laboratory, California Institute of Technology, under
contract with the National Aeronautics and Space Administration.  We thank
Alexey Vikhlinin for providing some of the software used in this work and
Diab Jerius for help assembling the list of clusters. BJM is supported by
NASA through Chandra Postdoctoral Fellowship Award Number PF4-50034 issued
by the Chandra X-ray Observatory Center, which is operated by the
Smithsonian Astrophysical Observatory for and on behalf of NASA under
contract NAS8-03060.


\begin{thebibliography}{}

\bibitem[\protect\citeauthoryear{{Akritas} \& {Bershady}}{{Akritas} \&
  {Bershady}}{1996}]{akr96}
{Akritas} M.~G.,  {Bershady} M.~A.,  1996, \apj, 470, 706

\bibitem[\protect\citeauthoryear{{Anders} \& {Grevesse}}{{Anders} \&
  {Grevesse}}{1989}]{and89}
{Anders} E.,  {Grevesse} N.,  1989, \gca, 53, 197

\bibitem[\protect\citeauthoryear{{Balestra}, {Tozzi}, {Ettori}, {Rosati},
  {Borgani}, {Mainieri}, {Norman} \& {Viola}}{{Balestra} et~al.}{2006}]{bal06a}
{Balestra} I.,  {Tozzi} P.,  {Ettori} S.,  {Rosati} P.,  {Borgani} S.,
  {Mainieri} V.,  {Norman} C.,    {Viola} M.,  2006, ArXiv Astrophysics
  e-prints

\bibitem[\protect\citeauthoryear{{Birkinshaw}}{{Birkinshaw}}{1999}]{bir99}
{Birkinshaw} M.,  1999, \physrep, 310, 97

\bibitem[\protect\citeauthoryear{{Bonamente}, {Joy}, {LaRoque}, {Carlstrom},
  {Reese} \& {Dawson}}{{Bonamente} et~al.}{2006}]{bon06}
{Bonamente} M.,  {Joy} M.~K.,  {LaRoque} S.~J.,  {Carlstrom} J.~E.,  {Reese}
  E.~D.,    {Dawson} K.~S.,  2006, \apj, 647, 25

\bibitem[\protect\citeauthoryear{{Buote} \& {Tsai}}{{Buote} \&
  {Tsai}}{1995}]{buo95}
{Buote} D.~A.,  {Tsai} J.~C.,  1995, \apj, 452, 522

\bibitem[\protect\citeauthoryear{Cavaliere \& Fusco-Femiano}{Cavaliere \&
  Fusco-Femiano}{1976}]{cav76}
Cavaliere A.,  Fusco-Femiano R.,  1976, \aap, 49, L137

\bibitem[\protect\citeauthoryear{{Cohn} \& {White}}{{Cohn} \&
  {White}}{2005}]{coh05}
{Cohn} J.~D.,  {White} M.,  2005, Astroparticle Physics, 24, 316

\bibitem[\protect\citeauthoryear{{De Grandi} \& {Molendi}}{{De Grandi} \&
  {Molendi}}{2001}]{deg01}
{De Grandi} S.,  {Molendi} S.,  2001, \apj, 551, 153

\bibitem[\protect\citeauthoryear{Dickey \& Lockman}{Dickey \&
  Lockman}{1990}]{dic90}
Dickey J.~M.,  Lockman F.~J.,  1990, \araa, 28, 215

\bibitem[\protect\citeauthoryear{{Ebeling}, {Edge}, {Bohringer}, {Allen},
  {Crawford}, {Fabian}, {Voges} \& {Huchra}}{{Ebeling} et~al.}{1998}]{ebe98}
{Ebeling} H.,  {Edge} A.~C.,  {Bohringer} H.,  {Allen} S.~W.,  {Crawford}
  C.~S.,  {Fabian} A.~C.,  {Voges} W.,    {Huchra} J.~P.,  1998, \mnras, 301,
  881

\bibitem[\protect\citeauthoryear{{Ebeling}, {White} \& {Rangarajan}}{{Ebeling}
  et~al.}{2006}]{ebe06a}
{Ebeling} H.,  {White} D.~A.,    {Rangarajan} F.~V.~N.,  2006, \mnras, 368, 65

\bibitem[\protect\citeauthoryear{{Ettori}}{{Ettori}}{2005}]{ett05}
{Ettori} S.,  2005, \mnras, 362, 110

\bibitem[\protect\citeauthoryear{Fabian}{Fabian}{1994}]{fab94b}
Fabian A.~C.,  1994, \araa, 32, 277

\bibitem[\protect\citeauthoryear{{Grevesse} \& {Sauval}}{{Grevesse} \&
  {Sauval}}{1998}]{gre98}
{Grevesse} N.,  {Sauval} A.~J.,  1998, Space Science Reviews, 85, 161

\bibitem[\protect\citeauthoryear{{Gutierrez} \& {Krawczynski}}{{Gutierrez} \&
  {Krawczynski}}{2005}]{gut05}
{Gutierrez} K.,  {Krawczynski} H.,  2005, \apj, 619, 161

\bibitem[\protect\citeauthoryear{{Hicks} \& {Mushotzky}}{{Hicks} \&
  {Mushotzky}}{2005}]{hic05}
{Hicks} A.~K.,  {Mushotzky} R.,  2005, \apjl, 635, L9

\bibitem[\protect\citeauthoryear{{Jeltema}, {Canizares}, {Bautz} \&
  {Buote}}{{Jeltema} et~al.}{2005}]{jel05}
{Jeltema} T.~E.,  {Canizares} C.~R.,  {Bautz} M.~W.,    {Buote} D.~A.,  2005,
  \apj, 624, 606

\bibitem[\protect\citeauthoryear{{Jones} \& {Forman}}{{Jones} \&
  {Forman}}{1984}]{jon84}
{Jones} C.,  {Forman} W.,  1984, \apj, 276, 38

\bibitem[\protect\citeauthoryear{Jones \& Forman}{Jones \&
  Forman}{1999}]{jon99}
Jones C.,  Forman W.,  1999, \apj, 511, 65

\bibitem[\protect\citeauthoryear{Kaastra \& Mewe}{Kaastra \&
  Mewe}{1993}]{kaa93}
Kaastra J.~S.,  Mewe R.,  1993, {\aaps}, 97, 443

\bibitem[\protect\citeauthoryear{{Kodama}, {Tanaka}, {Tamura}, {Yahagi},
  {Nagashima}, {Tanaka} \& {Arimoto}}{{Kodama} et~al.}{2005}]{kod05short}
{Kodama} T.,  {Tanaka} M.,  {Tamura} T.,  {Yahagi} H.,  {Nagashima} M.,
  {Tanaka} I.,    {Arimoto} N.,  2005, \pasj, 57, 309

\bibitem[\protect\citeauthoryear{{Kravtsov}, {Vikhlinin} \& {Nagai}}{{Kravtsov}
  et~al.}{2006}]{kra06a}
{Kravtsov} A.~V.,  {Vikhlinin} A.,    {Nagai} D.,  2006, \apj, 650, 128

\bibitem[\protect\citeauthoryear{{Lynds} \& {Petrosian}}{{Lynds} \&
  {Petrosian}}{1989}]{lyn89}
{Lynds} R.,  {Petrosian} V.,  1989, \apj, 336, 1

\bibitem[\protect\citeauthoryear{Markevitch}{Markevitch}{1998}]{mar98a}
Markevitch M.,  1998, \apj, 504, 27

\bibitem[\protect\citeauthoryear{{Markevitch}, {Gonzalez}, {Clowe},
  {Vikhlinin}, {Forman}, {Jones}, {Murray} \& {Tucker}}{{Markevitch}
  et~al.}{2004}]{mar04}
{Markevitch} M.,  {Gonzalez} A.~H.,  {Clowe} D.,  {Vikhlinin} A.,  {Forman} W.,
   {Jones} C.,  {Murray} S.,    {Tucker} W.,  2004, \apj, 606, 819

\bibitem[\protect\citeauthoryear{{Markevitch}, {Ponman}, {Nulsen}, {Bautz},
  {Burke}, {David}, {Davis}, {Donnelly}, {Forman}, {Jones} \& et.
  al.}{{Markevitch} et~al.}{2000}]{mar00c}
{Markevitch} M.,  {Ponman} T.~J.,  {Nulsen} P.~E.~J.,  {Bautz} M.~W.,  {Burke}
  D.~J.,  {David} L.~P.,  {Davis} D.,  {Donnelly} R.~H.,  {Forman} W.~R.,
  {Jones} C.,    et. al. 2000, \apj, 541, 542

\bibitem[\protect\citeauthoryear{{Maughan}, {Ellis}, {Jones}, {Mason},
  {C{\'o}rdova} \& {Priedhorsky}}{{Maughan} et~al.}{2006}]{mau06b}
{Maughan} B.~J.,  {Ellis} S.~C.,  {Jones} L.~R.,  {Mason} K.~O.,  {C{\'o}rdova}
  F.~A.,    {Priedhorsky} W.,  2006, \apj, 640, 219

\bibitem[\protect\citeauthoryear{{Maughan}, {Jones}, {Jones} \& {Van
  Speybroeck}}{{Maughan} et~al.}{2007}]{mau07a}
{Maughan} B.~J.,  {Jones} C.,  {Jones} L.~R.,    {Van Speybroeck} L.,  2007,
  \apj, 659, 1125

\bibitem[\protect\citeauthoryear{{Maughan}}{{Maughan}}{2007}]{mau07c}
{Maughan} B.~J.,  2007, \apj, in press

\bibitem[\protect\citeauthoryear{{McNamara}, {Wise}, {Nulsen}, {David},
  {Sarazin}, {Bautz}, {Markevitch}, {Vikhlinin}, {Forman}, {Jones} \&
  {Harris}}{{McNamara} et~al.}{2000}]{mcn00}
{McNamara} B.~R.,  {Wise} M.,  {Nulsen} P.~E.~J.,  {David} L.~P.,  {Sarazin}
  C.~L.,  {Bautz} M.,  {Markevitch} M.,  {Vikhlinin} A.,  {Forman} W.~R.,
  {Jones} C.,    {Harris} D.~E.,  2000, \apjl, 534, L135

\bibitem[\protect\citeauthoryear{{Mohr}, {Fabricant} \& {Geller}}{{Mohr}
  et~al.}{1993}]{moh93}
{Mohr} J.~J.,  {Fabricant} D.~G.,    {Geller} M.~J.,  1993, \apj, 413, 492

\bibitem[\protect\citeauthoryear{{Nulsen}, {McNamara}, {Wise} \&
  {David}}{{Nulsen} et~al.}{2005}]{nul05}
{Nulsen} P.~E.~J.,  {McNamara} B.~R.,  {Wise} M.~W.,    {David} L.~P.,  2005,
  \apj, 628, 629

\bibitem[\protect\citeauthoryear{{O'Hara}, {Mohr}, {Bialek} \&
  {Evrard}}{{O'Hara} et~al.}{2006}]{oha06}
{O'Hara} T.~B.,  {Mohr} J.~J.,  {Bialek} J.~J.,    {Evrard} A.~E.,  2006, \apj,
  639, 64

\bibitem[\protect\citeauthoryear{{Peres}, {Fabian}, {Edge}, {Allen},
  {Johnstone} \& {White}}{{Peres} et~al.}{1998}]{per98}
{Peres} C.~B.,  {Fabian} A.~C.,  {Edge} A.~C.,  {Allen} S.~W.,  {Johnstone}
  R.~M.,    {White} D.~A.,  1998, \mnras, 298, 416

\bibitem[\protect\citeauthoryear{{Peterson}, {Paerels}, {Kaastra}, {Arnaud},
  {Reiprich}, {Fabian}, {Mushotzky}, {Jernigan} \& {Sakelliou}}{{Peterson}
  et~al.}{2001}]{pet01}
{Peterson} J.~R.,  {Paerels} F.~B.~S.,  {Kaastra} J.~S.,  {Arnaud} M.,
  {Reiprich} T.~H.,  {Fabian} A.~C.,  {Mushotzky} R.~F.,  {Jernigan} J.~G.,
  {Sakelliou} I.,  2001, \aap, 365, L104

\bibitem[\protect\citeauthoryear{{Ponman}, {Cannon} \& {Navarro}}{{Ponman}
  et~al.}{1999}]{pon99}
{Ponman} T.~J.,  {Cannon} D.~B.,    {Navarro} J.~F.,  1999, \nat, 397, 135

\bibitem[\protect\citeauthoryear{{Poole}, {Fardal}, {Babul}, {McCarthy},
  {Quinn} \& {Wadsley}}{{Poole} et~al.}{2006}]{poo06}
{Poole} G.~B.,  {Fardal} M.~A.,  {Babul} A.,  {McCarthy} I.~G.,  {Quinn} T.,
  {Wadsley} J.,  2006, \mnras, pp 1285--+

\bibitem[\protect\citeauthoryear{{Pratt} \& {Arnaud}}{{Pratt} \&
  {Arnaud}}{2002}]{pra02}
{Pratt} G.~W.,  {Arnaud} M.,  2002, \aap, 394, 375

\bibitem[\protect\citeauthoryear{{Sanderson}, {Ponman}, {Finoguenov},
  {Lloyd-Davies} \& {Markevitch}}{{Sanderson} et~al.}{2003}]{san03}
{Sanderson} A.~J.~R.,  {Ponman} T.~J.,  {Finoguenov} A.,  {Lloyd-Davies} E.~J.,
     {Markevitch} M.,  2003, \mnras, 340, 989

\bibitem[\protect\citeauthoryear{{Sanderson}, {Ponman} \&
  {O'Sullivan}}{{Sanderson} et~al.}{2006}]{san06a}
{Sanderson} A.~J.~R.,  {Ponman} T.~J.,    {O'Sullivan} E.,  2006, \mnras, 372,
  1496

\bibitem[\protect\citeauthoryear{{Sarazin}}{{Sarazin}}{1986}]{sar86}
{Sarazin} C.~L.,  1986, Reviews of Modern Physics, 58, 1

\bibitem[\protect\citeauthoryear{{Smith}, {Brickhouse}, {Liedahl} \&
  {Raymond}}{{Smith} et~al.}{2001}]{smi01}
{Smith} R.~K.,  {Brickhouse} N.~S.,  {Liedahl} D.~A.,    {Raymond} J.~C.,
  2001, \apjl, 556, L91

\bibitem[\protect\citeauthoryear{{Snowden}, {Egger}, {Freyberg}, {McCammon},
  {Plucinsky}, {Sanders}, {Schmitt}, {Truemper} \& {Voges}}{{Snowden}
  et~al.}{1997}]{sno97}
{Snowden} S.~L.,  {Egger} R.,  {Freyberg} M.~J.,  {McCammon} D.,  {Plucinsky}
  P.~P.,  {Sanders} W.~T.,  {Schmitt} J.~H.~M.~M.,  {Truemper} J.,    {Voges}
  W.,  1997, \apj, 485, 125

\bibitem[\protect\citeauthoryear{{Tyson}, {Wenk} \& {Valdes}}{{Tyson}
  et~al.}{1990}]{tys90}
{Tyson} J.~A.,  {Wenk} R.~A.,    {Valdes} F.,  1990, \apjl, 349, L1

\bibitem[\protect\citeauthoryear{{Vikhlinin}, {Burenin}, {Forman}, {Jones},
  {Hornstrup}, {Murray} \& {Quintana}}{{Vikhlinin} et~al.}{2006}]{vik06c}
{Vikhlinin} A.,  {Burenin} R.,  {Forman} W.~R.,  {Jones} C.,  {Hornstrup} A.,
  {Murray} S.~S.,    {Quintana} H.,  2006, ArXiv Astrophysics e-prints

\bibitem[\protect\citeauthoryear{{Vikhlinin}, {Forman} \& {Jones}}{{Vikhlinin}
  et~al.}{1999}]{vik99}
{Vikhlinin} A.,  {Forman} W.,    {Jones} C.,  1999, \apj, 525, 47

\bibitem[\protect\citeauthoryear{{Vikhlinin}, {Kravtsov}, {Forman}, {Jones},
  {Markevitch}, {Murray} \& {Van Speybroeck}}{{Vikhlinin}
  et~al.}{2006}]{vik06a}
{Vikhlinin} A.,  {Kravtsov} A.,  {Forman} W.,  {Jones} C.,  {Markevitch} M.,
  {Murray} S.~S.,    {Van Speybroeck} L.,  2006, \apj, 640, 691

\bibitem[\protect\citeauthoryear{{Vikhlinin}, {Markevitch}, {Murray}, {Jones},
  {Forman} \& {Van Speybroeck}}{{Vikhlinin} et~al.}{2005}]{vik05a}
{Vikhlinin} A.,  {Markevitch} M.,  {Murray} S.~S.,  {Jones} C.,  {Forman} W.,
   {Van Speybroeck} L.,  2005, \apj, 628, 655

\bibitem[\protect\citeauthoryear{{Vikhlinin}, {McNamara}, {Forman}, {Jones},
  {Quintana} \& {Hornstrup}}{{Vikhlinin} et~al.}{1998}]{vik98b}
{Vikhlinin} A.,  {McNamara} B.~R.,  {Forman} W.,  {Jones} C.,  {Quintana} H.,
   {Hornstrup} A.,  1998, \apj, 502, 558

\bibitem[\protect\citeauthoryear{{Zwicky}}{{Zwicky}}{1937}]{zwi37}
{Zwicky} F.,  1937, \apj, 86, 217

\end{thebibliography}

\LongTables
\clearpage

\begin{deluxetable}{lccccccc}
\tablecaption{Summary of the observations used. Column 2
gives the \Chandra\ observation identifier, and columns 3 and 4 are the
ICRS equatorial coordinates of the cluster X-ray centroid. Column
5 gives the redshift of each cluster, and columns 6 and 7 give the
date of each observation and corresponding ``blank-sky'' background
period. Finally, column 8 lists the cleaned exposure time of each
observation. \label{t.cat}}
\tabletypesize{\small}
\tablehead{
\colhead{Cluster} & \colhead{ObsID} & \colhead{RA} & \colhead{DEC} & \colhead{z} & \colhead{Date} & \colhead{BG} & \colhead{Exposure (ks)}
}
\startdata
MS0015.9+1609	& 520	& 00:18:33.63	& +16:26:14.6	& 0.541	& 2000-08-18	& C	& 55.7\\
RXJ0027.6+2616	& 3249	& 00:27:45.92	& +26:16:19.9	& 0.367	& 2002-06-26	& D	& 9.1\\
CLJ0030+2618	& 5762	& 00:30:33.77	& +26:18:09.7	& 0.500	& 2005-05-28	& D	& 16.1\\
A68	& 3250	& 00:37:06.11	& +09:09:33.6	& 0.255	& 2002-09-07	& D	& 7.3\\
A115	& 3233	& 00:55:50.69	& +26:24:37.4	& 0.197	& 2002-10-07	& D	& 43.6\\
A209	& 3579	& 01:31:53.47	& -13:36:46.1	& 0.206	& 2003-08-03	& D	& 8.8\\
A209	& 522	& 01:31:53.59	& -13:36:46.1	& 0.206	& 2000-09-09	& C	& 7.5\\
CLJ0152.7-1357	& 913	& 01:52:40.03	& -13:58:24.2	& 0.831	& 2000-09-08	& C	& 30.6\\
A267	& 1448	& 01:52:42.12	& +01:00:41.4	& 0.230	& 1999-10-16	& B	& 6.5\\
MACSJ0159.8-0849	& 6106	& 01:59:49.32	& -08:50:00.7	& 0.405	& 2004-12-04	& D	& 30.3\\
MACSJ0159.8-0849	& 3265	& 01:59:49.44	& -08:50:00.6	& 0.405	& 2002-10-02	& D	& 15.0\\
CLJ0216-1747	& 6393	& 02:16:32.93	& -17:47:30.5	& 0.578	& 2005-10-04	& D	& 20.0\\
RXJ0232.2-4420	& 4993	& 02:32:18.34	& -44:20:50.3	& 0.284	& 2004-06-08	& D	& 15.0\\
MACSJ0242.5-2132	& 3266	& 02:42:35.86	& -21:32:25.8	& 0.314	& 2002-02-07	& D	& 8.8\\
A383	& 2320	& 02:48:03.43	& -03:31:46.2	& 0.187	& 2000-11-16	& C	& 15.8\\
A383	& 524	& 02:48:03.65	& -03:31:44.9	& 0.187	& 2000-09-08	& C	& 8.8\\
MACSJ0257.6-2209	& 3267	& 02:57:41.33	& -22:09:14.4	& 0.322	& 2001-11-12	& D	& 16.6\\
MS0302.7+1658	& 525	& 03:05:31.63	& +17:10:10.6	& 0.424	& 2000-10-03	& C	& 9.1\\
CLJ0318-0302	& 5775	& 03:18:33.60	& -03:02:55.3	& 0.370	& 2005-03-15	& D	& 12.7\\
MACSJ0329.6-0211	& 3582	& 03:29:41.54	& -02:11:45.9	& 0.450	& 2002-12-24	& D	& 14.5\\
MACSJ0404.6+1109	& 3269	& 04:04:32.66	& +11:08:16.8	& 0.355	& 2002-02-20	& D	& 18.7\\
MACSJ0429.6-0253	& 3271	& 04:29:35.95	& -02:53:06.1	& 0.399	& 2002-02-07	& D	& 20.0\\
RXJ0439.0+0715	& 3583	& 04:39:00.67	& +07:16:03.8	& 0.230	& 2003-01-04	& D	& 16.1\\
RXJ0439+0520	& 527	& 04:39:02.35	& +05:20:43.7	& 0.208	& 2000-08-29	& C	& 8.0\\
MACSJ0451.9+0006	& 5815	& 04:51:54.41	& +00:06:19.2	& 0.430	& 2005-01-08	& D	& 8.0\\
A521	& 901	& 04:54:06.58	& -10:13:15.2	& 0.253	& 1999-12-23	& B	& 35.8\\
A520	& 528	& 04:54:09.60	& +02:55:20.8	& 0.199	& 2000-10-10	& C	& 8.5\\
A520	& 4215	& 04:54:09.72	& +02:55:17.8	& 0.199	& 2003-12-04	& D	& 48.7\\
MS0451.6-0305	& 529	& 04:54:11.26	& -03:00:52.6	& 0.550	& 2000-01-14	& B	& 12.7\\
CLJ0522-3625	& 4926	& 05:22:14.66	& -36:24:58.7	& 0.472	& 2004-06-17	& D	& 15.8\\
CLJ0542.8-4100	& 914	& 05:42:50.14	& -41:00:02.2	& 0.634	& 2000-07-26	& C	& 44.0\\
MACSJ0647.7+7015	& 3584	& 06:47:49.92	& +70:14:56.0	& 0.584	& 2003-10-07	& D	& 17.9\\
MACSJ0647.7+7015	& 3196	& 06:47:50.16	& +70:14:55.7	& 0.584	& 2002-10-31	& D	& 15.3\\
1E0657-56	& 3184	& 06:58:29.52	& -55:56:39.1	& 0.296	& 2002-07-12	& D	& 69.5\\
1E0657-56	& 554	& 06:58:30.00	& -55:56:39.5	& 0.296	& 2000-10-16	& C	& 22.3\\
MACSJ0717.5+3745	& 4200	& 07:17:31.68	& +37:45:32.0	& 0.546	& 2003-01-08	& D	& 51.8\\
A586	& 530	& 07:32:20.16	& +31:37:54.5	& 0.171	& 2000-09-05	& C	& 8.6\\
MACSJ0744.9+3927	& 3585	& 07:44:52.32	& +39:27:25.2	& 0.697	& 2003-01-04	& D	& 14.8\\
MACSJ0744.9+3927	& 6111	& 07:44:52.32	& +39:27:27.0	& 0.697	& 2004-12-03	& D	& 39.9\\
MACSJ0744.9+3927	& 3197	& 07:44:52.56	& +39:27:26.3	& 0.697	& 2001-11-12	& D	& 16.6\\
A665	& 531	& 08:30:53.04	& +65:49:55.2	& 0.182	& 1999-12-29	& B	& 7.8\\
A665	& 3586	& 08:30:59.04	& +65:50:44.2	& 0.182	& 2002-12-28	& D	& 26.2\\
A697	& 4217	& 08:42:57.84	& +36:21:57.2	& 0.282	& 2002-12-15	& D	& 16.1\\
CLJ0848.7+4456	& 1708	& 08:48:47.76	& +44:56:12.5	& 0.574	& 2000-05-03	& C	& 44.3\\
CLJ0848.7+4456	& 927	& 08:48:50.88	& +44:55:27.8	& 0.574	& 2000-05-04	& C	& 95.1\\
ZWCLJ1953	& 1659	& 08:50:06.96	& +36:04:18.1	& 0.320	& 2000-10-22	& C	& 17.9\\
CLJ0853+5759	& 5765	& 08:53:14.64	& +57:59:47.8	& 0.475	& 2005-02-19	& D	& 22.0\\
CLJ0853+5759	& 4925	& 08:53:14.88	& +58:00:02.9	& 0.475	& 2004-09-19	& D	& 14.3\\
MS0906.5+1110	& 924	& 09:09:12.72	& +10:58:33.6	& 0.180	& 2000-10-02	& C	& 26.4\\
RXJ0910+5422	& 2452	& 09:10:44.40	& +54:22:04.4	& 1.110	& 2001-04-24	& D	& 54.4\\
RXJ0910+5422	& 2227	& 09:10:45.36	& +54:22:07.3	& 1.110	& 2001-04-29	& D	& 84.2\\
A773	& 5006	& 09:17:52.80	& +51:43:40.4	& 0.217	& 2004-01-21	& D	& 15.8\\
A773	& 3588	& 09:17:53.04	& +51:43:37.9	& 0.217	& 2003-01-25	& D	& 8.8\\
A773	& 533	& 09:17:53.04	& +51:43:39.4	& 0.217	& 2000-09-05	& C	& 9.8\\
A781	& 534	& 09:20:26.16	& +30:30:04.7	& 0.298	& 2000-10-03	& C	& 8.6\\
CLJ0926+1242	& 4929	& 09:26:36.48	& +12:43:04.8	& 0.489	& 2004-02-06	& D	& 16.3\\
CLJ0926+1242	& 5838	& 09:26:36.48	& +12:43:02.3	& 0.489	& 2005-02-21	& D	& 27.2\\
RBS797	& 2202	& 09:47:13.20	& +76:23:13.6	& 0.354	& 2000-10-20	& C	& 9.8\\
MACSJ0949.8+1708	& 3274	& 09:49:51.84	& +17:07:08.0	& 0.384	& 2002-11-06	& D	& 12.9\\
CLJ0956+4107	& 5759	& 09:56:03.12	& +41:07:14.2	& 0.587	& 2005-01-28	& D	& 34.7\\
CLJ0956+4107	& 5294	& 09:56:03.36	& +41:07:13.1	& 0.587	& 2003-12-30	& D	& 13.5\\
A907	& 535	& 09:58:21.84	& -11:03:49.0	& 0.153	& 2000-06-29	& C	& 9.1\\
A907	& 3205	& 09:58:22.08	& -11:03:49.3	& 0.153	& 2002-10-30	& D	& 37.1\\
A907	& 3185	& 09:58:22.08	& -11:03:50.0	& 0.153	& 2002-06-14	& D	& 38.4\\
MS1006.0+1202	& 925	& 10:08:47.52	& +11:47:40.6	& 0.221	& 2000-06-22	& C	& 23.3\\
MS1008.1-1224	& 926	& 10:10:32.16	& -12:39:30.2	& 0.301	& 2000-06-11	& C	& 14.5\\
ZW3146	& 909	& 10:23:39.60	& +04:11:11.3	& 0.291	& 2000-05-10	& C	& 40.4\\
CLJ1113.1-2615	& 915	& 11:13:05.28	& -26:15:41.0	& 0.725	& 2000-08-13	& C	& 44.6\\
A1204	& 2205	& 11:13:20.40	& +17:35:39.1	& 0.171	& 2001-06-01	& D	& 19.4\\
CLJ1117+1745	& 4933	& 11:17:30.00	& +17:44:52.1	& 0.305	& 2004-06-18	& D	& 16.6\\
CLJ1117+1745	& 5836	& 11:17:30.00	& +17:44:52.8	& 0.548	& 2005-02-15	& D	& 39.9\\
CLJ1120+4318	& 5771	& 11:20:07.20	& +43:18:06.1	& 0.600	& 2005-01-11	& D	& 16.6\\
RXJ1121+2327	& 1660	& 11:20:57.36	& +23:26:29.0	& 0.562	& 2001-04-23	& D	& 62.7\\
A1240	& 4961	& 11:23:37.68	& +43:05:44.5	& 0.159	& 2005-02-05	& D	& 43.5\\
MACSJ1131.8-1955	& 3276	& 11:31:55.20	& -19:55:52.7	& 0.307	& 2002-06-14	& D	& 11.1\\
MS1137.5+6625	& 536	& 11:40:22.32	& +66:08:16.1	& 0.782	& 1999-09-30	& B	& 103.1\\
MACSJ1149.5+2223	& 3589	& 11:49:35.28	& +22:24:09.0	& 0.545	& 2003-02-07	& D	& 15.0\\
A1413	& 5003	& 11:55:18.00	& +23:24:17.6	& 0.143	& 2004-03-06	& D	& 60.9\\
A1413	& 1661	& 11:55:18.00	& +23:24:14.4	& 0.143	& 2001-05-16	& D	& 9.3\\
CLJ1213+0253	& 4934	& 12:13:35.04	& +02:53:46.4	& 0.409	& 2004-07-17	& D	& 16.1\\
CLJ1216+2633	& 4931	& 12:16:19.68	& +26:33:13.7	& 0.428	& 2004-05-12	& D	& 15.3\\
RXJ1221+4918	& 1662	& 12:21:26.40	& +49:18:28.1	& 0.700	& 2001-08-05	& D	& 66.4\\
CLJ1226.9+3332	& 5014	& 12:26:57.84	& +33:32:47.8	& 0.890	& 2004-08-07	& D	& 25.1\\
CLJ1226.9+3332	& 3180	& 12:26:57.84	& +33:32:47.8	& 0.890	& 2003-01-27	& D	& 25.4\\
RXJ1234.2+0947	& 539	& 12:34:18.00	& +09:46:18.8	& 0.229	& 2000-03-23	& C	& 8.6\\
RDCS1252-29	& 4198	& 12:52:54.96	& -29:27:20.5	& 1.237	& 2003-03-20	& D	& 133.2\\
A1682	& 3244	& 13:06:51.12	& +46:33:29.5	& 0.234	& 2002-10-19	& D	& 7.3\\
MACSJ1311.0-0310	& 6110	& 13:11:01.68	& -03:10:37.6	& 0.494	& 2005-04-20	& D	& 52.6\\
MACSJ1311.0-0310	& 3258	& 13:11:01.92	& -03:10:36.0	& 0.494	& 2002-12-15	& D	& 13.0\\
A1689	& 5004	& 13:11:29.52	& -01:20:29.8	& 0.183	& 2004-02-28	& D	& 16.6\\
A1689	& 1663	& 13:11:29.52	& -01:20:28.7	& 0.183	& 2001-01-07	& D	& 8.8\\
A1689	& 540	& 13:11:29.52	& -01:20:30.4	& 0.183	& 2000-04-15	& C	& 8.8\\
RXJ1317.4+2911	& 2228	& 13:17:20.88	& +29:11:15.0	& 0.805	& 2001-05-04	& D	& 88.4\\
CLJ1334+5031	& 5772	& 13:34:18.72	& +50:31:02.3	& 0.620	& 2005-08-05	& D	& 15.3\\
A1763	& 3591	& 13:35:18.24	& +40:59:59.3	& 0.223	& 2003-08-28	& D	& 17.4\\
RXJ1347.5-1145	& 3592	& 13:47:30.72	& -11:45:10.4	& 0.451	& 2003-09-03	& D	& 45.9\\
RXJ1350.0+6007	& 2229	& 13:50:48.72	& +60:07:02.6	& 0.804	& 2001-08-29	& D	& 47.7\\
CLJ1354-0221	& 5835	& 13:54:17.04	& -02:21:52.3	& 0.546	& 2005-05-17	& D	& 29.8\\
CLJ1354-0221	& 4932	& 13:54:17.28	& -02:21:44.8	& 0.546	& 2004-12-07	& D	& 15.3\\
CLJ1415.1+3612	& 4163	& 14:15:11.04	& +36:12:03.6	& 1.030	& 2003-09-16	& D	& 72.9\\
RXJ1416+4446	& 541	& 14:16:28.08	& +44:46:42.6	& 0.400	& 1999-12-02	& B	& 27.7\\
MACSJ1423.8+2404	& 1657	& 14:23:47.76	& +24:04:40.8	& 0.543	& 2001-06-01	& D	& 16.9\\
A1914	& 3593	& 14:26:01.20	& +37:49:34.0	& 0.171	& 2003-09-03	& D	& 16.1\\
A1914	& 542	& 14:26:01.20	& +37:49:35.4	& 0.171	& 1999-11-21	& B	& 7.2\\
A1942	& 3290	& 14:38:22.08	& +03:40:06.2	& 0.224	& 2002-03-13	& D	& 52.6\\
MS1455.0+2232	& 543	& 14:57:15.12	& +22:20:34.4	& 0.258	& 2000-05-19	& C	& 9.6\\
MS1455.0+2232	& 4192	& 14:57:15.12	& +22:20:34.8	& 0.258	& 2003-09-05	& D	& 73.1\\
RXJ1504-0248	& 4935	& 15:04:07.44	& -02:48:18.4	& 0.215	& 2004-01-07	& D	& 12.2\\
A2034	& 2204	& 15:10:12.48	& +33:30:28.4	& 0.113	& 2001-05-05	& D	& 51.6\\
A2069	& 4965	& 15:24:09.36	& +29:53:10.0	& 0.116	& 2004-05-31	& D	& 53.4\\
RXJ1525+0958	& 1664	& 15:24:39.84	& +09:57:42.6	& 0.516	& 2002-04-01	& D	& 43.6\\
RXJ1532.9+3021	& 1665	& 15:32:53.76	& +30:20:59.3	& 0.345	& 2001-09-06	& D	& 8.8\\
A2111	& 544	& 15:39:41.28	& +34:25:10.2	& 0.229	& 2000-03-22	& C	& 9.1\\
A2125	& 2207	& 15:41:08.64	& +66:15:51.8	& 0.246	& 2001-08-24	& D	& 71.3\\
A2163	& 545	& 16:15:45.60	& -06:08:56.5	& 0.203	& 2000-07-29	& C	& 9.1\\
MACSJ1621.3+3810	& 3254	& 16:21:24.72	& +38:10:07.3	& 0.463	& 2002-10-18	& D	& 8.3\\
MACSJ1621.3+3810	& 6109	& 16:21:24.72	& +38:10:10.2	& 0.463	& 2004-12-11	& D	& 32.9\\
MACSJ1621.3+3810	& 6172	& 16:21:24.72	& +38:10:09.5	& 0.463	& 2004-12-25	& D	& 23.1\\
MS1621.5+2640	& 546	& 16:23:35.28	& +26:34:19.9	& 0.426	& 2000-04-24	& C	& 26.2\\
A2204	& 6104	& 16:32:47.04	& +05:34:32.5	& 0.152	& 2004-09-20	& D	& 9.1\\
A2218	& 1666	& 16:35:52.08	& +66:12:34.6	& 0.176	& 2001-08-30	& D	& 37.3\\
A2218	& 1454	& 16:35:52.08	& +66:12:34.2	& 0.176	& 1999-10-19	& B	& 10.1\\
CLJ1641+4001	& 3575	& 16:41:52.80	& +40:01:40.4	& 0.464	& 2003-09-24	& D	& 36.0\\
RXJ1701+6414	& 547	& 17:01:23.04	& +64:14:11.4	& 0.453	& 2000-10-31	& C	& 38.4\\
RXJ1716.9+6708	& 548	& 17:16:49.44	& +67:08:26.9	& 0.813	& 2000-02-27	& C	& 42.5\\
A2259	& 3245	& 17:20:08.64	& +27:40:09.8	& 0.164	& 2002-09-16	& D	& 7.8\\
RXJ1720.1+2638	& 1453	& 17:20:10.08	& +26:37:30.0	& 0.164	& 1999-10-19	& B	& 7.0\\
RXJ1720.1+2638	& 3224	& 17:20:10.08	& +26:37:29.3	& 0.164	& 2002-10-03	& D	& 18.4\\
RXJ1720.1+2638	& 4361	& 17:20:10.08	& +26:37:29.3	& 0.164	& 2002-08-19	& D	& 20.2\\
MACSJ1720.2+3536	& 3280	& 17:20:16.56	& +35:36:21.6	& 0.387	& 2002-11-03	& D	& 17.4\\
A2261	& 5007	& 17:22:27.12	& +32:07:56.6	& 0.224	& 2004-01-14	& D	& 19.7\\
A2261	& 550	& 17:22:27.12	& +32:07:57.4	& 0.224	& 1999-12-11	& B	& 8.3\\
A2294	& 3246	& 17:24:09.60	& +85:53:11.0	& 0.178	& 2001-12-24	& D	& 8.3\\
MACSJ1824.3+4309	& 3255	& 18:24:18.96	& +43:09:49.0	& 0.487	& 2002-09-14	& D	& 12.4\\
MACSJ1931.8-2634	& 3282	& 19:31:49.68	& -26:34:32.9	& 0.352	& 2002-10-20	& D	& 12.4\\
RXJ2011.3-5725	& 4995	& 20:11:27.12	& -57:25:09.8	& 0.279	& 2004-06-08	& D	& 21.8\\
MS2053.7-0449	& 1667	& 20:56:21.12	& -04:37:47.2	& 0.583	& 2001-10-07	& D	& 37.1\\
MS2053.7-0449	& 551	& 20:56:21.12	& -04:37:46.6	& 0.583	& 2000-05-13	& C	& 38.4\\
MACSJ2129.4-0741	& 3595	& 21:29:25.92	& -07:41:30.0	& 0.594	& 2003-10-18	& D	& 16.9\\
RXJ2129.6+0005	& 552	& 21:29:40.08	& +00:05:19.6	& 0.235	& 2000-10-21	& C	& 8.5\\
A2409	& 3247	& 22:00:53.04	& +20:58:27.8	& 0.148	& 2002-10-08	& D	& 9.1\\
MACSJ2228.5+2036	& 3285	& 22:28:32.88	& +20:37:11.6	& 0.412	& 2003-01-22	& D	& 16.1\\
MACSJ2229.7-2755	& 3286	& 22:29:45.36	& -27:55:36.5	& 0.324	& 2002-11-13	& D	& 12.7\\
MACSJ2245.0+2637	& 3287	& 22:45:04.80	& +26:38:03.1	& 0.301	& 2002-11-24	& D	& 12.7\\
RXJ2247+0337	& 911	& 22:47:28.08	& +03:36:59.5	& 0.200	& 2000-05-11	& C	& 41.5\\
AS1063	& 4966	& 22:48:44.88	& -44:31:44.4	& 0.252	& 2004-05-17	& D	& 22.3\\
CLJ2302.8+0844	& 918	& 23:02:48.00	& +08:43:53.0	& 0.722	& 2000-08-05	& C	& 87.9\\
A2631	& 3248	& 23:37:37.92	& +00:16:07.5	& 0.273	& 2002-07-08	& D	& 8.0\\
\enddata
\end{deluxetable}

\clearpage

\begin{deluxetable}{lcccc}
\tablecaption{Structural properties of the clusters. $R_w$ is the maximum
radius within which the centroid shift ($\langle w \rangle$) could be
measured. $\beta_{500}$ is the logarithmic slope of the surface brightness
profile at \rf. For clusters marked with a $\dagger$, the emission was not
detected to a large enough radius to measure $\beta_{500}$.\label{t.morpho}}
\tablehead{
\colhead{Cluster} & \colhead{e} & \colhead{$\langle w \rangle$
$(10^{-2}\rf)$} & \colhead{$R_w$ $(\rf)$} & \colhead{$\beta_{500}$}
}
\startdata
MS0015.9+1609	& $0.18^{+0.01}_{-0.01}$	& $0.90\pm0.08$	& 1.00	& $0.58^{+0.04}_{-0.04}$\\
RXJ0027.6+2616	& $0.20^{+0.05}_{-0.05}$	& $1.62\pm0.14$	& 1.00	& $0.62^{+0.14}_{-0.13}$\\
CLJ0030+2618	& $0.30^{+0.06}_{-0.06}$	& $3.63\pm0.33$	& 1.00	& $0.68^{+0.47}_{-0.30}$\\
A68	& $0.31^{+0.02}_{-0.02}$	& $1.19\pm0.10$	& 1.00	& $0.78^{+0.18}_{-0.16}$\\
A115	& $0.25^{+0.01}_{-0.01}$	& $6.95\pm0.59$	& 1.00	& $0.64^{+0.02}_{-0.02}$\\
A209	& $0.21^{+0.01}_{-0.01}$	& $0.55\pm0.05$	& 1.00	& $0.73^{+0.04}_{-0.04}$\\
CLJ0152.7-1357S	& $0.42^{+0.05}_{-0.06}$	& $4.99\pm0.47$	& 1.00	& $0.44^{+0.29}_{-0.28}$\\
A267	& $0.28^{+0.02}_{-0.02}$	& $3.21\pm0.28$	& 1.00	& $0.79^{+0.10}_{-0.10}$\\
CLJ0152.7-1357N	& $0.06^{+0.07}_{-0.06}$	& $2.34\pm0.22$	& 1.00	& $0.91^{+0.21}_{-0.18}$\\
MACSJ0159.8-0849	& $0.06^{+0.01}_{-0.01}$	& $0.34\pm0.03$	& 1.00	& $0.69^{+0.06}_{-0.05}$\\
CLJ0216-1747$^\dagger$	& $0.46^{+0.07}_{-0.09}$	& $1.18\pm0.10$	& 1.00	& $\cdots$\\
RXJ0232.2-4420	& $0.17^{+0.01}_{-0.01}$	& $2.24\pm0.19$	& 1.00	& $0.96^{+0.15}_{-0.14}$\\
MACSJ0242.5-2132	& $0.00^{+0.18}_{-0.00}$	& $0.33\pm0.03$	& 1.00	& $0.91^{+0.18}_{-0.16}$\\
A383	& $0.04^{+0.01}_{-0.01}$	& $0.18\pm0.02$	& 1.00	& $0.68^{+0.09}_{-0.08}$\\
MACSJ0257.6-2209	& $0.14^{+0.02}_{-0.02}$	& $0.43\pm0.04$	& 1.00	& $0.83^{+0.13}_{-0.12}$\\
MS0302.7+1658	& $0.36^{+0.05}_{-0.05}$	& $2.00\pm0.18$	& 1.00	& $0.28^{+0.27}_{-0.29}$\\
CLJ0318-0302	& $0.16^{+0.05}_{-0.05}$	& $3.09\pm0.28$	& 1.00	& $0.57^{+0.26}_{-0.22}$\\
MACSJ0329.6-0211	& $0.15^{+0.03}_{-0.03}$	& $1.38\pm0.13$	& 1.00	& $0.93^{+0.11}_{-0.10}$\\
MACSJ0404.6+1109	& $0.28^{+0.03}_{-0.03}$	& $2.92\pm0.25$	& 1.00	& $0.43^{+0.06}_{-0.06}$\\
MACSJ0429.6-0253	& $0.21^{+0.02}_{-0.02}$	& $0.38\pm0.03$	& 1.00	& $0.79^{+0.16}_{-0.14}$\\
RXJ0439.0+0715	& $0.20^{+0.01}_{-0.01}$	& $0.57\pm0.05$	& 1.00	& $0.88^{+0.09}_{-0.09}$\\
RXJ0439+0520	& $0.11^{+0.02}_{-0.03}$	& $0.30\pm0.03$	& 1.00	& $0.47^{+0.14}_{-0.14}$\\
MACSJ0451.9+0006	& $0.00^{+0.61}_{-0.00}$	& $1.59\pm0.14$	& 1.00	& $0.70^{+0.13}_{-0.12}$\\
A521	& $0.20^{+0.01}_{-0.01}$	& $4.18\pm0.36$	& 1.00	& $0.78^{+0.02}_{-0.02}$\\
A520	& $0.07^{+0.01}_{-0.01}$	& $6.43\pm0.54$	& 1.00	& $0.94^{+0.03}_{-0.03}$\\
MS0451.6-0305	& $0.25^{+0.02}_{-0.02}$	& $0.94\pm0.08$	& 1.00	& $0.73^{+0.09}_{-0.09}$\\
CLJ0522-3625	& $0.32^{+0.06}_{-0.07}$	& $1.29\pm0.12$	& 1.00	& $0.88^{+0.31}_{-0.23}$\\
CLJ0542.8-4100	& $0.27^{+0.03}_{-0.03}$	& $3.78\pm0.34$	& 1.00	& $0.56^{+0.16}_{-0.15}$\\
MACSJ0647.7+7015	& $0.36^{+0.02}_{-0.02}$	& $0.62\pm0.06$	& 1.00	& $0.58^{+0.10}_{-0.10}$\\
1E0657-56	& $0.13^{+0.00}_{-0.00}$	& $2.29\pm0.19$	& 1.00	& $1.04^{+0.04}_{-0.03}$\\
MACSJ0717.5+3745	& $0.30^{+0.01}_{-0.01}$	& $2.11\pm0.18$	& 1.00	& $0.84^{+0.04}_{-0.04}$\\
A586	& $0.08^{+0.02}_{-0.02}$	& $0.26\pm0.02$	& 0.81	& $0.80^{+0.14}_{-0.12}$\\
MACSJ0744.9+3927	& $0.11^{+0.02}_{-0.02}$	& $1.41\pm0.13$	& 1.00	& $0.65^{+0.04}_{-0.04}$\\
A665	& $0.07^{+0.01}_{-0.01}$	& $3.40\pm0.29$	& 0.77	& $0.72^{+0.02}_{-0.02}$\\
A697	& $0.23^{+0.01}_{-0.01}$	& $0.41\pm0.03$	& 1.00	& $0.73^{+0.05}_{-0.05}$\\
CLJ0848.7+4456	& $0.34^{+0.05}_{-0.05}$	& $1.06\pm0.10$	& 1.00	& $0.28^{+0.17}_{-0.18}$\\
ZWCLJ1953	& $0.18^{+0.02}_{-0.02}$	& $2.39\pm0.20$	& 1.00	& $0.84^{+0.12}_{-0.11}$\\
CLJ0853+5759	& $0.36^{+0.05}_{-0.05}$	& $1.54\pm0.14$	& 1.00	& $0.67^{+0.54}_{-0.38}$\\
MS0906.5+1110	& $0.19^{+0.01}_{-0.01}$	& $4.82\pm0.42$	& 0.95	& $0.54^{+0.04}_{-0.04}$\\
RXJ0910+5422$^\dagger$	& $0.07^{+0.10}_{-0.07}$	& $1.68\pm0.16$	& 1.00	& $\cdots$\\
A773	& $0.21^{+0.01}_{-0.01}$	& $1.05\pm0.09$	& 1.00	& $0.60^{+0.04}_{-0.04}$\\
A781	& $0.13^{+0.03}_{-0.03}$	& $3.88\pm0.34$	& 1.00	& $0.55^{+0.08}_{-0.08}$\\
CLJ0926+1242	& $0.37^{+0.04}_{-0.04}$	& $0.36\pm0.03$	& 1.00	& $0.60^{+0.17}_{-0.16}$\\
RBS797	& $0.27^{+0.01}_{-0.01}$	& $0.21\pm0.02$	& 1.00	& $0.88^{+0.16}_{-0.14}$\\
MACSJ0949.8+1708	& $0.15^{+0.02}_{-0.02}$	& $1.07\pm0.09$	& 1.00	& $0.61^{+0.06}_{-0.06}$\\
CLJ0956+4107	& $0.42^{+0.04}_{-0.04}$	& $3.90\pm0.36$	& 1.00	& $0.66^{+0.22}_{-0.20}$\\
A907	& $0.29^{+0.00}_{-0.00}$	& $0.81\pm0.07$	& 1.00	& $0.92^{+0.05}_{-0.04}$\\
MS1006.0+1202	& $0.24^{+0.02}_{-0.02}$	& $2.19\pm0.19$	& 1.00	& $0.70^{+0.09}_{-0.09}$\\
MS1008.1-1224	& $0.18^{+0.03}_{-0.03}$	& $4.87\pm0.43$	& 1.00	& $0.75^{+0.15}_{-0.13}$\\
ZW3146	& $0.18^{+0.01}_{-0.01}$	& $0.23\pm0.02$	& 1.00	& $0.78^{+0.08}_{-0.07}$\\
CLJ1113.1-2615$^\dagger$	& $0.00^{+0.38}_{-0.00}$	& $0.88\pm0.08$	& 1.00	& $\cdots$\\
A1204	& $0.16^{+0.01}_{-0.01}$	& $0.34\pm0.03$	& 1.00	& $0.57^{+0.08}_{-0.08}$\\
CLJ1117+1745	& $0.28^{+0.06}_{-0.07}$	& $1.43\pm0.13$	& 1.00	& $0.76^{+0.23}_{-0.19}$\\
CLJ1120+4318	& $0.14^{+0.05}_{-0.05}$	& $0.93\pm0.08$	& 1.00	& $1.01^{+0.27}_{-0.21}$\\
RXJ1121+2327	& $0.16^{+0.03}_{-0.03}$	& $1.39\pm0.13$	& 1.00	& $0.84^{+0.13}_{-0.12}$\\
A1240	& $0.28^{+0.02}_{-0.02}$	& $1.66\pm0.15$	& 1.00	& $1.05^{+0.06}_{-0.06}$\\
MACSJ1131.8-1955	& $0.31^{+0.02}_{-0.02}$	& $2.91\pm0.25$	& 1.00	& $0.72^{+0.05}_{-0.05}$\\
MS1137.5+6625	& $0.14^{+0.02}_{-0.03}$	& $0.34\pm0.03$	& 1.00	& $0.41^{+0.10}_{-0.10}$\\
MACSJ1149.5+2223	& $0.25^{+0.02}_{-0.02}$	& $1.21\pm0.10$	& 1.00	& $0.66^{+0.07}_{-0.07}$\\
A1413	& $0.32^{+0.00}_{-0.00}$	& $0.34\pm0.03$	& 0.90	& $0.67^{+0.03}_{-0.03}$\\
CLJ1213+0253$^\dagger$	& $0.06^{+0.08}_{-0.06}$	& $1.72\pm0.16$	& 1.00	& $\cdots$\\
CLJ1216+2633$^\dagger$	& $0.18^{+0.08}_{-0.09}$	& $0.67\pm0.06$	& 1.00	& $\cdots$\\
RXJ1221+4918	& $0.18^{+0.03}_{-0.03}$	& $1.28\pm0.12$	& 1.00	& $0.69^{+0.06}_{-0.06}$\\
CLJ1226.9+3332$^\dagger$	& $0.10^{+0.03}_{-0.03}$	& $1.30\pm0.11$	& 1.00	& $\cdots$\\
RXJ1234.2+0947	& $0.10^{+0.05}_{-0.05}$	& $4.71\pm0.40$	& 0.95	& $1.00^{+0.35}_{-0.25}$\\
RDCS1252-29	& $0.12^{+0.07}_{-0.12}$	& $0.83\pm0.08$	& 1.00	& $0.91^{+0.27}_{-0.21}$\\
A1682	& $0.22^{+0.04}_{-0.04}$	& $3.92\pm0.34$	& 0.70	& $0.41^{+0.10}_{-0.09}$\\
MACSJ1311.0-0310	& $0.08^{+0.02}_{-0.02}$	& $0.39\pm0.03$	& 1.00	& $1.00^{+0.30}_{-0.23}$\\
A1689	& $0.13^{+0.01}_{-0.01}$	& $0.36\pm0.03$	& 1.00	& $0.84^{+0.06}_{-0.06}$\\
RXJ1317.4+2911	& $0.52^{+0.07}_{-0.08}$	& $1.26\pm0.12$	& 1.00	& $1.22^{+0.70}_{-0.39}$\\
CLJ1334+5031	& $0.22^{+0.08}_{-0.09}$	& $1.70\pm0.16$	& 1.00	& $0.53^{+0.46}_{-0.40}$\\
A1763	& $0.31^{+0.01}_{-0.01}$	& $0.69\pm0.06$	& 0.95	& $0.76^{+0.05}_{-0.05}$\\
RXJ1347.5-1145	& $0.26^{+0.01}_{-0.01}$	& $0.63\pm0.05$	& 1.00	& $0.80^{+0.08}_{-0.07}$\\
RXJ1350.0+6007	& $0.37^{+0.05}_{-0.05}$	& $2.67\pm0.24$	& 1.00	& $0.62^{+0.20}_{-0.18}$\\
CLJ1354-0221	& $0.23^{+0.05}_{-0.06}$	& $2.58\pm0.24$	& 1.00	& $0.65^{+0.14}_{-0.13}$\\
CLJ1415.1+3612	& $0.13^{+0.05}_{-0.05}$	& $1.20\pm0.11$	& 1.00	& $0.62^{+0.15}_{-0.13}$\\
RXJ1416+4446	& $0.25^{+0.03}_{-0.03}$	& $0.82\pm0.07$	& 1.00	& $0.52^{+0.10}_{-0.10}$\\
MACSJ1423.8+2404	& $0.17^{+0.03}_{-0.03}$	& $0.25\pm0.02$	& 1.00	& $0.59^{+0.14}_{-0.13}$\\
A1914	& $0.16^{+0.01}_{-0.01}$	& $1.94\pm0.16$	& 0.81	& $1.37^{+0.19}_{-0.16}$\\
A1942	& $0.11^{+0.02}_{-0.02}$	& $0.93\pm0.08$	& 1.00	& $0.57^{+0.03}_{-0.03}$\\
MS1455.0+2232	& $0.21^{+0.01}_{-0.01}$	& $0.38\pm0.03$	& 1.00	& $0.55^{+0.03}_{-0.03}$\\
RXJ1504-0248	& $0.23^{+0.01}_{-0.01}$	& $0.15\pm0.01$	& 1.00	& $0.80^{+0.11}_{-0.10}$\\
A2034	& $0.15^{+0.01}_{-0.01}$	& $0.86\pm0.07$	& 0.81	& $0.90^{+0.04}_{-0.04}$\\
A2069	& $0.37^{+0.01}_{-0.01}$	& $0.85\pm0.07$	& 0.74	& $0.82^{+0.05}_{-0.05}$\\
RXJ1525+0958	& $0.23^{+0.03}_{-0.04}$	& $2.04\pm0.19$	& 1.00	& $0.73^{+0.10}_{-0.09}$\\
RXJ1532.9+3021	& $0.18^{+0.02}_{-0.02}$	& $0.07\pm0.01$	& 1.00	& $0.69^{+0.09}_{-0.09}$\\
A2111	& $0.28^{+0.02}_{-0.02}$	& $2.08\pm0.18$	& 1.00	& $0.56^{+0.08}_{-0.08}$\\
A2125	& $0.32^{+0.02}_{-0.02}$	& $1.65\pm0.15$	& 1.00	& $0.56^{+0.03}_{-0.03}$\\
A2163	& $0.17^{+0.01}_{-0.01}$	& $2.51\pm0.20$	& 0.63	& $1.48^{+0.15}_{-0.13}$\\
MACSJ1621.3+3810	& $0.18^{+0.02}_{-0.02}$	& $0.25\pm0.02$	& 1.00	& $0.88^{+0.14}_{-0.12}$\\
MS1621.5+2640	& $0.10^{+0.03}_{-0.03}$	& $1.94\pm0.17$	& 1.00	& $0.81^{+0.12}_{-0.11}$\\
A2204	& $0.09^{+0.01}_{-0.01}$	& $0.25\pm0.02$	& 0.86	& $0.89^{+0.11}_{-0.10}$\\
A2218	& $0.21^{+0.01}_{-0.01}$	& $1.87\pm0.16$	& 1.00	& $0.87^{+0.04}_{-0.04}$\\
CLJ1641+4001	& $0.18^{+0.05}_{-0.06}$	& $0.45\pm0.04$	& 1.00	& $0.34^{+0.17}_{-0.16}$\\
RXJ1701+6414	& $0.27^{+0.03}_{-0.03}$	& $1.39\pm0.13$	& 1.00	& $0.60^{+0.10}_{-0.10}$\\
RXJ1716.9+6708	& $0.32^{+0.04}_{-0.04}$	& $0.60\pm0.05$	& 1.00	& $0.73^{+0.18}_{-0.15}$\\
A2259	& $0.20^{+0.02}_{-0.02}$	& $0.77\pm0.07$	& 0.95	& $1.04^{+0.20}_{-0.16}$\\
RXJ1720.1+2638	& $0.14^{+0.01}_{-0.01}$	& $0.16\pm0.01$	& 1.00	& $0.73^{+0.05}_{-0.05}$\\
MACSJ1720.2+3536	& $0.17^{+0.02}_{-0.02}$	& $0.57\pm0.05$	& 1.00	& $0.97^{+0.17}_{-0.15}$\\
A2261	& $0.10^{+0.01}_{-0.01}$	& $0.71\pm0.06$	& 1.00	& $0.68^{+0.04}_{-0.04}$\\
A2294	& $0.07^{+0.02}_{-0.02}$	& $1.23\pm0.10$	& 0.77	& $1.22^{+0.32}_{-0.25}$\\
MACSJ1824.3+4309	& $0.20^{+0.08}_{-0.08}$	& $3.57\pm0.32$	& 1.00	& $0.71^{+0.12}_{-0.12}$\\
MACSJ1931.8-2634	& $0.30^{+0.01}_{-0.01}$	& $0.28\pm0.02$	& 1.00	& $0.69^{+0.10}_{-0.09}$\\
RXJ2011.3-5725	& $0.00^{+0.44}_{-0.00}$	& $0.37\pm0.03$	& 1.00	& $0.73^{+0.12}_{-0.11}$\\
MS2053.7-0449	& $0.14^{+0.04}_{-0.04}$	& $0.95\pm0.09$	& 1.00	& $0.53^{+0.19}_{-0.18}$\\
MACSJ2129.4-0741	& $0.19^{+0.03}_{-0.03}$	& $1.56\pm0.14$	& 1.00	& $1.22^{+0.20}_{-0.17}$\\
RXJ2129.6+0005	& $0.26^{+0.02}_{-0.02}$	& $0.55\pm0.05$	& 1.00	& $0.82^{+0.15}_{-0.13}$\\
A2409	& $0.13^{+0.02}_{-0.02}$	& $1.08\pm0.09$	& 0.70	& $0.92^{+0.14}_{-0.12}$\\
MACSJ2228.5+2036	& $0.03^{+0.03}_{-0.01}$	& $3.00\pm0.26$	& 1.00	& $0.79^{+0.06}_{-0.05}$\\
MACSJ2229.7-2755	& $0.19^{+0.02}_{-0.02}$	& $0.27\pm0.02$	& 1.00	& $0.77^{+0.12}_{-0.11}$\\
MACSJ2245.0+2637	& $0.23^{+0.02}_{-0.02}$	& $0.36\pm0.03$	& 1.00	& $0.58^{+0.15}_{-0.14}$\\
RXJ2247+0337	& $0.21^{+0.05}_{-0.06}$	& $0.53\pm0.05$	& 1.00	& $0.17^{+0.34}_{-0.39}$\\
AS1063	& $0.20^{+0.01}_{-0.01}$	& $0.74\pm0.06$	& 1.00	& $0.74^{+0.07}_{-0.07}$\\
CLJ2302.8+0844	& $0.12^{+0.05}_{-0.06}$	& $1.11\pm0.10$	& 1.00	& $0.42^{+0.11}_{-0.11}$\\
A2631	& $0.22^{+0.02}_{-0.02}$	& $5.01\pm0.43$	& 1.00	& $0.76^{+0.11}_{-0.10}$\\
\enddata
\end{deluxetable}

\clearpage

\begin{deluxetable}{lp{0.4cm}cp{1.5cm}ccccp{1.5cm}cc}
\setlength{\tabcolsep}{0.01in}
\tabletypesize{\scriptsize}
\tablecaption{Summary of the cluster properties. Clusters marked with a
$\dagger$ were too faint to permit a full spectral fit in the
$(0.15<r<1)\rf$ aperture. For these clusters \Yx\ was derived using the
temperature measured in the $r<\rf$ aperture, and \Lx\ was measured by a
spectral fit in the $(0.15<r<1)\rf$ aperture with kT fixed at the global
temperature.\label{t.ap}}
\tablehead{
\colhead{ } & \colhead{ } & \multicolumn{4}{c}{$r<\rf$} & \colhead{ } & \multicolumn{3}{c}{$(0.15<r<1)\rf$} & \colhead{ }\\
\cline{3-6}
\cline{8-10}\\
\colhead{Cluster} & \colhead{\rf} & \colhead{kT} & \colhead{\Lx} &
\colhead{Z} & \colhead{\Mgas} & \colhead{ } & \colhead{kT} & \colhead{\Lx} & \colhead{Z} & \colhead{\Yx}\\
\colhead{ } & \colhead{(Mpc)} & \colhead{(keV)} & \colhead{($10^{44}\ergps$)} & \colhead{($\Zsol$)} & \colhead{($10^{13}\Msol$)} & \colhead{ } & \colhead{(keV)} & \colhead{($10^{44}\ergps$)} & \colhead{($\Zsol$)} & \colhead{($10^{13}\Msol\keV$)}
}
\startdata
MS0015.9+1609	& 1.24	& $9.0^{+0.5}_{-0.5}$	& $47.5\pm0.7$	& $0.32^{+0.06}_{-0.06}$	& $15.17^{+0.08}_{-0.11}$	& & $8.9^{+0.6}_{-0.7}$	& $34.5\pm0.6$	& $0.24^{+0.07}_{-0.07}$	& $135.6^{+9.8}_{-10.1}$\\
RXJ0027.6+2616	& 0.93	& $4.7^{+0.7}_{-0.7}$	& $6.6\pm0.6$	& $0.47^{+0.32}_{-0.28}$	& $4.60^{+0.14}_{-0.14}$	& & $4.2^{+0.9}_{-0.5}$	& $5.0\pm0.6$	& $0.65^{+0.40}_{-0.33}$	& $19.1^{+4.3}_{-2.5}$\\
CLJ0030+2618	& 0.93	& $4.8^{+1.3}_{-0.8}$	& $5.2\pm0.6$	& $0.82^{+0.54}_{-0.43}$	& $3.80^{+0.17}_{-0.18}$	& & $4.7^{+1.5}_{-0.9}$	& $3.9\pm0.6$	& $1.02^{+0.74}_{-0.56}$	& $18.0^{+5.9}_{-3.6}$\\
A68	& 1.19	& $8.1^{+0.9}_{-0.8}$	& $16.7\pm0.6$	& $0.39^{+0.17}_{-0.16}$	& $7.83^{+0.15}_{-0.13}$	& & $6.5^{+1.0}_{-0.7}$	& $9.3\pm0.5$	& $0.46^{+0.22}_{-0.21}$	& $51.0^{+7.8}_{-5.7}$\\
A115	& 1.25	& $5.3^{+0.1}_{-0.1}$	& $12.6\pm0.1$	& $0.36^{+0.04}_{-0.04}$	& $8.79^{+0.08}_{-0.06}$	& & $7.1^{+0.4}_{-0.4}$	& $9.3\pm0.1$	& $0.30^{+0.07}_{-0.07}$	& $62.4^{+3.5}_{-3.4}$\\
A209	& 1.30	& $7.1^{+0.4}_{-0.4}$	& $17.6\pm0.3$	& $0.26^{+0.07}_{-0.07}$	& $10.83^{+0.08}_{-0.09}$	& & $7.3^{+0.5}_{-0.5}$	& $12.2\pm0.3$	& $0.29^{+0.10}_{-0.09}$	& $78.7^{+5.4}_{-5.3}$\\
CLJ0152.7-1357S	& 0.72	& $4.8^{+1.3}_{-0.8}$	& $7.6\pm0.8$	& $0.55^{+0.45}_{-0.37}$	& $3.38^{+0.17}_{-0.15}$	& & $6.0^{+2.4}_{-1.8}$	& $6.0\pm0.7$	& $0.10^{+0.42}_{-0.10}$	& $20.2^{+8.2}_{-6.0}$\\
A267	& 1.04	& $4.9^{+0.3}_{-0.3}$	& $11.1\pm0.5$	& $0.49^{+0.18}_{-0.17}$	& $5.74^{+0.10}_{-0.07}$	& & $4.6^{+0.5}_{-0.4}$	& $7.0\pm0.5$	& $0.64^{+0.27}_{-0.24}$	& $26.5^{+2.6}_{-2.4}$\\
CLJ0152.7-1357N	& 0.78	& $5.3^{+0.8}_{-0.8}$	& $10.3\pm0.6$	& $0.00^{+0.13}_{-0.00}$	& $4.63^{+0.15}_{-0.14}$	& & $5.3^{+1.0}_{-1.0}$	& $8.7\pm0.7$	& $0.03^{+0.27}_{-0.03}$	& $24.6^{+4.7}_{-4.5}$\\
MACSJ0159.8-0849	& 1.28	& $7.9^{+0.3}_{-0.3}$	& $39.3\pm0.5$	& $0.31^{+0.05}_{-0.05}$	& $11.41^{+0.11}_{-0.07}$	& & $9.9^{+0.9}_{-0.9}$	& $18.0\pm0.4$	& $0.05^{+0.10}_{-0.05}$	& $113.4^{+10.7}_{-10.5}$\\
CLJ0216-1747	& 0.71	& $8.2^{+16.0}_{-5.1}$	& $2.8\pm0.5$	& $0.00^{+0.51}_{-0.00}$	& $1.87^{+0.09}_{-0.12}$	& & $3.0^{+9.2}_{-1.3}$	& $1.4\pm0.5$	& $0.00^{+0.74}_{-0.00}$	& $5.6^{+17.3}_{-2.5}$\\
RXJ0232.2-4420	& 1.34	& $7.9^{+0.4}_{-0.4}$	& $31.7\pm0.6$	& $0.49^{+0.09}_{-0.09}$	& $11.96^{+0.11}_{-0.19}$	& & $9.2^{+1.1}_{-1.0}$	& $17.6\pm0.5$	& $0.41^{+0.16}_{-0.16}$	& $110.3^{+13.2}_{-12.6}$\\
MACSJ0242.5-2132	& 1.09	& $4.8^{+0.2}_{-0.2}$	& $27.3\pm0.7$	& $0.48^{+0.10}_{-0.09}$	& $6.74^{+0.07}_{-0.21}$	& & $5.9^{+0.9}_{-0.7}$	& $7.9\pm0.4$	& $0.08^{+0.18}_{-0.08}$	& $39.9^{+6.0}_{-5.1}$\\
A383	& 0.98	& $3.9^{+0.1}_{-0.1}$	& $9.1\pm0.2$	& $0.48^{+0.06}_{-0.06}$	& $4.07^{+0.04}_{-0.04}$	& & $4.2^{+0.4}_{-0.2}$	& $3.6\pm0.1$	& $0.35^{+0.12}_{-0.11}$	& $17.1^{+1.7}_{-0.8}$\\
MACSJ0257.6-2209	& 1.14	& $7.4^{+0.6}_{-0.6}$	& $16.3\pm0.5$	& $0.41^{+0.12}_{-0.12}$	& $7.04^{+0.16}_{-0.07}$	& & $6.9^{+1.1}_{-0.7}$	& $8.1\pm0.3$	& $0.17^{+0.17}_{-0.17}$	& $48.6^{+7.6}_{-5.2}$\\
MS0302.7+1658	& 0.79	& $3.5^{+0.5}_{-0.4}$	& $6.0\pm0.8$	& $0.56^{+0.43}_{-0.34}$	& $2.98^{+0.13}_{-0.13}$	& & $3.3^{+0.8}_{-0.5}$	& $3.2\pm0.7$	& $0.70^{+0.82}_{-0.56}$	& $10.0^{+2.4}_{-1.7}$\\
CLJ0318-0302	& 0.88	& $4.0^{+0.6}_{-0.5}$	& $5.1\pm0.3$	& $0.00^{+0.12}_{-0.00}$	& $3.19^{+0.19}_{-0.13}$	& & $4.3^{+0.7}_{-0.8}$	& $3.4\pm0.2$	& $0.00^{+0.08}_{-0.00}$	& $13.7^{+2.4}_{-2.5}$\\
MACSJ0329.6-0211	& 0.99	& $4.5^{+0.3}_{-0.3}$	& $27.1\pm0.9$	& $0.58^{+0.12}_{-0.11}$	& $7.43^{+0.16}_{-0.17}$	& & $4.4^{+0.5}_{-0.4}$	& $12.1\pm0.6$	& $0.35^{+0.18}_{-0.15}$	& $32.5^{+3.5}_{-3.4}$\\
MACSJ0404.6+1109	& 1.07	& $5.9^{+0.7}_{-0.6}$	& $9.7\pm0.4$	& $0.19^{+0.14}_{-0.13}$	& $6.90^{+0.10}_{-0.14}$	& & $5.9^{+0.8}_{-0.7}$	& $8.1\pm0.4$	& $0.13^{+0.15}_{-0.13}$	& $40.5^{+5.5}_{-4.7}$\\
MACSJ0429.6-0253	& 1.12	& $5.4^{+0.4}_{-0.2}$	& $23.1\pm0.6$	& $0.51^{+0.10}_{-0.09}$	& $7.02^{+0.10}_{-0.10}$	& & $6.8^{+1.1}_{-0.6}$	& $9.2\pm0.4$	& $0.39^{+0.18}_{-0.18}$	& $47.6^{+7.5}_{-4.3}$\\
RXJ0439.0+0715	& 1.14	& $5.6^{+0.3}_{-0.3}$	& $15.7\pm0.3$	& $0.38^{+0.07}_{-0.07}$	& $7.19^{+0.09}_{-0.10}$	& & $5.6^{+0.4}_{-0.4}$	& $8.7\pm0.2$	& $0.28^{+0.11}_{-0.10}$	& $40.2^{+3.2}_{-2.8}$\\
RXJ0439+0520	& 0.94	& $3.8^{+0.2}_{-0.2}$	& $8.4\pm0.4$	& $0.56^{+0.14}_{-0.12}$	& $3.43^{+0.08}_{-0.06}$	& & $3.8^{+0.5}_{-0.4}$	& $3.0\pm0.3$	& $0.35^{+0.26}_{-0.22}$	& $13.2^{+1.6}_{-1.5}$\\
MACSJ0451.9+0006	& 0.97	& $5.6^{+0.8}_{-0.8}$	& $15.1\pm0.8$	& $0.35^{+0.23}_{-0.18}$	& $6.87^{+0.23}_{-0.16}$	& & $4.8^{+1.0}_{-0.7}$	& $10.9\pm0.8$	& $0.44^{+0.32}_{-0.28}$	& $32.9^{+6.8}_{-4.8}$\\
A521	& 1.18	& $5.1^{+0.2}_{-0.2}$	& $13.9\pm0.3$	& $0.42^{+0.07}_{-0.07}$	& $10.56^{+0.05}_{-0.09}$	& & $5.0^{+0.2}_{-0.2}$	& $12.2\pm0.2$	& $0.43^{+0.08}_{-0.08}$	& $53.1^{+1.9}_{-1.9}$\\
A520	& 1.31	& $7.1^{+0.2}_{-0.2}$	& $17.6\pm0.2$	& $0.42^{+0.04}_{-0.04}$	& $11.11^{+0.05}_{-0.07}$	& & $7.2^{+0.3}_{-0.3}$	& $14.2\pm0.1$	& $0.43^{+0.05}_{-0.05}$	& $80.0^{+2.9}_{-3.0}$\\
MS0451.6-0305	& 1.11	& $6.7^{+0.6}_{-0.5}$	& $39.9\pm1.4$	& $0.55^{+0.15}_{-0.15}$	& $12.08^{+0.16}_{-0.23}$	& & $6.6^{+0.7}_{-0.6}$	& $28.8\pm1.2$	& $0.33^{+0.16}_{-0.15}$	& $80.1^{+8.7}_{-7.5}$\\
CLJ0522-3625	& 0.77	& $3.8^{+0.9}_{-0.8}$	& $3.1\pm0.3$	& $0.00^{+0.27}_{-0.00}$	& $2.43^{+0.11}_{-0.14}$	& & $3.4^{+0.9}_{-0.8}$	& $2.3\pm0.3$	& $0.00^{+0.16}_{-0.00}$	& $8.2^{+2.3}_{-1.9}$\\
CLJ0542.8-4100	& 0.90	& $6.7^{+1.2}_{-0.9}$	& $9.9\pm0.4$	& $0.05^{+0.16}_{-0.05}$	& $4.92^{+0.10}_{-0.08}$	& & $6.2^{+1.2}_{-0.9}$	& $7.5\pm0.4$	& $0.13^{+0.19}_{-0.13}$	& $30.7^{+5.8}_{-4.7}$\\
MACSJ0647.7+7015	& 1.18	& $10.5^{+1.4}_{-1.0}$	& $42.4\pm1.0$	& $0.21^{+0.12}_{-0.13}$	& $11.59^{+0.16}_{-0.12}$	& & $10.6^{+2.0}_{-1.4}$	& $22.1\pm0.6$	& $0.00^{+0.16}_{-0.00}$	& $122.4^{+23.4}_{-16.4}$\\
1E0657-56	& 1.58	& $11.7^{+0.4}_{-0.4}$	& $75.8\pm0.5$	& $0.31^{+0.03}_{-0.03}$	& $23.06^{+0.09}_{-0.10}$	& & $11.7^{+0.5}_{-0.5}$	& $54.0\pm0.4$	& $0.33^{+0.04}_{-0.04}$	& $270.5^{+10.8}_{-10.6}$\\
MACSJ0717.5+3745	& 1.36	& $10.5^{+0.6}_{-0.4}$	& $73.3\pm0.9$	& $0.28^{+0.06}_{-0.06}$	& $22.35^{+0.12}_{-0.12}$	& & $10.1^{+0.5}_{-0.5}$	& $57.1\pm0.8$	& $0.32^{+0.06}_{-0.06}$	& $225.5^{+11.7}_{-11.6}$\\
A586	& 1.16	& $6.6^{+0.4}_{-0.3}$	& $13.3\pm0.3$	& $0.60^{+0.12}_{-0.12}$	& $6.24^{+0.07}_{-0.07}$	& & $6.4^{+0.6}_{-0.5}$	& $6.5\pm0.3$	& $0.84^{+0.23}_{-0.21}$	& $40.1^{+3.6}_{-3.2}$\\
MACSJ0744.9+3927	& 1.04	& $7.6^{+0.4}_{-0.4}$	& $46.2\pm1.3$	& $0.36^{+0.07}_{-0.06}$	& $10.75^{+0.12}_{-0.14}$	& & $7.7^{+0.6}_{-0.6}$	& $26.3\pm1.0$	& $0.31^{+0.09}_{-0.09}$	& $83.1^{+6.6}_{-6.5}$\\
A665	& 1.37	& $7.5^{+0.2}_{-0.2}$	& $21.0\pm0.2$	& $0.34^{+0.04}_{-0.04}$	& $12.77^{+0.05}_{-0.07}$	& & $7.5^{+0.3}_{-0.3}$	& $15.2\pm0.2$	& $0.34^{+0.06}_{-0.05}$	& $96.0^{+3.7}_{-3.7}$\\
A697	& 1.41	& $9.0^{+0.6}_{-0.5}$	& $36.5\pm0.6$	& $0.43^{+0.08}_{-0.08}$	& $15.57^{+0.17}_{-0.12}$	& & $8.8^{+0.7}_{-0.6}$	& $24.4\pm0.5$	& $0.47^{+0.10}_{-0.10}$	& $137.1^{+11.2}_{-10.0}$\\
CLJ0848.7+4456	& 0.58	& $2.5^{+0.5}_{-0.4}$	& $1.1\pm0.1$	& $0.07^{+0.24}_{-0.07}$	& $1.06^{+0.04}_{-0.03}$	& & $2.5^{+0.5}_{-0.5}$	& $0.8\pm0.2$	& $0.13^{+0.15}_{-0.13}$	& $2.6^{+0.5}_{-0.5}$\\
ZWCLJ1953	& 1.12	& $7.3^{+0.7}_{-0.7}$	& $15.6\pm0.4$	& $0.09^{+0.10}_{-0.09}$	& $7.31^{+0.10}_{-0.08}$	& & $6.2^{+0.7}_{-0.6}$	& $8.7\pm0.3$	& $0.15^{+0.14}_{-0.13}$	& $45.3^{+4.9}_{-4.8}$\\
CLJ0853+5759	& 0.85	& $6.1^{+2.2}_{-1.4}$	& $3.2\pm0.3$	& $0.71^{+0.46}_{-0.41}$	& $2.69^{+0.14}_{-0.06}$	& & $7.7^{+3.7}_{-2.5}$	& $3.0\pm0.3$	& $0.84^{+0.62}_{-0.51}$	& $20.7^{+10.1}_{-6.7}$\\
MS0906.5+1110	& 1.06	& $5.3^{+0.2}_{-0.2}$	& $8.4\pm0.1$	& $0.31^{+0.07}_{-0.07}$	& $5.27^{+0.05}_{-0.04}$	& & $4.8^{+0.3}_{-0.3}$	& $4.7\pm0.1$	& $0.22^{+0.09}_{-0.09}$	& $25.2^{+1.5}_{-1.4}$\\
RXJ0910+5422$^\dagger$	& 0.53	& $3.9^{+1.2}_{-1.0}$	& $2.9\pm0.4$	& $0.00^{+0.13}_{-0.00}$	& $1.41^{+0.07}_{-0.07}$	& & $\cdots$ 	& $2.2\pm0.6$	& $\cdots$ 	& $5.8^{+2.1}_{-1.2}$\\
A773	& 1.25	& $7.4^{+0.3}_{-0.3}$	& $17.2\pm0.3$	& $0.48^{+0.06}_{-0.06}$	& $9.09^{+0.06}_{-0.06}$	& & $7.0^{+0.4}_{-0.4}$	& $10.9\pm0.3$	& $0.57^{+0.09}_{-0.08}$	& $63.9^{+3.8}_{-3.5}$\\
A781	& 1.10	& $5.6^{+0.6}_{-0.6}$	& $10.8\pm0.5$	& $0.47^{+0.18}_{-0.17}$	& $7.80^{+0.14}_{-0.10}$	& & $5.3^{+0.7}_{-0.4}$	& $9.2\pm0.5$	& $0.53^{+0.21}_{-0.19}$	& $41.7^{+5.9}_{-3.5}$\\
CLJ0926+1242	& 0.81	& $4.5^{+0.8}_{-0.7}$	& $4.6\pm0.3$	& $0.24^{+0.20}_{-0.21}$	& $3.10^{+0.07}_{-0.07}$	& & $5.1^{+1.0}_{-1.0}$	& $3.2\pm0.2$	& $0.00^{+0.25}_{-0.00}$	& $15.7^{+3.1}_{-3.2}$\\
RBS797	& 1.13	& $6.0^{+0.3}_{-0.3}$	& $45.9\pm0.9$	& $0.35^{+0.07}_{-0.07}$	& $7.89^{+0.31}_{-0.18}$	& & $6.3^{+0.9}_{-0.7}$	& $11.7\pm0.6$	& $0.54^{+0.23}_{-0.21}$	& $50.0^{+7.2}_{-5.9}$\\
MACSJ0949.8+1708	& 1.23	& $8.0^{+0.7}_{-0.6}$	& $27.7\pm0.8$	& $0.40^{+0.12}_{-0.12}$	& $11.00^{+0.15}_{-0.09}$	& & $7.7^{+0.9}_{-0.9}$	& $17.2\pm0.6$	& $0.27^{+0.15}_{-0.15}$	& $84.3^{+9.8}_{-9.4}$\\
CLJ0956+4107	& 0.78	& $4.0^{+0.7}_{-0.5}$	& $5.3\pm0.4$	& $0.46^{+0.27}_{-0.23}$	& $3.15^{+0.09}_{-0.09}$	& & $3.9^{+0.8}_{-0.6}$	& $4.2\pm0.4$	& $0.20^{+0.29}_{-0.20}$	& $12.3^{+2.6}_{-1.9}$\\
A907	& 1.12	& $5.3^{+0.1}_{-0.1}$	& $10.4\pm0.1$	& $0.49^{+0.03}_{-0.03}$	& $5.44^{+0.03}_{-0.03}$	& & $5.6^{+0.3}_{-0.3}$	& $4.8\pm0.1$	& $0.36^{+0.05}_{-0.05}$	& $30.6^{+1.4}_{-1.5}$\\
MS1006.0+1202	& 1.11	& $5.9^{+0.4}_{-0.4}$	& $7.3\pm0.2$	& $0.16^{+0.10}_{-0.10}$	& $5.29^{+0.06}_{-0.06}$	& & $6.3^{+0.6}_{-0.6}$	& $4.9\pm0.2$	& $0.09^{+0.14}_{-0.09}$	& $33.4^{+3.3}_{-3.2}$\\
MS1008.1-1224	& 1.02	& $5.0^{+0.4}_{-0.4}$	& $9.7\pm0.4$	& $0.27^{+0.12}_{-0.12}$	& $5.60^{+0.10}_{-0.11}$	& & $4.6^{+0.5}_{-0.4}$	& $6.4\pm0.3$	& $0.34^{+0.18}_{-0.16}$	& $26.0^{+2.7}_{-2.6}$\\
ZW3146	& 1.30	& $6.4^{+0.1}_{-0.1}$	& $44.4\pm0.4$	& $0.37^{+0.03}_{-0.03}$	& $10.26^{+0.08}_{-0.10}$	& & $8.2^{+0.4}_{-0.4}$	& $15.6\pm0.3$	& $0.22^{+0.08}_{-0.08}$	& $84.1^{+4.6}_{-4.6}$\\
CLJ1113.1-2615	& 0.66	& $3.8^{+0.9}_{-0.7}$	& $3.7\pm0.6$	& $0.36^{+0.51}_{-0.36}$	& $1.80^{+0.09}_{-0.11}$	& & $4.5^{+1.0}_{-1.1}$	& $2.1\pm0.5$	& $0.16^{+0.64}_{-0.16}$	& $8.2^{+1.9}_{-2.1}$\\
A1204	& 0.92	& $3.4^{+0.1}_{-0.1}$	& $9.2\pm0.2$	& $0.37^{+0.05}_{-0.05}$	& $3.10^{+0.21}_{-0.06}$	& & $3.8^{+0.3}_{-0.3}$	& $2.7\pm0.1$	& $0.13^{+0.10}_{-0.10}$	& $11.8^{+1.1}_{-0.8}$\\
CLJ1117+1745	& 0.73	& $3.4^{+1.2}_{-0.6}$	& $2.0\pm0.3$	& $0.15^{+0.40}_{-0.15}$	& $2.00^{+0.05}_{-0.07}$	& & $3.6^{+1.1}_{-0.8}$	& $1.7\pm0.2$	& $0.00^{+0.34}_{-0.00}$	& $7.2^{+2.2}_{-1.6}$\\
CLJ1120+4318	& 0.88	& $4.9^{+0.7}_{-0.6}$	& $12.1\pm0.8$	& $0.35^{+0.21}_{-0.19}$	& $5.20^{+0.15}_{-0.11}$	& & $4.9^{+1.1}_{-0.7}$	& $8.6\pm0.7$	& $0.19^{+0.27}_{-0.19}$	& $25.4^{+5.6}_{-3.9}$\\
RXJ1121+2327	& 0.77	& $3.8^{+0.4}_{-0.3}$	& $4.7\pm0.3$	& $0.28^{+0.16}_{-0.14}$	& $3.29^{+0.05}_{-0.07}$	& & $3.5^{+0.4}_{-0.3}$	& $4.2\pm0.3$	& $0.28^{+0.17}_{-0.15}$	& $11.7^{+1.3}_{-1.1}$\\
A1240	& 0.92	& $3.9^{+0.3}_{-0.3}$	& $1.7\pm0.1$	& $0.19^{+0.10}_{-0.09}$	& $2.75^{+0.04}_{-0.03}$	& & $3.9^{+0.3}_{-0.3}$	& $1.6\pm0.1$	& $0.18^{+0.11}_{-0.10}$	& $10.8^{+0.9}_{-0.9}$\\
MACSJ1131.8-1955	& 1.35	& $8.3^{+0.7}_{-0.5}$	& $29.5\pm0.7$	& $0.47^{+0.12}_{-0.11}$	& $14.20^{+0.17}_{-0.16}$	& & $9.1^{+1.1}_{-1.0}$	& $20.9\pm0.6$	& $0.41^{+0.15}_{-0.15}$	& $129.2^{+15.2}_{-14.0}$\\
MS1137.5+6625	& 0.79	& $5.8^{+0.7}_{-0.6}$	& $12.2\pm0.5$	& $0.20^{+0.16}_{-0.15}$	& $3.99^{+0.06}_{-0.06}$	& & $5.5^{+1.0}_{-0.6}$	& $7.1\pm0.4$	& $0.20^{+0.22}_{-0.20}$	& $21.8^{+4.1}_{-2.3}$\\
MACSJ1149.5+2223	& 1.22	& $8.4^{+0.9}_{-0.7}$	& $40.8\pm1.3$	& $0.28^{+0.11}_{-0.11}$	& $15.49^{+0.27}_{-0.15}$	& & $8.2^{+1.0}_{-0.8}$	& $32.6\pm1.1$	& $0.27^{+0.13}_{-0.13}$	& $127.8^{+16.3}_{-12.4}$\\
A1413	& 1.26	& $7.2^{+0.2}_{-0.2}$	& $15.9\pm0.1$	& $0.41^{+0.03}_{-0.03}$	& $7.95^{+0.04}_{-0.05}$	& & $7.0^{+0.3}_{-0.3}$	& $7.4\pm0.1$	& $0.34^{+0.05}_{-0.05}$	& $55.8^{+2.3}_{-2.1}$\\
CLJ1213+0253	& 0.76	& $3.4^{+0.6}_{-0.5}$	& $1.9\pm0.6$	& $1.94^{+1.79}_{-0.97}$	& $1.99^{+0.11}_{-0.09}$	& & $4.1^{+0.9}_{-0.9}$	& $1.4\pm0.6$	& $2.03^{+2.97}_{-1.36}$	& $8.2^{+1.8}_{-1.9}$\\
CLJ1216+2633$^\dagger$	& 0.87	& $6.8^{+6.6}_{-2.2}$	& $3.2\pm0.5$	& $0.76^{+0.82}_{-0.65}$	& $2.48^{+0.13}_{-0.13}$	& & $\cdots$ 	& $3.1\pm0.9$	& $\cdots$ 	& $15.5^{+7.2}_{-3.9}$\\
RXJ1221+4918	& 0.86	& $5.8^{+0.6}_{-0.6}$	& $11.8\pm0.5$	& $0.48^{+0.14}_{-0.13}$	& $5.44^{+0.08}_{-0.07}$	& & $5.7^{+0.7}_{-0.6}$	& $9.8\pm0.4$	& $0.48^{+0.15}_{-0.14}$	& $31.1^{+3.7}_{-3.5}$\\
CLJ1226.9+3332	& 0.94	& $10.4^{+1.4}_{-1.0}$	& $43.5\pm1.3$	& $0.30^{+0.14}_{-0.14}$	& $7.94^{+0.22}_{-0.11}$	& & $9.0^{+1.6}_{-1.3}$	& $21.7\pm1.1$	& $0.52^{+0.22}_{-0.22}$	& $71.3^{+13.1}_{-10.5}$\\
RXJ1234.2+0947	& 1.17	& $6.7^{+2.3}_{-1.0}$	& $6.7\pm0.3$	& $0.12^{+0.25}_{-0.12}$	& $6.21^{+0.11}_{-0.12}$	& & $7.7^{+1.7}_{-2.2}$	& $6.1\pm0.2$	& $0.00^{+0.26}_{-0.00}$	& $47.8^{+10.9}_{-13.9}$\\
RDCS1252-29	& 0.52	& $4.8^{+1.2}_{-0.8}$	& $5.9\pm0.7$	& $0.27^{+0.27}_{-0.24}$	& $1.93^{+0.09}_{-0.07}$	& & $4.3^{+0.8}_{-0.8}$	& $5.1\pm0.6$	& $0.36^{+0.37}_{-0.31}$	& $8.3^{+1.5}_{-1.5}$\\
A1682	& 1.13	& $6.2^{+0.8}_{-0.8}$	& $10.0\pm0.6$	& $0.42^{+0.27}_{-0.25}$	& $7.07^{+0.13}_{-0.12}$	& & $6.1^{+1.1}_{-0.9}$	& $7.5\pm0.5$	& $0.33^{+0.34}_{-0.30}$	& $42.9^{+7.9}_{-6.5}$\\
MACSJ1311.0-0310	& 0.95	& $5.6^{+0.3}_{-0.3}$	& $16.2\pm0.3$	& $0.39^{+0.08}_{-0.07}$	& $4.59^{+0.11}_{-0.08}$	& & $6.2^{+0.7}_{-0.7}$	& $6.7\pm0.2$	& $0.27^{+0.14}_{-0.14}$	& $28.6^{+3.2}_{-3.1}$\\
A1689	& 1.37	& $9.0^{+0.3}_{-0.3}$	& $36.7\pm0.3$	& $0.42^{+0.04}_{-0.04}$	& $11.26^{+0.11}_{-0.06}$	& & $8.4^{+0.4}_{-0.3}$	& $14.3\pm0.2$	& $0.40^{+0.07}_{-0.07}$	& $95.1^{+4.7}_{-3.7}$\\
RXJ1317.4+2911	& 0.50	& $2.0^{+0.7}_{-0.5}$	& $1.2\pm0.3$	& $0.01^{+0.68}_{-0.01}$	& $0.95^{+0.06}_{-0.07}$	& & $2.1^{+0.6}_{-0.5}$	& $1.0\pm1.2$	& $0.28^{+2.70}_{-0.28}$	& $2.0^{+0.6}_{-0.5}$\\
CLJ1334+5031$^\dagger$	& 1.00	& $16.9^{+33.6}_{-7.0}$	& $11.8\pm1.4$	& $0.01^{+1.02}_{-0.01}$	& $4.83^{+0.25}_{-0.19}$	& & $\cdots$ 	& $12.4\pm5.3$	& $\cdots$ 	& $38.1^{+35.5}_{-12.3}$\\
A1763	& 1.32	& $7.8^{+0.4}_{-0.4}$	& $19.6\pm0.3$	& $0.29^{+0.07}_{-0.07}$	& $11.47^{+0.10}_{-0.07}$	& & $7.7^{+0.5}_{-0.5}$	& $13.7\pm0.3$	& $0.26^{+0.09}_{-0.09}$	& $87.8^{+6.0}_{-5.9}$\\
RXJ1347.5-1145	& 1.45	& $12.2^{+0.5}_{-0.5}$	& $145.4\pm1.3$	& $0.41^{+0.04}_{-0.04}$	& $19.45^{+0.11}_{-0.37}$	& & $11.7^{+1.1}_{-1.1}$	& $38.3\pm0.8$	& $0.48^{+0.10}_{-0.10}$	& $227.6^{+21.3}_{-21.4}$\\
RXJ1350.0+6007$^\dagger$	& 0.70	& $4.8^{+1.3}_{-0.9}$	& $5.6\pm0.6$	& $0.60^{+0.39}_{-0.33}$	& $2.78^{+0.11}_{-0.10}$	& & $\cdots$ 	& $4.6\pm0.7$	& $\cdots$ 	& $12.2^{+2.4}_{-2.2}$\\
CLJ1354-0221	& 0.76	& $3.8^{+0.8}_{-0.6}$	& $3.5\pm0.3$	& $0.26^{+0.33}_{-0.26}$	& $2.66^{+0.08}_{-0.05}$	& & $3.1^{+0.5}_{-0.4}$	& $2.9\pm0.4$	& $0.55^{+0.44}_{-0.34}$	& $8.2^{+1.4}_{-1.1}$\\
CLJ1415.1+3612	& 0.64	& $5.7^{+0.8}_{-0.7}$	& $11.3\pm0.6$	& $0.00^{+0.14}_{-0.00}$	& $3.15^{+0.08}_{-0.07}$	& & $4.4^{+0.5}_{-0.6}$	& $7.4\pm0.4$	& $0.00^{+0.04}_{-0.00}$	& $13.8^{+1.6}_{-1.8}$\\
RXJ1416+4446	& 0.86	& $3.7^{+0.3}_{-0.3}$	& $5.2\pm0.4$	& $0.91^{+0.30}_{-0.26}$	& $3.23^{+0.08}_{-0.05}$	& & $4.3^{+0.5}_{-0.4}$	& $3.3\pm0.4$	& $1.07^{+0.48}_{-0.41}$	& $13.8^{+1.8}_{-1.3}$\\
MACSJ1423.8+2404	& 0.99	& $5.9^{+0.5}_{-0.4}$	& $30.1\pm1.0$	& $0.45^{+0.12}_{-0.11}$	& $6.96^{+0.11}_{-0.17}$	& & $5.7^{+0.9}_{-0.7}$	& $11.3\pm0.7$	& $0.43^{+0.23}_{-0.21}$	& $39.5^{+6.3}_{-5.1}$\\
A1914	& 1.37	& $9.8^{+0.3}_{-0.3}$	& $32.6\pm0.3$	& $0.34^{+0.05}_{-0.05}$	& $10.69^{+0.09}_{-0.07}$	& & $8.9^{+0.6}_{-0.6}$	& $12.4\pm0.2$	& $0.28^{+0.09}_{-0.09}$	& $94.8^{+6.4}_{-5.9}$\\
A1942	& 0.94	& $4.3^{+0.3}_{-0.2}$	& $3.9\pm0.1$	& $0.27^{+0.08}_{-0.08}$	& $3.59^{+0.03}_{-0.03}$	& & $3.9^{+0.2}_{-0.2}$	& $2.9\pm0.1$	& $0.20^{+0.09}_{-0.08}$	& $13.8^{+0.9}_{-0.9}$\\
MS1455.0+2232	& 1.04	& $4.5^{+0.1}_{-0.1}$	& $20.2\pm0.2$	& $0.48^{+0.03}_{-0.03}$	& $5.62^{+0.10}_{-0.09}$	& & $4.7^{+0.2}_{-0.2}$	& $6.3\pm0.1$	& $0.40^{+0.06}_{-0.06}$	& $26.2^{+1.2}_{-1.1}$\\
RXJ1504-0248	& 1.34	& $6.8^{+0.2}_{-0.2}$	& $61.1\pm0.6$	& $0.35^{+0.04}_{-0.04}$	& $10.90^{+0.04}_{-0.47}$	& & $8.3^{+0.8}_{-0.7}$	& $14.3\pm0.4$	& $0.35^{+0.14}_{-0.13}$	& $90.2^{+8.5}_{-8.1}$\\
A2034	& 1.22	& $6.7^{+0.2}_{-0.2}$	& $9.0\pm0.1$	& $0.38^{+0.04}_{-0.04}$	& $6.88^{+0.02}_{-0.03}$	& & $6.4^{+0.2}_{-0.2}$	& $6.2\pm0.1$	& $0.36^{+0.05}_{-0.05}$	& $44.1^{+1.4}_{-1.4}$\\
A2069	& 1.20	& $6.3^{+0.2}_{-0.2}$	& $6.1\pm0.1$	& $0.29^{+0.05}_{-0.05}$	& $6.56^{+0.03}_{-0.03}$	& & $6.2^{+0.3}_{-0.3}$	& $5.1\pm0.1$	& $0.26^{+0.06}_{-0.05}$	& $40.8^{+1.7}_{-1.7}$\\
RXJ1525+0958	& 0.80	& $5.2^{+0.9}_{-0.5}$	& $5.9\pm0.3$	& $0.43^{+0.18}_{-0.17}$	& $3.86^{+0.10}_{-0.06}$	& & $5.4^{+1.1}_{-0.6}$	& $5.3\pm0.3$	& $0.39^{+0.20}_{-0.19}$	& $20.7^{+4.2}_{-2.4}$\\
RXJ1532.9+3021	& 1.12	& $5.1^{+0.2}_{-0.2}$	& $35.9\pm0.9$	& $0.60^{+0.10}_{-0.10}$	& $7.56^{+0.31}_{-0.15}$	& & $6.1^{+0.8}_{-0.7}$	& $12.3\pm0.6$	& $0.42^{+0.19}_{-0.18}$	& $46.3^{+6.0}_{-5.4}$\\
A2111	& 1.18	& $6.8^{+0.9}_{-0.5}$	& $10.7\pm0.4$	& $0.23^{+0.15}_{-0.15}$	& $7.44^{+0.10}_{-0.05}$	& & $6.6^{+0.9}_{-0.7}$	& $7.7\pm0.3$	& $0.20^{+0.19}_{-0.18}$	& $49.0^{+6.7}_{-5.1}$\\
A2125	& 0.77	& $2.6^{+0.1}_{-0.1}$	& $1.7\pm0.1$	& $0.21^{+0.08}_{-0.07}$	& $2.19^{+0.03}_{-0.02}$	& & $2.5^{+0.2}_{-0.2}$	& $1.5\pm0.1$	& $0.20^{+0.08}_{-0.07}$	& $5.4^{+0.4}_{-0.4}$\\
A2163	& 1.86	& $15.5^{+0.9}_{-0.9}$	& $87.8\pm1.3$	& $0.43^{+0.09}_{-0.09}$	& $32.45^{+0.18}_{-0.24}$	& & $15.5^{+1.2}_{-1.2}$	& $54.1\pm1.2$	& $0.48^{+0.13}_{-0.13}$	& $502.4^{+39.5}_{-39.1}$\\
MACSJ1621.3+3810	& 1.01	& $6.3^{+0.3}_{-0.3}$	& $17.8\pm0.6$	& $0.39^{+0.08}_{-0.08}$	& $6.08^{+0.11}_{-0.10}$	& & $6.1^{+0.6}_{-0.6}$	& $8.7\pm0.5$	& $0.26^{+0.12}_{-0.12}$	& $37.1^{+3.6}_{-3.5}$\\
MS1621.5+2640	& 1.03	& $6.1^{+0.6}_{-0.6}$	& $10.0\pm0.4$	& $0.58^{+0.17}_{-0.16}$	& $6.10^{+0.11}_{-0.13}$	& & $6.1^{+0.8}_{-0.7}$	& $7.8\pm0.4$	& $0.54^{+0.20}_{-0.19}$	& $37.1^{+4.7}_{-4.6}$\\
A2204	& 1.37	& $6.8^{+0.3}_{-0.2}$	& $38.0\pm0.4$	& $0.50^{+0.05}_{-0.05}$	& $11.50^{+0.09}_{-0.20}$	& & $7.4^{+0.6}_{-0.6}$	& $12.3\pm0.3$	& $0.32^{+0.12}_{-0.11}$	& $85.6^{+6.7}_{-6.8}$\\
A2218	& 1.21	& $6.8^{+0.3}_{-0.2}$	& $13.1\pm0.1$	& $0.31^{+0.05}_{-0.04}$	& $7.78^{+0.02}_{-0.07}$	& & $6.3^{+0.2}_{-0.2}$	& $8.3\pm0.1$	& $0.35^{+0.06}_{-0.06}$	& $49.3^{+1.9}_{-2.0}$\\
CLJ1641+4001	& 0.74	& $3.0^{+0.4}_{-0.4}$	& $2.6\pm0.4$	& $0.89^{+0.49}_{-0.37}$	& $2.06^{+0.06}_{-0.06}$	& & $2.9^{+0.3}_{-0.7}$	& $1.7\pm0.4$	& $0.41^{+0.48}_{-0.33}$	& $5.9^{+0.7}_{-1.4}$\\
RXJ1701+6414	& 0.93	& $5.2^{+0.6}_{-0.4}$	& $7.5\pm0.4$	& $0.60^{+0.18}_{-0.17}$	& $4.21^{+0.08}_{-0.06}$	& & $6.0^{+1.0}_{-0.9}$	& $5.3\pm0.3$	& $0.54^{+0.26}_{-0.23}$	& $25.4^{+4.1}_{-3.7}$\\
RXJ1716.9+6708	& 0.78	& $6.5^{+1.1}_{-0.9}$	& $12.0\pm0.8$	& $0.57^{+0.25}_{-0.23}$	& $4.08^{+0.11}_{-0.13}$	& & $5.4^{+1.1}_{-0.9}$	& $7.8\pm0.7$	& $0.74^{+0.38}_{-0.31}$	& $22.1^{+4.6}_{-3.6}$\\
A2259	& 1.09	& $5.6^{+0.4}_{-0.4}$	& $8.5\pm0.2$	& $0.28^{+0.11}_{-0.10}$	& $5.29^{+0.08}_{-0.07}$	& & $5.2^{+0.6}_{-0.4}$	& $5.2\pm0.2$	& $0.41^{+0.17}_{-0.16}$	& $27.7^{+3.3}_{-2.1}$\\
RXJ1720.1+2638	& 1.24	& $6.1^{+0.1}_{-0.1}$	& $19.8\pm0.2$	& $0.48^{+0.03}_{-0.03}$	& $7.46^{+0.31}_{-0.01}$	& & $7.2^{+0.4}_{-0.4}$	& $7.8\pm0.2$	& $0.45^{+0.07}_{-0.07}$	& $53.5^{+3.5}_{-2.8}$\\
MACSJ1720.2+3536	& 1.16	& $6.1^{+0.4}_{-0.4}$	& $24.0\pm0.6$	& $0.35^{+0.09}_{-0.09}$	& $7.92^{+0.18}_{-0.13}$	& & $7.8^{+1.0}_{-1.0}$	& $11.2\pm0.5$	& $0.44^{+0.20}_{-0.18}$	& $61.7^{+8.2}_{-7.8}$\\
A2261	& 1.31	& $7.2^{+0.3}_{-0.3}$	& $26.6\pm0.3$	& $0.39^{+0.05}_{-0.05}$	& $11.15^{+0.06}_{-0.08}$	& & $7.4^{+0.4}_{-0.4}$	& $13.5\pm0.2$	& $0.33^{+0.08}_{-0.08}$	& $82.2^{+4.6}_{-4.6}$\\
A2294	& 1.30	& $9.0^{+0.8}_{-0.7}$	& $14.6\pm0.4$	& $0.44^{+0.14}_{-0.14}$	& $8.64^{+0.11}_{-0.09}$	& & $8.6^{+1.2}_{-0.7}$	& $8.7\pm0.3$	& $0.23^{+0.19}_{-0.18}$	& $74.7^{+10.6}_{-6.1}$\\
MACSJ1824.3+4309	& 0.84	& $3.7^{+0.8}_{-0.6}$	& $5.2\pm0.6$	& $0.12^{+0.35}_{-0.12}$	& $3.46^{+0.24}_{-0.11}$	& & $3.9^{+1.0}_{-0.7}$	& $4.1\pm0.6$	& $0.40^{+0.53}_{-0.38}$	& $13.6^{+3.4}_{-2.4}$\\
MACSJ1931.8-2634	& 1.13	& $5.4^{+0.3}_{-0.2}$	& $43.1\pm0.8$	& $0.35^{+0.06}_{-0.06}$	& $9.31^{+0.15}_{-0.13}$	& & $5.8^{+0.6}_{-0.5}$	& $15.3\pm0.5$	& $0.11^{+0.10}_{-0.10}$	& $53.6^{+5.4}_{-5.0}$\\
RXJ2011.3-5725	& 0.85	& $3.8^{+0.2}_{-0.2}$	& $6.4\pm0.3$	& $0.59^{+0.14}_{-0.12}$	& $2.98^{+0.05}_{-0.05}$	& & $3.3^{+0.3}_{-0.3}$	& $3.0\pm0.2$	& $0.42^{+0.21}_{-0.18}$	& $9.9^{+0.9}_{-0.8}$\\
MS2053.7-0449	& 0.75	& $4.1^{+0.5}_{-0.4}$	& $4.4\pm0.3$	& $0.18^{+0.17}_{-0.16}$	& $2.68^{+0.04}_{-0.06}$	& & $4.2^{+0.9}_{-0.6}$	& $3.0\pm0.3$	& $0.10^{+0.21}_{-0.10}$	& $11.2^{+2.3}_{-1.6}$\\
MACSJ2129.4-0741	& 1.12	& $9.4^{+1.3}_{-1.0}$	& $36.2\pm1.3$	& $0.30^{+0.16}_{-0.16}$	& $10.58^{+0.23}_{-0.21}$	& & $8.4^{+1.3}_{-1.2}$	& $21.0\pm1.0$	& $0.40^{+0.22}_{-0.21}$	& $88.5^{+13.7}_{-13.1}$\\
RXJ2129.6+0005	& 1.20	& $5.6^{+0.3}_{-0.3}$	& $20.2\pm0.5$	& $0.50^{+0.10}_{-0.10}$	& $8.02^{+0.14}_{-0.14}$	& & $7.0^{+0.9}_{-0.7}$	& $9.7\pm0.4$	& $0.40^{+0.19}_{-0.18}$	& $56.3^{+7.2}_{-5.5}$\\
A2409	& 1.16	& $5.5^{+0.3}_{-0.2}$	& $10.6\pm0.2$	& $0.46^{+0.09}_{-0.08}$	& $6.42^{+0.06}_{-0.07}$	& & $5.7^{+0.4}_{-0.4}$	& $6.3\pm0.2$	& $0.32^{+0.12}_{-0.11}$	& $36.5^{+2.7}_{-2.5}$\\
MACSJ2228.5+2036	& 1.23	& $7.7^{+0.6}_{-0.6}$	& $29.9\pm0.7$	& $0.40^{+0.10}_{-0.10}$	& $12.58^{+0.20}_{-0.18}$	& & $7.5^{+0.8}_{-0.7}$	& $20.8\pm0.6$	& $0.41^{+0.13}_{-0.13}$	& $94.5^{+9.6}_{-9.4}$\\
MACSJ2229.7-2755	& 1.06	& $4.2^{+0.3}_{-0.2}$	& $19.5\pm0.6$	& $0.61^{+0.11}_{-0.10}$	& $5.86^{+0.11}_{-0.15}$	& & $5.9^{+0.8}_{-0.8}$	& $6.9\pm0.3$	& $0.25^{+0.18}_{-0.17}$	& $34.6^{+4.8}_{-4.8}$\\
MACSJ2245.0+2637	& 1.04	& $5.1^{+0.3}_{-0.3}$	& $16.5\pm0.5$	& $0.68^{+0.13}_{-0.12}$	& $5.62^{+0.12}_{-0.09}$	& & $5.3^{+0.7}_{-0.5}$	& $6.9\pm0.4$	& $0.62^{+0.23}_{-0.21}$	& $29.6^{+3.9}_{-2.7}$\\
RXJ2247+0337	& 0.62	& $2.2^{+0.5}_{-0.4}$	& $0.3\pm0.1$	& $0.48^{+0.49}_{-0.28}$	& $0.60^{+0.02}_{-0.03}$	& & $2.2^{+0.9}_{-0.5}$	& $0.2\pm0.1$	& $0.35^{+0.81}_{-0.28}$	& $1.3^{+0.6}_{-0.3}$\\
AS1063	& 1.42	& $11.1^{+0.8}_{-0.9}$	& $43.2\pm0.6$	& $0.08^{+0.06}_{-0.06}$	& $12.11^{+0.08}_{-0.09}$	& & $10.4^{+1.4}_{-0.9}$	& $16.4\pm0.4$	& $0.13^{+0.13}_{-0.13}$	& $126.5^{+16.7}_{-10.9}$\\
CLJ2302.8+0844	& 0.73	& $5.0^{+1.0}_{-0.7}$	& $4.7\pm0.3$	& $0.03^{+0.17}_{-0.03}$	& $2.61^{+0.06}_{-0.05}$	& & $4.9^{+1.4}_{-1.0}$	& $3.3\pm0.3$	& $0.04^{+0.23}_{-0.04}$	& $12.7^{+3.6}_{-2.5}$\\
A2631	& 1.22	& $6.5^{+0.5}_{-0.5}$	& $17.9\pm0.6$	& $0.44^{+0.13}_{-0.13}$	& $9.91^{+0.15}_{-0.21}$	& & $6.5^{+0.6}_{-0.6}$	& $13.5\pm0.5$	& $0.43^{+0.16}_{-0.15}$	& $64.1^{+6.5}_{-5.7}$\\
\enddata
\end{deluxetable}

\clearpage












\end{document}